\documentclass[ %
3p, %
]{elsarticle}

\usepackage{graphicx}
\usepackage{amssymb,amsmath}
\usepackage{hyperref}
\usepackage{bm}


\biboptions{sort&compress} 

\newcommand{\V}[1]{{\bm #1}}
\newcommand{\T}[1]{{\tt #1}} 
\newcommand\Alfven{Alfv\'en }

\newcommand{\xhat}{\hat{\bm{x}}}
\newcommand{\yhat}{\hat{\bm{y}}}
\newcommand{\zhat}{\hat{\bm{z}}}
\newcommand{\Xhat}{\hat{\bm{X}}_{s}}
\newcommand{\Yhat}{\hat{\bm{Y}}_{s}}
\newcommand{\Zhat}{\hat{\bm{Z}}_{s}}

\newcommand{\M}{\mathcal M}

\newcommand{\sgn}{\mathop{\mathrm{sgn}}\nolimits}

\newcommand{\kpar}{k_{\parallel}}

\newcommand{\Bg}{B_0}

\newcommand{\p}{\partial}
\newcommand{\EB}{\bm{E} \times \bm{B}}
\newcommand{\imag}{\mathrm{i}}
\newcommand{\diff}{\mathrm{d}}

\newcommand{\pdf}[2]{\frac{\partial #1}{\partial #2}}

\usepackage{color}

\journal{Journal of Computational Physics}
 
\usepackage{amsmath}	
\begin{document}

\begin{frontmatter}

\title{\T{AstroGK}: Astrophysical Gyrokinetics Code}

\author[umd,wpi]{Ryusuke~Numata\corref{cor_numata}}
\ead{rnumata@umd.edu}
\author[ui]{Gregory~G.~Howes}
\author[umd]{Tomoya~Tatsuno}
\author[ox,cul]{Michael~Barnes}
\author[umd]{William~Dorland}
\address[umd]{CSCAMM \& IREAP, University of Maryland, College Park, MD
 20742, USA.}
\address[wpi]{Wolfgang Pauli Institute, University of Vienna, A-1090
 Vienna, Austria}
\address[ui]{Department of Physics and Astronomy, University of Iowa,
 Iowa City, IA 52242, USA.}
\address[ox]{Rudolf Peierls Centre for Theoretical Physics, University
 of Oxford, Oxford OX1 3NP, UK.}
 \address[cul]{Euratom/CCFE Fusion Association, Culham Science Centre,
 Abingdon OX14 3DB, UK.}
\cortext[cor_numata]{Corresponding author at: CSCAMM \& IREAP,
 University of Maryland, College Park, MD 20742, USA. Tel.: +1 301 405
 1608; fax: +1 301 405 1678.}

\begin{abstract}
 The gyrokinetic simulation code \T{AstroGK} is developed to study
 fundamental aspects of kinetic plasmas and for applications mainly to
 astrophysical problems. \T{AstroGK} is an Eulerian slab code that
 solves the electromagnetic gyrokinetic-Maxwell equations in
 five-dimensional phase space, and is derived from the existing
 gyrokinetics code \T{GS2} by removing magnetic geometry
 effects. Algorithms used in the code are described.  The code is
 benchmarked using linear and nonlinear problems. Serial and parallel
 performance scalings are also presented.
\end{abstract}

\begin{keyword}
 Gyrokinetic simulation \sep Eulerian \sep Numerical Methods 
 \PACS 52.30.Gz 
 \sep 52.65.Tt 
 \sep 94.05.-a 
 \sep 95.30.Qd 
\end{keyword}

\end{frontmatter}




\section{Introduction}
\label{sec:intro}

Gyrokinetics is a limit of kinetic theory that describes the
low-frequency dynamics of weakly collisional plasmas in a mean
magnetic field. Developed for the study of magnetically confined
fusion plasmas, it has proven to be a valuable tool in understanding
the dynamics of drift-wave turbulence, a key cause of the enhanced
transport measured in modern fusion experiments that leads to poor
device performance.  Tremendous theoretical, computational, and 
experimental efforts have been devoted to this problem, with steady 
progress over three decades.

It has been recently recognized that the gyrokinetic approach is also
well-suited to the study of astrophysical plasmas, including galaxy
clusters, accretion disks around compact objects, the interstellar
medium, and the solar corona and solar wind~\cite{HowesCowleyDorland_06,HowesCowleyDorland_08,SchekochihinCowleyDorland_09}.
Taking advantage of the knowledge and computational techniques
developed in the simulation of turbulence in fusion plasmas, 
\T{AstroGK}~\cite{sourceforge_gyrokinetics}, a gyrokinetic simulation
code, is developed specifically for the study of astrophysical problems.
In this paper, we describe the algorithms employed in
\T{AstroGK} and present verification tests and performance results.

Gyrokinetics describes the low-frequency fluctuations of magnetized
plasmas by exploiting the timescale separation between the
low-frequency dynamics of interest and the fast cyclotron motion of
particles, $\omega\ll\Omega$, where $\omega$ is the typical frequency
of fluctuations and $\Omega$ is the cyclotron frequency. By averaging
the kinetic Vlasov--Landau (or Boltzmann) equation and Maxwell's
equations over 
the fast cyclotron motion, a self-consistent gyrokinetic-Maxwell
(GK-M) system is defined in five-dimensional phase space. This system
orders out the fast MHD waves and the cyclotron resonance, but retains
finite Larmor radius (FLR) effects and collisionless wave-particle
interactions via the Landau resonance~\cite{Landau_46,Barnes_66}.

The theoretical foundation of gyrokinetics has been developed
extensively over the past four decades~\cite{RutherfordFrieman_68,TaylorHastie_68,Catto_78,AntonsenLane_80,CattoTangBaldwin_81,FriemanChen_82,DubinKrommesOberman_83,HahmLeeBrizard_88,Brizard_92,Sugama_00,HowesCowleyDorland_06,BrizardHahm_07},
and gyrokinetics is now broadly employed for numerical studies of 
turbulence driven by microinstabilities in laboratory plasmas.  The
reduction of the phase-space dimensionality and the relaxed timestep
constraints under the gyrokinetic approximation have made possible
gyrokinetic simulations of fusion devices, yet it is still
computationally demanding. The simulation of the kinetic dynamics of
magnetized plasmas in both laboratory and astrophysical settings
presents a challenge to the scientific community requiring the most
advanced computing technology available.

A number of computational codes for gyrokinetics have been developed
and are actively refined worldwide by the fusion community~\cite{KotschenreutherRewoldtTang_95,LinHahmLee_98,DorlandJenkoKotschenreuther_00,JenkoDorlandKotschenreuther_00,CandyWaltz_03,ParkerChenWan_04,WatanabeSugama_04,GrandgirardSarazinAngelino_07,IdomuraIdaKano_09}.
These codes are typically classified by the following characteristics:
Eulerian continuum vs. Lagrangian particle (PIC), local flux tube
vs.~global toroidal, $\delta f$ vs.~ full$-f$, and electromagnetic
vs.~electrostatic. \T{AstroGK} has been derived from the Eulerian
continuum, local flux tube, $\delta f$, electromagnetic code
\T{GS2}~\cite{KotschenreutherRewoldtTang_95,DorlandJenkoKotschenreuther_00}.
\T{GS2} was designed to simulate plasma dynamics in fusion devices
where the magnetic geometry plays a central role. The handling in the
code of the toroidal magnetic geometry and of particle trapping
requires a rather complicated numerical implementation.  In contrast,
the study of the fundamental properties of kinetic plasmas for
application to astrophysical situations demands the simulation of the
dynamics at scales of order the particle Larmor radii on which the
magnetic field is well approximated as straight or gently curved, so
the coding to handle complicated magnetic geometries is
unnecessary. Therefore 
\T{AstroGK} was created by stripping \T{GS2} of the cumbersome coding
necessary to describe the magnetic geometry effects, leading to a
simplified, faster code ideally suited to study weakly collisional
astrophysical plasmas.  In addition to this, we believe the code is
also an ideal testbed for new ideas of additional physics,
diagnostics, numerical algorithms, and optimizations which may be exported
back to
\T{GS2}.

\T{AstroGK} has already proven its usefulness in a number of studies.
For example, it has produced the first kinetic simulations of
turbulence describing the transition from \Alfven to kinetic \Alfven
wave turbulence at the scale of the ion Larmor radius in an attempt to
understand solar wind turbulence~\cite{HowesDorlandCowley_08}, revealed
nonlinear phase-mixing properties of
turbulence~\cite{TatsunoDorlandSchekochihin_09}, enabled the study of
the statistical properties of phase-space structures of plasma
turbulence~\cite{TatsunoBarnesCowley_10}, explored the transition from
collisional to collisionless tearing
instabilities~\cite{NumataDorlandHowes_10}, and described the
Alfv\'en wave dynamics in the LAPD
experiment~\cite{NielsonHowesTatsuno_10}.

This paper is organized as follows: In Section~\ref{sec:gkeqns}, the set
of GK-M equations solved in the code is
given. Section~\ref{sec:algorithm} describes algorithms employed in the
code, including the velocity-space discretization and integration, the
finite-difference formalism of the GK-M system along the mean field
direction and its solution by the special technique developed by
Kotschenreuther et al.~\cite{KotschenreutherRewoldtTang_95}, the 
parallelization scheme, and some additional features of the
code. Section~\ref{sec:examples} presents tests of 
\T{AstroGK}, ranging from linear electrostatic problems to
nonlinear fully electromagnetic problems. Comparisons to analytic
solutions of the linear problems enable the verification of the code and
nonlinear examples show potential ability to apply the code to
complicated problems. Serial and parallel performance measurements of
the code on cutting edge supercomputers are given in
Section~\ref{sec:performance}. In Section~\ref{sec:summary}, we present
a summary of the paper.


\section{Gyrokinetic-Maxwell equations}
\label{sec:gkeqns}

In this section, we present the gyrokinetic-Maxwell (GK-M) system of
equations solved in \T{AstroGK}. For notational simplicity, we summarize
all the symbols and their definitions in \ref{sec:symbols}.

We first assume that scale separations in space and time are well
satisfied such that small fluctuations are locally embedded in a
background plasma which is slowly varying spatially and temporally.
We consider a temporally constant mean magnetic field
$\bm{B}_{0}=B_{0}\hat{\bm{b}}_0$. The mean field is pointing almost
parallel to $\hat{\bm{z}}$, but is allowed to have curvature
$\kappa=|(\hat{\bm{b}}_{0}\cdot\nabla)\hat{\bm{b}}_{0}|$. We also assume
the amplitude of the mean field is constant along its direction,
$\hat{\bm{b}}_{0}\cdot\nabla B_{0}=0$, but has finite gradient
perpendicular to its direction, $\hat{\bm{b}}_{0}\times\nabla
B_{0}\neq0$. Under this assumption, particle trapping do not take
place. In the presence of a mean magnetic field, we can adopt 
the gyrokinetic ordering and average over the fast cyclotron motion to
reduce the Vlasov--Maxwell equations to the GK-M equations; see Howes
et al.~\cite{HowesCowleyDorland_06} and Schekochihin et 
al.~\cite{SchekochihinCowleyDorland_09} for derivations of these
equations expressly intended for the study of astrophysical plasmas.

Under the gyrokinetic ordering, the distribution function of particles
up to the first order is given by
\begin{equation}
 f_{s} = \left(1-\frac{q_{s}\phi}{T_{0s}}
	  + \frac{\bm{v}\times\zhat}{\Omega_{s}} \cdot\nabla_{\perp}
	 \right) f_{0s} + h_{s},
  \label{eq:decompose_distribution_func}
\end{equation}
where $f_{0s} = n_{0s}/(\sqrt{\pi}v_{\mathrm{th},s})^{3}
\exp(-v^{2}/v_{\mathrm{th},s}^{2})$ is the zeroth-order, equilibrium
Maxwellian distribution function. The first-order part of the
distribution function is composed of the Boltzmann response term, a
term due to gradients of the equilibrium,
and the gyro-center distribution function $h_{s}$ defined in the
gyro-center coordinate $(\bm{R}_s,\bm{V}_{s})$.\footnote{
$\bm{V}_{s}=\bm{v}$, and, normally, is not distinguished from
$\bm{v}$. See
\ref{sec:coordinate}.}  Upon averaging over the gyro-phase, the
gyrokinetic equation evolves
$h_{s}=h_{s}(X_{s},Y_{s},Z_{s},V_{\parallel,s},V_{\perp,s},t)$:
\begin{equation}
\pdf{h_{s}}{t}
 +  V_{\parallel,s} \frac{\p h_{s}}{\p Z_{s}}
 + \V{v}_{\mathrm{D},s} \cdot
 \left(\pdf{h_{s}}{\bm{R}_{s}}+\Xhat\pdf{f_{0s}}{X_{s}}\right)
 = \frac{q_{s} f_{0s}}{T_{0s}} 
 \pdf{\langle \chi \rangle_{\V{R}_{s}}}{t}
 + C(h_{s}),
\label{eq:gkeq}
\end{equation}
where parallel and perpendicular subscripts refer to directions with
respect to the mean magnetic field. The perpendicular drift velocity is
given by 
\begin{equation}
 \V{v}_{\mathrm{D},s} =
  -\Yhat \frac{V_{\perp,s}^{2}}{2\Omega_{s}} L_{B_{0}}^{-1} 
  - \Yhat \frac{V_{\parallel,s}^{2}}{\Omega_{s}} \kappa
  - \pdf{\langle \chi \rangle_{\bm{R}_s}}{\bm{R}_{s}} \times
  \frac{\Zhat}{B_{0}},
  \label{eq:drift_velocity}
\end{equation}
where the terms correspond, from left to right, to the gradient-$B$
drift ($L_{\Bg}^{-1}\equiv-\p \ln \Bg/\p X_{s}$), the curvature drift,
and a {\it nonlinear} drift.\footnote{
$\phi$, $V_{\parallel}A_{\parallel}$, and
$\bm{V}_{\perp}\cdot\bm{A}_{\perp}$ terms in $\chi$ yield the $\EB$
drift, a parallel streaming along the perturbed magnetic field, and the
$\nabla B$ drift due to $\delta B_{\parallel}$, respectively.
} We have taken the direction of the gradients of
$\Bg$ and $f_{0s}$ and the curvature in the $-\Xhat$ direction.
The gyrokinetic potential is given by
$\chi=\phi-\V{v}\cdot\V{A}$, and the linear collision term is
represented by $C(h_{s})$. The angle bracket $\langle \;\cdot\;
\rangle_{\bm{R}_{s}}$ denotes the gyro-average at fixed gyro-center
coordinate $\bm{R}_{s}$:
\begin{equation}
 \langle F(\bm{r}) \rangle_{\bm{R}_{s}}
  = \frac{1}{2\pi} \oint
  F\left(\bm{R}_{s}+\frac{\bm{V}_{s}\times\Zhat}{\Omega_{s}}\right)
  \diff \Theta_{s},
\end{equation}
where $\bm{V}_{s}=(V_{\perp,s},V_{\parallel,s},\Theta_{s})$. (The
gyro-average at fixed particle coordinate $\langle\;\cdot\;
\rangle_{\bm{r}}$ can also be defined by switching roles of $\bm{r}$ and
$\bm{R}_{s}$.)

In the GK-M system, the electromagnetic fields are specified by the
three scalar functions $\phi(\bm{r},t)$, $A_{\parallel}(\bm{r},t)$,
and $\delta B_{\parallel}(\bm{r},t)$
\footnote{$\delta
B_{\parallel}=(\nabla_{\perp}\times\bm{A}_{\perp})_{z}$. We use the
Coulomb gauge, which leads to $\nabla_{\perp}\cdot\bm{A}_{\perp}=0$ with
the ordering. Then, we can write $\bm{A}_{\perp}=\nabla_{\perp}\varsigma
\times \zhat$, and $\delta B_{\parallel}=-\nabla_{\perp}^{2}\varsigma$ in
terms of a single scalar function $\varsigma$.
} according to:
\begin{align}
 \bm{B} = & \nabla_{\perp} A_{\parallel} \times \zhat
 + \delta B_{\parallel} \zhat, &
 \bm{E} = & - \nabla \phi - \pdf{\bm{A}}{t}.
\end{align}
Maxwell's equations in the gyrokinetic limit reduce to the
quasi-neutrality condition, and the parallel and perpendicular
components of Amp\`ere's law:
\begin{align}
 &
 \sum_{s} 
 \left[ - \frac{q_{s}^{2} n_{0s}}{T_{0s}} \phi 
 + q_{s} \int \langle h_s \rangle_{\V{r}} \diff \V{v} \right] = 0,
 \label{eq:quasi_neutrality_physicalunit}
 \\
 &
 - \nabla^{2}_{\perp} A_\parallel
 = \mu_{0} \sum_{s} q_{s} 
 \int \langle V_{\parallel,s} h_{s} \rangle_{\V{r}} \diff \V{v},
 \label{eq:ampere_para_physicalunit}
 \\
 & B_{0} \nabla_{\perp} \delta B_{\parallel}
 = - \mu_{0} \nabla_{\perp} \cdot \sum_{s} q_{s}
 \int \langle m\V{V}_{\perp,s}\V{V}_{\perp,s} h_{s} \rangle_{\V{r}} \diff \V{v}.
 \label{eq:ampere_perp_physicaunit}
\end{align}
To lowest order, the quasi-neutrality condition and the perpendicular
Amp\`ere's law imply the constraints on the background plasma of
quasi-neutrality and total pressure balance:
\begin{align}
 \sum_{s} n_{0s} q_{s} = & 0, &
 \pdf{}{X_{s}} \left(\frac{B_{0}^{2}}{2\mu_{0}} +
	       \sum_{s} n_{0s}T_{0s}\right) = & 0.
 \label{eq:equilibrium_quasi_neutrality_pressure_balance_physicalunit}
\end{align}

It is advantageous, both analytically and in the code, to
Fourier transform the equations only in the plane perpendicular to the
mean magnetic field. The distribution function and the fields are 
decomposed as
\begin{align}
 h_{s}(X_{s},Y_{s},Z_{s},V_{\parallel,s},V_{\perp,s},t) = &
 \sum_{\bm{k_{\perp}}}
 h_{s,\bm{k}_{\perp}}(Z_{s},V_{\parallel,s},V_{\perp,s},t)
 e^{\imag \bm{k}_{\perp}\cdot\bm{R}_{s}}, 
 \label{eq:fourier_transform_dist}\\
 \phi(x,y,z,t) = & \sum_{\bm{k}_{\perp}}
 \phi_{\bm{k}_{\perp}}(z,t) e^{\imag \bm{k}_{\perp}\cdot\bm{r}}
 \label{eq:fourier_transform_field}
\end{align}
with $\bm{k}_{\perp}=(k_{x},k_{y},0)$. The virtue of expressing
variables in Fourier space is that the gyro-averaging operation
becomes a multiplication by a Bessel function. For example, the
gyro-average of the gyrokinetic potential is given by
\begin{equation}
 \langle \chi \rangle_{\bm{R}_{s}}
  = \sum_{\bm{k}_{\perp}}
  \left[
   J_{0}(\alpha_{s}) \left(\phi_{\bm{k}_{\perp}} - V_{\parallel,s}
		      A_{\parallel,\bm{k}_{\perp}}\right)
   + \frac{T_{0s}}{q_{s}}
   \frac{2V_{\perp,s}^{2}}{v_{\mathrm{th},s}^{2}}
    \frac{J_{1}(\alpha_{s})}{\alpha_{s}}
    \frac{\delta B_{\parallel,\bm{k}_{\perp}}}{B_{0}}
  \right]
  e^{\imag \bm{k}_{\perp}\cdot\bm{R}_{s}},
\end{equation}
where $J_{n}$ is the Bessel function of the first kind with the argument
$\alpha_{s}=k_{\perp}V_{\perp,s}/\Omega_{s}$, taking $k_{\perp}=|\bm{k}_{\perp}|$. The Fourier coefficients of the fields
are now functions of $Z_{s}$ and $t$, for example 
$\phi_{\bm{k}_{\perp}}=\phi_{\bm{k}_{\perp}}(Z_{s},t)$.

In the large-scale limit $k_\perp \rho_{\mathrm{i}} \ll 1$, Alfv\'enic
perturbations have a gyro-center distribution function $h_s$ that is
largely canceled by the Boltzmann response term $(q_s\phi/T_{0s})f_{0s}$
(see, for example,  Section~5 of~\cite{SchekochihinCowleyDorland_09}). To
avoid numerical error 
arising from this near cancellation, a complementary distribution
function $g_{s}$ is introduced, given by
\begin{equation}
 g_{s}= h_{s} - \frac{q_s f_{0s}}{T_{0s}}
  \left\langle \phi - \bm{v}_{\perp}\cdot\bm{A}_{\perp} \right\rangle_{\V{R}_s}.
\label{eq:define_g}
\end{equation}
After normalization described in \ref{sec:normalization}, we finally
obtain a normalized version of the GK-M equations as follows.
The normalized gyrokinetic equation is
\begin{multline}
 \pdf{g_{\bm{k}_{\perp},s}}{t} +
 \sqrt{\frac{T_{0s}}{m_{s}}} V_{\parallel,s}
 \pdf{h_{\bm{k}_{\perp},s}}{Z_{s}}
 + \frac{1}{2}
 {\mathcal F}
 \left(\left\{ \langle \chi \rangle_{\bm{R}_{s}}, h_{s} \right\}\right)
 \\
 + \frac{\imag k_{y}}{2}
 \left[
 -\frac{T_{0s}}{q_{s}}
 \left(
 V_{\perp,s}^{2} L_{\Bg}^{-1} 
 + 2 V_{\parallel,s}^{2} \kappa 
 \right) h_{\bm{k}_{\perp},s}
 + 
 \left(L_{n_{0s}}^{-1}+\left(V_{s}^{2}-\frac{3}{2}\right)L_{T_{0s}}^{-1}\right) 
 \langle \chi \rangle_{\bm{R}_{s},\bm{k}_{\perp}}
 \right]
 \\
 = - \frac{q_{s}}{\sqrt{m_{s}T_{0s}}} V_{\parallel,s} J_{0}(\alpha_{s})
 \pdf{A_{\parallel,\bm{k}_{\perp}}}{t}
 + C_{\bm{k}_{\perp}}(h_{\bm{k}_{\perp},s})
 \label{eq:gyrokinetic_eq_normalized},
\end{multline}
where $V_{s}^{2}=V_{\perp,s}^{2}+V_{\parallel,s}^{2}$ and the normalized
gyrokinetic potential is
\begin{align}
 \langle \chi \rangle_{\bm{R}_{s}} = &
 \sum_{\bm{k}_{\perp}}
 \left[
 J_{0}(\alpha_{s}) \phi_{\bm{k}_{\perp}}
 -
 \sqrt{\frac{T_{0s}}{m_{s}}}V_{\parallel,s} J_{0}(\alpha_{s})
 A_{\parallel,\bm{k}_{\perp}}
 + \frac{T_{0s}}{q_{s}} 2V_{\perp,s}^{2}
 \frac{J_{1}(\alpha_{s})}{\alpha_{s}}
 \delta B_{\parallel,\bm{k}_{\perp}}
 \right] e^{\imag \bm{k}_{\perp}\cdot\bm{R}_{s}}.
\end{align}
The nonlinear term, given by the Poisson bracket
$\left\{a,b\right\}=(\p a/\p X_{s})(\p b/\p Y_{s})-(\p b/\p X_{s})(\p
a/\p Y_{s})$, is evaluated in real space and then transformed into
Fourier space (denoted by ${\mathcal F}$). The terms in the gyrokinetic
equation due to gradients of the background plasma, given in square  
brackets,  contribute
at the same order as the other terms, and are characterized
by the parameters $L_{\Bg}^{-1}$, $L_{n_{0s}}^{-1}\equiv-\p \ln
n_{0s}/\p X_{s}$, $L_{T_{0s}}^{-1}\equiv-\p \ln T_{0s}/\p X_{s}$, and
the curvature $\kappa$.
The normalized Maxwell's equations are
\begin{align}
 \phi_{\bm{k}_{\perp}}
 \sum_{s}
 \frac{n_{0s}q_{s}^{2}}{T_{0s}}\left(1-\Gamma_{0s}\right)
 - \delta B_{\parallel,\bm{k}_{\perp}}
 \sum_{s} q_{s}n_{0s} \Gamma_{1s}
 = & \sum_{s} q_{s} {\mathcal M}^{(0)}(g_{\bm{k}_{\perp},s}), 
 \label{eq:quasi_neutrality_normalized}\\
 \frac{k_{\perp}^{2}}{2\beta_{0}} A_{\parallel,\bm{k}_{\perp}}
 = & \sum_{s} q_{s} n_{0s} {\mathcal
 M}^{(1)}(g_{\bm{k}_{\perp},s}),
 \label{eq:ampparallel_normalized}\\
 \phi_{\bm{k}_{\perp}}
 \sum_{s}
 q_{s} n_{0s} \Gamma_{1s}
 + 
 \delta B_{\parallel,\bm{k}_{\perp}}
 \left(\frac{2}{\beta_{0}} +
 \sum_{s} n_{0s}T_{0s}
 \Gamma_{2s}
 \right)
 = & - \sum_{s} {\mathcal M}^{(2)}(g_{\bm{k}_{\perp},s})
 \label{eq:ampperp_normalized}.
\end{align}
The plasma beta of the  reference species, defined by the ratio of the
thermal pressure of the reference species to the magnetic pressure of
the mean magnetic  field, is given by $\beta_{0}$.
The operators ${\mathcal M}^{(n)}$ to take the $n$th order moment of
the distribution function are:
\begin{align}
 {\mathcal M}^{(0)} (g_{\bm{k}_{\perp},s})= & n_{0s} \int g_{\bm{k}_{\perp},s}
 J_{0}(\alpha_{s}) \frac{e^{-v_{s}^{2}}}{\pi^{3/2}} \diff \bm{v}_{s},
 \label{eq:zero_moment_normalized}\\
 {\mathcal M}^{(1)}(g_{\bm{k}_{\perp},s}) = &
 \sqrt{\frac{T_{0s}}{m_{s}}} \int  g_{\bm{k}_{\perp},s}
 v_{\parallel,s}J_{0}(\alpha_{s})
 \frac{e^{-v_s^{2}}}{\pi^{3/2}} \diff \bm{v}_{s},
 \label{eq:first_moment_normalized}\\
 {\mathcal M}^{(2)} (g_{\bm{k}_{\perp},s})= &
 n_{0s}T_{0s} \int g_{\bm{k}_{\perp},s}
 v_{\perp,s}^{2} \frac{2J_{1}(\alpha_{s})}{\alpha_{s}}
 \frac{e^{-v_{s}^{2}}}{\pi^{3/2}} \diff \bm{v}_{s}.
 \label{eq:second_moment_normalized}
\end{align}
The function $\Gamma_{ns}=\Gamma_{n}(b_{s})$ arises from the integration
over perpendicular velocity space of products of two Bessel functions.  For
$n=0,1,2$, it is given by
\begin{align}
 \Gamma_{0}(b_{s}) = & I_{0}(b_{s}) e^{-b_{s}}, &
 \Gamma_{1}(b_{s}) = &
 \left(I_{0}(b_{s})-I_{1}(b_{s})\right) e^{-b_{s}}, &
 \Gamma_{2}(b_{s}) = & 2\Gamma_{1}(b_{s})
\end{align}
where $I_{n}$ is the modified Bessel function of the first kind, and 
the argument is $b_{s}=(k_{\perp}\rho_{s})^{2}/2$~\cite{HowesCowleyDorland_06}.

The background plasma must also satisfy the normalized
quasi-neutrality and total pressure balance constraints:
\begin{align}
 \sum_{s} n_{0s} q_{s} = & 0, &
 L_{\Bg}^{-1} + \frac{\beta_{0}}{2} \sum_{s} n_{0s}T_{0s}
 \left(L_{n_{0s}}^{-1} + L_{T_{0s}}^{-1}\right) & = 0.
 \label{eq:background_quasi_neutrality_pressure_balance_normalized}
\end{align}

We defer description of the explicit form of the collision operator used in
the code to \ref{sec:collisions}, as it has a rather cumbersome form 
and is fully documented in~\cite{BarnesAbelDorland_09}.  We mention here
the basic properties of the operator. The operator is based on the
linearized Landau collision operator transformed into the gyro-center
coordinate. It has second-order velocity derivatives providing
diffusion in velocity space and {\it conserving terms} which include
integrations over velocity space. It is constructed to satisfy
Boltzmann's {\it H}-theorem and the conservation of particles,
momentum, and energy. It contains both like-species collisions and
inter-species collisions, but the inter-species collisions account
only for the collisions of electrons with one species of ions with
large mass. Note that the linearized collision operator for a given
species can be made independent of the first-order evolution of any other
species. The  theoretical basis of the collision operator is discussed
in detail in~\cite{AbelBarnesCowley_08}.


\section{Algorithm description}
\label{sec:algorithm}

This section describes the numerical algorithms used in \T{AstroGK} to
evolve the GK-M system of Eqs.~\eqref{eq:gyrokinetic_eq_normalized}, and
\eqref{eq:quasi_neutrality_normalized}--\eqref{eq:ampperp_normalized}.  
The gyrokinetic equation combined with
the field equations together comprise a set of integro-differential equations for the
evolution of the distribution function $g$ defined in the
five-dimensional phase space $(X,Y,Z,V_{\parallel},V_{\perp})$. In
this section, the species subscript $s$ is omitted unless
necessary. Periodic boundary conditions are assumed for the spatial
dimensions $(X,Y,Z)$, and the derivatives in the plane perpendicular
to the mean field, $(X,Y)$, are handled using a Fourier-spectral
method. Except for the nonlinear term, each Fourier mode is
independent of the others in the gyrokinetic equation, so we omit the
$\bm{k}_{\perp}$ subscript for simplicity.  In the numerical
implementation described here, because fields are calculated
separately from the gyrokinetic equations, the gyrokinetic equation
for each species is essentially independent of that for the other
species. The mean field parallel direction $Z$ and the time $t$ are
discretized by $Z_{i}=i \Delta Z$ ($i=0,\cdots,N_{Z}$) and
$t_{n}=\sum_{j=1}^{n} \Delta 
t_{j}$, where  $\Delta Z$ is fixed and $\Delta t_{j}$ may vary to satisfy
the Courant--Friedrichs--Lewy (CFL)
condition~\cite{CourantFriedrichsLewy_28} for the nonlinear
term. Velocity space is discretized with grid points chosen by Gaussian
quadrature rules for optimal integration, generating nonuniform meshes.

\subsection{Velocity-space integration}
\label{sec:vspace}

The velocity grid in \T{AstroGK} is specified by $(\lambda,E,\sigma)$,
where the pitch
angle\footnote{The pitch-angle parameter $\lambda$ is used for
historical reasons.  If the magnitude of the mean magnetic field
changes along its direction, particles may be trapped in magnetic
wells. Trapped and untrapped regions in velocity space are
conveniently described by the $\lambda$
coordinate~\cite{KotschenreutherRewoldtTang_95}. Since \T{AstroGK} does
not 
contain this physics, the use of $\lambda$ is not necessary, but is
inherited from \T{GS2}.  The pitch-angle variable
$\xi= v_\parallel/v$ could be used instead.} is
$\lambda=v_{\perp}^{2}/v^{2}$, the energy is
$E=v_{\perp}^{2}+v_{\parallel}^{2}$, and the sign of parallel velocity
is $\sigma=\sgn(v_{\parallel})$.  The velocity-space integral to
calculate moments is represented by an integration of some function
$F(\lambda,E,\sigma,\dots)$ multiplied by the Maxwellian $e^{-E}$:
\begin{equation}
 \int F e^{-E} \diff \bm{v} = \pi \sum_{\sigma=\pm1}
  \int_{0}^{\infty} e^{-E}\sqrt{E}\diff E
  \int_{0}^{1} \frac{\diff \lambda}{\sqrt{1-\lambda}}
  F(\lambda,E,\sigma,\dots).
  \label{eq:integration_in_lambda_E}
\end{equation}

Gaussian quadrature evaluates an integral of a function $F(x)$ with
weight $W(x)$ by
\begin{equation}
 \int_{a}^{b} W(x) F(x) \diff x \approx \sum_{j=1}^{N} w_{j}
  F(x_{j}),
  \label{eq:Gaussian_quadrature}
\end{equation}
where $x_{j}$ is the $j$th root of the $N$th order polynomial, and $w_{j}$
is the corresponding discretized weight. The weights for the
Gauss--Legendre and Gauss--Laguerre rules are given by
\begin{align}
 w_{j}^{\mathrm{Leg}} = &
 \frac{2}{\left(1-\left(x_{j}^{\mathrm{Leg}}\right)^{2}\right)
 \left(P_{N}'\left(x_{j}^{\mathrm{Leg}}\right)\right)^{2}}, 
 &
 w_{j}^{\mathrm{Lag}} = & \frac{x_{j}^{\mathrm{Lag}}}
 {(N+1)^{2}\left(L_{N+1}\left(x_{j}^{\mathrm{Lag}}\right)\right)^{2}},
 \label{eq:weigth_legendre_laguerre}
\end{align}
where $P_{N}$ and $L_{N}$ are the $N$th order Legendre and Laguerre
polynomials, respectively~\cite{AbramowitzStegun_72a}. The superscripts
`Leg' and `Lag' to $x_{j}$ and $w_{j}$ explicitly denote that they are
associated with the Gauss--Legendre and Gauss--Laguerre rules. The weight
function $W(x)$ and the integration range are chosen according to the
polynomial.

For the pitch-angle integration, the Gauss--Legendre quadrature ($a=-1$,
$b=1$, $W(x)=1$) is immediately applied by defining
$\xi=\sqrt{1-\lambda}$:
\begin{align}
 \int_{0}^{1} F(\lambda,\dots) \frac{\diff \lambda}{\sqrt{1-\lambda}}
  = & 2 \int_{0}^{1} F(1-\xi^{2},\dots) \diff \xi
 \nonumber 
 \\
  \approx & \sum_{j=1}^{N_{\lambda}}
  w_{j}^{\mathrm{Leg}}
  F\left(1-\left(\frac{x_{j}^{\mathrm{Leg}}+1}{2}\right)^{2},\dots\right). 
  \label{eq:lambda_integ_approx}
\end{align}
The integration range is changed using a linear transformation to fit
the range of the Gauss--Legendre rule.
$N_{\lambda}$ is the number of grid points describing the $\lambda$ grid
and is specified by the user input, \T{ngauss}:
$N_{\lambda}=2\times\T{ngauss}$.

A rather careful treatment of the energy integral is necessary because
there are singularities of the integrand at $E=0$ and $E= \infty$
which may prevent a simple approximation from achieving spectral
convergence. To avoid the problem, we split the integration range at
$E_{\mathrm{cut}}=v_{\mathrm{cut}}^{2}$ into a lower and an upper range
and change the variable to $v$ from $E$ for the lower range integration:
the Gauss--Legendre scheme is used for the lower range; and the
Gauss--Laguerre ($a=0$, $b=\infty$, $W(x)=e^{-x}$) is used for the
upper range. Therefore, the integration is approximated by
\begin{align}
 \int_{0}^{\infty} F(E,\dots) e^{-E}\sqrt{E}\diff E = &
 \int_{0}^{v_{\mathrm{cut}}} F(v^{2},\dots) e^{-v^{2}} 2v^{2} \diff v
  + \int_{E_{\mathrm{cut}}}^{\infty}
  F(E,\dots) e^{-E}\sqrt{E} \diff E
  \nonumber \\
 \approx & \frac{v_{\mathrm{cut}}}{2} \sum_{j=1}^{N_{E}^{-}}
 w_{j}^{\mathrm{Leg}} G_{1}
 \left(\frac{v_{\mathrm{cut}}}{2}\left(x_{j}^{\mathrm{Leg}}+1\right),\dots\right)
 \nonumber \\
 & + e^{-E_{\mathrm{cut}}}
 \sum_{j=1}^{N_{E}^{+}} w_{j}^{\mathrm{Lag}}
 G_{2}\left(x_{j}^{\mathrm{Lag}}+E_{\mathrm{cut}},\dots\right),
 \label{eq:E_integ_approx}
\end{align}
where
\begin{align}
 G_{1}(x) = & F(x^{2},\dots) e^{-x^{2}} 2x^{2}, &
 G_{2}(x) = & F(x,\dots) \sqrt{x},
 \label{eq:integrand_E_integ}
\end{align}
and the integration ranges are shifted appropriately.
We allow users to specify $v_{\mathrm{cut}}=\T{vcut}$,
$N_{E}^{-}=\T{nesub}$, $N_{E}^{+}=\T{nesup}$, and
$N_{E}=N_{E}^{+}+N_{E}^{-}=\T{negrid}$ in the code input. 

We also have another mode to evaluate the energy integral in the code
called the `\T{egrid} mode', whereas the above method is called the
`\T{vgrid} mode'. In the \T{egrid} mode, the energy integral is
calculated by the method suggested by Candy and
Waltz~\cite{CandyWaltz_03}, which is not exponentially accurate. We
note, 
however, that if $N_{E}$ is very small ($\lesssim 8$), we find empirically
that the \T{egrid} mode may give better results than the
\T{vgrid} mode. Therefore, the optimal choice of energy grid mode 
is governed by the simulation parameters.

Further discussion of our velocity-space coordinates can be found
in~\cite{BarnesDorlandTatsuno_10}.

\subsection{Time integration}
\label{sec:time_integ}

The gyrokinetic equation is symbolically denoted by
\begin{equation}
 \pdf{g}{t} = {\mathcal L}g + {\mathcal C}g + {\mathcal N}(g,g)
 \label{eq:gk_symbolic_time}
\end{equation}
where ${\mathcal L}$ is the linear term except the collision term,
${\mathcal C}$ is the collision term which is also linear, and
${\mathcal N}$ is the nonlinear term. We consider the time derivative
using first-order finite differentiation for the linear term, treating
the collision term by the implicit Euler method, and handling the
nonlinear term explicitly by the third-order Adams--Bashforth method
(AB3):
\begin{equation}
 \frac{g^{n+1}-g^{n}}{\Delta t} = 
 {\mathcal L} g^{n} r_{t} +{\mathcal L} g^{n+1}(1-r_{t}) +
 {\mathcal C} g^{n+1}
 + \frac{23}{12}{\mathcal N}^{n} -
  \frac{4}{3}{\mathcal N}^{n-1} +
  \frac{5}{12}{\mathcal N}^{n-2} + {\mathcal O}(\Delta t)
  \label{eq:gk_time_difference_1st}
\end{equation}
where ${\mathcal N}^{n}={\mathcal N}(g^{n},g^{n})$.  The
time-centering parameter $r_{t}$ (\T{fexp} in the code input) may be
chosen within the range $0\leq r_{t}\leq 1$, where $r_{t}=1$
($r_{t}=0$) represents a fully explicit (implicit) scheme. If
$r_{t}\leq1/2$, the scheme is stable for any $\Delta t$.  We mainly
use an implicit trapezoidal rule $r_{t}=1/2$ which is second-order
accurate and free from time step restrictions due to the linear
term. Hereafter, we fix $r_{t}=1/2$.

Note that Eq.~\eqref{eq:gk_time_difference_1st} is linear with
respect to $g^{n+1}$ due to the explicitness of the nonlinear term. We
then employ a Godunov splitting technique~\cite{Godunov_59},
which is first-order accurate in $\Delta t$, to separate the collision
term:
\begin{align}
 \frac{g^{(\ast)}-g^{n}}{\Delta t} = &
 {\mathcal L} \frac{g^{n}+g^{(\ast)}}{2}
 + \frac{23}{12}{\mathcal N}^{n} -
  \frac{4}{3}{\mathcal N}^{n-1} +
  \frac{5}{12}{\mathcal N}^{n-2},
 \label{eq:gk_time_difference_godunov_linearnonlinear}
 \\
 \frac{g^{n+1}-g^{(\ast)}}{\Delta t} = &
 {\mathcal C} g^{n+1}
 + {\mathcal O}(\Delta t),
 \label{eq:gk_time_difference_godunov_collision}
\end{align}
which greatly reduces the size of the matrix to be inverted. Solving for
$g^{n+1}$, we obtain:
\begin{multline}
 g^{n+1} = 
 \left(1-\Delta t{\mathcal C}\right)^{-1}
 \left(1-\frac{\Delta t}{2}{\mathcal L}\right)^{-1}
 \\
 \times \left[
 \left(1+\frac{\Delta t}{2} {\mathcal L}
 \right) g^{n} +
 \Delta t
 \left(
 \frac{23}{12}{\mathcal N}^{n} -
 \frac{4}{3}{\mathcal N}^{n-1} +
 \frac{5}{12}{\mathcal N}^{n-2}
 \right)
 \right]
 + {\mathcal O}(\Delta t^{2}).
 \label{eq:gk_time_difference_trapezoidal_godunov_linearnonlinear}
\end{multline}
The method is first-order accurate in time. The use of AB3 for the
nonlinear term is to make nonlinear runs stable. Note that the first
time step for the nonlinear term is evaluated by the (explicit) Euler
method (which is also first order in $\Delta t$), and the second timestep
is evaluated using the second-order Adams--Bashforth scheme (AB2).

A second-order accurate method may, in principle, be derived by applying
a higher-order scheme for the collision term as well, and by using a Strang
splitting~\cite{Strang_68} for the operator splitting. The first two steps
for the nonlinear term could also be computed using a higher-order
method.  These ideas are not implemented in the current version.

\subsection{Gyrokinetic solver}
\label{sec:gk_solver}

We now describe the implicit advance of the linear terms in the
gyrokinetic equation. Basically, the fields in the gyrokinetic
equation can be obtained by a separate procedure, and the gyrokinetic
equation becomes a differential equation with the given fields. Thus
the collisionless gyrokinetic equation is written as
\begin{equation}
 \pdf{g}{t} + a_{Z} \pdf{g}{Z} + a_{0} g = 
  \bm{b}_{t} \cdot \pdf{\bm{\Psi}}{t} + \bm{b}_{Z}\cdot\pdf{\bm{\Psi}}{Z}
  + \bm{b}_{0} \cdot \bm{\Psi} + S,
  \label{eq:gk_symbolic_linearnonlinear}
\end{equation}
where $\bm{\Psi}=(\phi,A_{\parallel},\delta B_{\parallel})$,
coefficients $a_{Z,0}$ and $\bm{b}_{t,Z,0}$ are functions of
$V_{\perp}$ and $V_{\parallel}$, and $S$ contains the AB3 nonlinear
term. The collision term is always consecutively applied after
\eqref{eq:gk_symbolic_linearnonlinear}, but separately
(Section~\ref{sec:alg_coll}).

We use a compact finite-difference method for evaluation of $\p/\p Z$ to
achieve up to second-order accuracy with keeping the matrix
bi-diagonal. To take into  
account global information, compact finite differencing schemes use the
derivatives at neighboring grids to evaluate the derivatives. A general
formula of the compact finite-difference scheme using two neighboring
grids is
\begin{equation}
 \frac{1}{2} \left(
	      \left.(1-r_{Z})\pdf{g}{Z}\right|_{i} + 
	      \left.(1+r_{Z})\pdf{g}{Z}\right|_{i+1}
	     \right)
 = \frac{g_{i+1}-g_{i}}{\Delta Z} + {\mathcal O}(\Delta Z),
 \label{eq:compact_finite_difference}
\end{equation}
for $V_{\parallel}>0$. ($V_{\parallel}$ is a coefficient of $\p g/\p
Z$. See \eqref{eq:gyrokinetic_eq_normalized}.) Information at $i-1$
instead of $i+1$ should be 
used for $V_{\parallel}<0$. (In the following discussions, we show 
equations for $V_{\parallel}>0$ only.) $0\leq r_{Z}\leq1$ is the
space-centering 
parameter specified by $\T{bakdif}$ in the code. We fix $r_{Z}=0$ in the
following discussion, and therefore second-order accuracy is achieved.

Combining the trapezoidal rule for the time derivatives with the compact
scheme for the space derivatives\footnote{
The combination of the trapezoidal rule and the second-order compact
finite differentiation was first suggested by Beam and
Warming~\cite{BeamWarming_76}. Note that the so-called Beam--Warming
scheme refers to a different scheme.
}
\begin{multline}
 \frac{g^{n+1}_{i+\frac{1}{2}}-g^{n}_{i+\frac{1}{2}}}{\Delta t}
  + a_{Z}
 \frac{g^{n+\frac{1}{2}}_{i+1}-g^{n+\frac{1}{2}}_{i}}{\Delta Z}
 + a_{0} g_{i+\frac{1}{2}}^{n+\frac{1}{2}}
 \\
 = \bm{b}_{t} \cdot
 \frac{\bm{\Psi}^{n+1}_{i+\frac{1}{2}}-\bm{\Psi}^{n}_{i+\frac{1}{2}}}{\Delta t}
 + \bm{b}_{Z} \cdot
 \frac{\bm{\Psi}^{n+\frac{1}{2}}_{i+1}-\bm{\Psi}^{n+\frac{1}{2}}_{i}}{\Delta Z}
 + \bm{b}_{0} \cdot \bm{\Psi}_{i+\frac{1}{2}}^{n+\frac{1}{2}}
 + S_{i+\frac{1}{2}}^{n}
 \label{eq:gk_linear_beam_warming}
\end{multline}
where $i+1/2$ denotes the average value of the variables at $i$ and $i+1$
grids, and similarly for $n+1/2$. Then, the gyrokinetic equation is cast into
the following symbolic form:
\begin{multline}
 A_{1} g_{i}^{n} + A_{2} g_{i+1}^{n} + 
 B_{1} g_{i}^{n+1} + B_{2} g_{i+1}^{n+1} = \\
 \bm{D}_{1}\cdot\bm{\Psi}_{i}^{n}
 + \bm{D}_{2}\cdot\bm{\Psi}_{i+1}^{n}
 + \bm{E}_{1}\cdot \bm{\Psi}_{i}^{n+1}
 + \bm{E}_{2}\cdot\bm{\Psi}_{i+1}^{n+1}
 + S_{i}^{n} + S_{i+1}^{n}.
 \label{eq:gk_linear_discrete_symbolic}
\end{multline}
 
\subsubsection{Kotschenreuther's Green's function approach}
\label{sec:alg_green}

Kotschenreuther et al. developed an efficient way to solve the gyrokinetic
equation by breaking the large matrix to be inverted into a number of
small matrices~\cite{KotschenreutherRewoldtTang_95}. In the method,
fields at the future timestep $n+1$ are obtained using a Green's
function formalism to decouple parts of the matrix related to
velocity-space integrations in the gyrokinetic equation. Consequently,
the matrix to be inverted becomes $N_{Z}\times N_{Z}$ at each velocity
grid for each species. (Due to the periodic boundary condition, $i=0,
N_{Z}$ modes are not independent.)
The actual scheme~\cite{Belli_06} that was originally
implemented in \T{GS2}, and has been inherited by \T{AstroGK}, differs
slightly from that described in the original paper by Kotschenreuther et
al.,
as described below.

Because \eqref{eq:gk_linear_discrete_symbolic} is linear with respect
to variables at timestep $n+1$, the solution to the equation may
consist of any linear combination of solutions to parts of the
equation.  Thus we may split the solution at timestep $n+1$ into two
pieces, $g^{n+1}=g^{(\mathrm{inh})}+g^{(\mathrm{h})}$, each satisfying
the following equations (here we ignore the spatial index $i$ in the
interest of clarity):
\begin{align}
 A g^{n} + B g^{(\mathrm{inh})} = &
 \bm{D}\cdot\bm{\Psi}^{n} + \bm{E}\cdot\bm{\Psi}^{n} + S^{n},
  \label{eq:gk_linear_discrete_symbolic_two1}
 \\
 B g^{(\mathrm{h})} = & \bm{E}\cdot\bm{\Psi}^{(\ast)},
  \label{eq:gk_linear_discrete_symbolic_two2}
\end{align}
where $\bm{\Psi}^{n+1}=\bm{\Psi}^{n}+\bm{\Psi}^{(\ast)}$.
An inhomogeneous piece $g^{(\mathrm{inh})}$ depends only on the known
quantities at timestep $n$, thus is immediately solved, while a
homogeneous piece $g^{(\mathrm{h})}$ depends on the fields at timestep
$n+1$. The unknown portion of the fields $\bm{\Psi}^{(\ast)}$ may be
solved as a separate step using a Green's function approach.

A formal homogeneous solution is given by 
\begin{align}
 g^{(\mathrm{h})} = & B^{-1} \bm{E}\cdot \bm{\Psi}^{(\ast)}
 = 
 \left(\frac{\delta g}{\delta \phi}\right) \phi^{(\ast)} +
 \left(\frac{\delta g}{\delta A_{\parallel}}\right) A_{\parallel}^{(\ast)} +
 \left(\frac{\delta g}{\delta \left(\delta B_{\parallel}\right)}\right)
 \delta B_{\parallel}^{(\ast)}
 \label{eq:formal_homogeneous_solution}
\end{align}
where $(\delta g/\delta \phi)$, $(\delta g/\delta A_{\parallel})$, and
$(\delta g/\delta (\delta B_{\parallel}))$ are called the plasma response
matrices. Using the response matrices, the field
Eqs.~\eqref{eq:quasi_neutrality_normalized}--\eqref{eq:ampperp_normalized}
can be written as
\begin{equation}
 \begin{pmatrix}
  P_{11} & P_{12} & P_{13} \\
  P_{21} & P_{22} & P_{23} \\
  P_{31} & P_{32} & P_{33} \\
 \end{pmatrix}
 \begin{pmatrix}
  \phi^{(\ast)} \\ A_{\parallel}^{(\ast)} \\ \delta B_{\parallel}^{(\ast)}
 \end{pmatrix}
 = 
 \begin{pmatrix}
  Q_{1} \\ Q_{2} \\ Q_{3}
 \end{pmatrix},
 \label{eq:field_equation_response_matrix}
\end{equation}
where 
\begin{align}
 P_{11} = & \sum_{s}
 \left[ q_{s} \M^{(0)}\left(\frac{\delta g_{s}}{\delta \phi}\right)
 - \frac{n_{0s}q_{s}^{2}}{T_{0s}} \left(1-\Gamma_{0s}\right) I
 \right],
 \label{eq:field_matrix_definition_p11} \\
 P_{12} = & \sum_{s}
 q_{s} \M^{(0)}\left(\frac{\delta g_{s}}{\delta A_{\parallel}}\right),
 \label{eq:field_matrix_definition_p12} \\
 P_{13} = & \sum_{s}
 \left[ q_{s} \M^{(0)}\left(\frac{\delta g_{s}}{\delta (\delta
 B_{\parallel})}\right) + q_{s}n_{0s} \Gamma_{1s} I
 \right],
 \label{eq:field_matrix_definition_p13} \\
 P_{21} = & \sum_{s}
 q_{s} n_{0s} \M^{(1)}\left(\frac{\delta g_{s}}{\delta
 \phi}\right),
 \label{eq:field_matrix_definition_p21} \\
 P_{22} = & - \frac{k_{\perp}^{2}}{2\beta_{0}} I
 + \sum_{s}
 q_{s} n_{0s} \M^{(1)}\left(\frac{\delta g_{s}}{\delta
 A_{\parallel}}\right),
 \label{eq:field_matrix_definition_p22} \\
 P_{23} = & \sum_{s}
 q_{s} n_{0s} \M^{(1)}\left(\frac{\delta g_{s}}{\delta (\delta
 B_{\parallel})}\right),
 \label{eq:field_matrix_definition_p23} \\
 P_{31} = & \sum_{s}
 \left[
 \M^{(2)}\left(\frac{\delta g_{s}}{\delta \phi}\right) +
 q_{s} n_{0s} \Gamma_{2s} I
 \right],
 \label{eq:field_matrix_definition_p31} \\
 P_{32} = & \sum_{s}
 \M^{(2)}\left(\frac{\delta g_{s}}{\delta A_{\parallel}}\right),
 \label{eq:field_matrix_definition_p32}
 \\
 P_{33} = & \frac{2}{\beta_{0}} I
 + \sum_{s}
 \left[
 \M^{(2)}\left(\frac{\delta g_{s}}{\delta(\delta B_{\parallel})} \right)
 + n_{0s}T_{0s} \Gamma_{2s} I
 \right],
 \label{eq:field_matrix_definition_p33}
\end{align}
with $I$ being the $N_{Z}\times N_{Z}$ identity matrix and 
\begin{align}
 Q_{1} = &  \phi^{n} \sum_{s}
 \frac{n_{0s}q_{s}^{2}}{T_{0s}}\left(1-\Gamma_{0s}\right)
 - \delta B_{\parallel}^{n} \sum_{s} q_{s}n_{0s}\Gamma_{1s}
 - \sum_{s}
 q_{s}\M^{(0)}\left(g_{s}^{(\mathrm{inh})}\right),
 \label{eq:field_matrix_definition_q1}
 \\
 Q_{2} = & \frac{k_{\perp}^{2}}{2\beta_{0}} A_{\parallel}^{n}
 - \sum_{s}
 q_{s} n_{0s} \M^{(1)}\left(g_{s}^{(\mathrm{inh})}\right),
 \label{eq:field_matrix_definition_q2}
 \\
 Q_{3} = &
 - \phi^{n} \sum_{s} q_{s}n_{0s}\Gamma_{1s}
 -  \delta B_{\parallel}^{n}
 \left(\frac{2}{\beta_{0}}
 + \sum_{s} n_{0s}T_{0s}\Gamma_{2s}\right)
 - \M^{(2)}\left(g_{s}^{(\mathrm{inh})}\right).
 \label{eq:field_matrix_definition_q3}
\end{align}
By solving the field Eq.~\eqref{eq:field_equation_response_matrix},
we obtain $\bm{\Psi}^{(\ast)}$, and ultimately $\bm{\Psi}^{n+1}$.
Given $\bm{\Psi}^{n+1}$, we solve \eqref{eq:gk_linear_discrete_symbolic}
for $g^{n+1}$. This is equivalent to solving
\eqref{eq:gk_linear_discrete_symbolic_two1} with
$\bm{E}\cdot\bm{\Psi}^{n}$ replaced by 
$\bm{E}\cdot\bm{\Psi}^{n+1}$. Finally, the full procedure to solve the
gyrokinetic equation becomes:
\begin{enumerate}
 \item Solve \eqref{eq:gk_linear_discrete_symbolic_two1} and apply
       the collision term to obtain the inhomogeneous part of the
       distribution function $g^{(\mathrm{inh})}$.
 \item Solve \eqref{eq:field_equation_response_matrix} for
       $\bm{\Psi}^{(\ast)}$, and obtain $\bm{\Psi}^{n+1}$.
 \item Replace $g^{(\mathrm{inh})}$ by $g^{n+1}$ and
       $\bm{E}\cdot\bm{\Psi}^{n}$ by $\bm{E}\cdot\bm{\Psi}^{n+1}$ in
       \eqref{eq:gk_linear_discrete_symbolic_two1}, solve it, and
       again apply the collision term to get $g^{n+1}$.
\end{enumerate}

Up to this point in the subsection, we have ignored the $i$ index for
notational simplicity. In fact, the gyrokinetic equation and the field
equations should be solved at all $Z$ grids simultaneously. The field
vector in \eqref{eq:field_equation_response_matrix} is actually a
vector of length $N_{f}N_{Z}$, where $N_{f}$ is the number of fields
evolved.  An electrostatic simulation requires only the evolution of
$\phi$ and so has $N_{f}=1$, whereas a fully electromagnetic
simulation evolves $\phi$, $A_\parallel$, and $\delta B_\parallel$ and
so has $N_{f}=3$.  Each of the  elements of the $P$ matrix in
\eqref{eq:field_equation_response_matrix}  represents an
$N_{Z}\times N_{Z}$ matrix, so the entire $P$ matrix is of size
$(N_fN_{Z})\times (N_fN_{Z})$. Similarly, the 
$Q$ vector is  of length $N_fN_{Z}$.

To evaluate the computational efficiency of Kotschenreuther's
approach, let us compare it to a brute-force approach to solving the
GK-M system.  A brute-force approach requires the inversion of a dense
$\left(N_{Z} N_{\lambda} N_{E} N_{s}\right)$-size square matrix (where
$N_{s}$ is the number of species), which generally takes ${\mathcal
O}\left(\left(N_{Z}N_{\lambda}N_{E}N_{s}\right)^{2}\right)$
operations. On the other hand, Kotschenreuther's method requires the
following: (a) for the gyrokinetic equation,
$\left(N_{\lambda}N_{E}N_{s}\right)$ inversions of bi-diagonal
$N_{Z}$-size square matrices, which costs ${\mathcal
O}(N_{Z}N_{\lambda}N_{E}N_{s})$; and, (b) for the fields, the
inversion of the matrix $P$ in
\eqref{eq:field_equation_response_matrix}, a dense $(N_{f}N_{Z})$-size
square matrix. For a fixed timestep $\Delta t$, however, the matrix
$P$ does not change, so this matrix inversion need only be performed
once during an initialization stage. Therefore, during each timestep
the field solver requires only an ${\mathcal
O}\left((N_{f}N_{Z})^{2}\right)$ matrix multiplication
operation. Ignoring the factor $N_{f}^{2}$ since $N_{f}\le 3$,
Kotschenreuther's approach requires ${\mathcal
O}\left(N_{Z}^{2}N_{\lambda}N_{E}N_{s}
\right)$ operations per timestep,  much more efficient than the
brute-force approach.

\subsubsection{Response matrix}
\label{sec:response_matrix}

A remaining task is to determine the response matrices, given in
\eqref{eq:formal_homogeneous_solution}. 
For clarity, we rewrite \eqref{eq:formal_homogeneous_solution} with the
$i$ index (in the code, the $i$ index is counted from $-\T{ntgrid}+1$ to
$\T{ntgrid}$ for $V_{\parallel}>0$ with $2 \times \T{ntgrid}=N_{Z}$):
\begin{equation}
 \begin{pmatrix}
  g_{1}^{(\mathrm{h})} \\
  g_{2}^{(\mathrm{h})} \\
  \vdots \\
  g_{N_{Z}}^{(\mathrm{h})} \\
 \end{pmatrix}
 =
 B^{-1}
 \left[
 E_{\phi}
 \begin{pmatrix}
  \phi_{1}^{(\ast)} \\
  \phi_{2}^{(\ast)} \\
  \vdots \\
  \phi_{N_{Z}}^{(\ast)} \\
 \end{pmatrix}
 + E_{A_{\parallel}}
 \begin{pmatrix}
  A_{\parallel,1}^{(\ast)} \\
  A_{\parallel,2}^{(\ast)} \\
  \vdots \\
  A_{\parallel,N_{Z}}^{(\ast)} \\
 \end{pmatrix}
 + E_{\delta B_{\parallel}}
 \begin{pmatrix}
  \delta B_{\parallel,1}^{(\ast)} \\
  \delta B_{\parallel,2}^{(\ast)} \\
  \vdots \\
  \delta B_{\parallel,N_{Z}}^{(\ast)} \\
 \end{pmatrix}
 \right],
 \label{eq:homogeneous_solution_matrix}
\end{equation}
where $B$ and $E_{\phi,A_{\parallel},\delta B_{\parallel}}$ are all
$N_{Z}\times N_{Z}$ matrices. To obtain the response matrices
$B^{-1}E_{\phi,A_{\parallel},\delta B_{\parallel}}$, we solve
\eqref{eq:homogeneous_solution_matrix} column by column.
If we put trial functions $\check{\phi}_{i}=\delta_{il}$ (where $\delta_{il}$ 
is the Kronecker's delta), $\check{A}_{\parallel,i}=0$, $\delta 
\check{B}_{\parallel,i}=0$ in the RHS of
\eqref{eq:homogeneous_solution_matrix}, a solution contains the 
$l$th column of $B^{-1}E_{\phi}$. By running $l$ from $1$ to $N_{Z}$,
the full response matrix is calculated. The same procedures are carried
out to obtain the other response matrices. This is fundamentally a
Green's function method.\footnote{The method can be trivially extended
to Fourier space along $Z$ when the coefficients of the linear terms are
independent of $Z$. This may be advantageous for some applications but
is not described here.}
Note that we can use the same routine to get
the response matrices as that to solve
\eqref{eq:gk_linear_discrete_symbolic} by replacing $g^{n}=0$,
$\bm{\Psi}^{n}=0$, and $\bm{\Psi}^{n+1}$ with the trial functions.

\subsubsection{Boundary conditions}
\label{sec:alg_bc}

The boundary condition in \T{AstroGK} in the $Z$ direction is always
periodic, thus it is rather trivial to implement the boundary condition
compared with \T{GS2}. For the field equation, periodicity is
immediately satisfied if the trial functions described in
Section~\ref{sec:response_matrix} satisfy periodicity. For the gyrokinetic
equation, periodicity is imposed in the matrix $B$. Let us write the
gyrokinetic equation as
\begin{equation}
 B \bm{x} = \bm{y} 
  ~~~ \textrm{where} ~~~
  B = 
  \begin{pmatrix}
   B_{2} & 0 & \cdots & 0 & B_{1} \\
   B_{1} & B_{2} & \ddots & & 0 \\
   0 & B_{1} & B_{2} & \ddots & \vdots \\
   \vdots & \ddots & \ddots & \ddots & 0 \\
   0 & \cdots & 0 & B_{1} & B_{2}
  \end{pmatrix}.
\end{equation}
The matrix $B$ is readily {\it LU} decomposed by forward elimination to
yield:
\begin{equation}
 B_{2}
  \begin{pmatrix}
   1 & 0 & \cdots & & 0 & - \left(-\frac{B_{1}}{B_{2}}\right)\\
   0 & 1 & \ddots & & \vdots & - \left(-\frac{B_{1}}{B_{2}}\right)^{2}\\
   & & \ddots & \ddots &  & \vdots \\
   \vdots & & \ddots & \ddots & 0 & - \left(-\frac{B_{1}}{B_{2}}\right)^{N_{Z}-2} \\
   & & & & 1 & - \left(-\frac{B_{1}}{B_{2}}\right)^{N_{Z}-1} \\
   0 & & \cdots & & 0 & - \left(-\frac{B_{1}}{B_{2}}\right)^{N_{Z}}
  \end{pmatrix}
  \bm{x} = L^{-1} \bm{y},
\end{equation}
thus the equation is solved.

\subsubsection{Collision term}
\label{sec:alg_coll}

The collision term is handled by operator splitting as shown in
\eqref{eq:gk_time_difference_godunov_collision}. After splitting,
one must compute $\left(1-\Delta t {\mathcal C}\right)^{-1}$,
where $\mathcal{C}$ is an advanced model collision operator designed for
gyrokinetics (Section~\ref{sec:gkeqns} and~\ref{sec:collisions}). 
The numerical implementation is given
in~\cite{BarnesAbelDorland_09}. Here we provide a brief overview of the
key features.

The collision operator can be written schematically as
\begin{equation}
\mathcal{C} \equiv {\mathcal C}_{\mathrm{L}} + {\mathcal C}_{\mathrm{D}} + 
 {\mathcal U}_{\mathrm{L}} + {\mathcal U}_{\mathrm{D}},
\end{equation}
where ${\mathcal C}_{\mathrm{L}}$ and ${\mathcal C}_{\mathrm{D}}$ are
second-order 
differential operators describing pitch-angle scattering and energy
diffusion, respectively, and ${\mathcal U}_{\mathrm{L}}$ and ${\mathcal
U}_{\mathrm{D}}$ 
are integral operators designed to make $\mathcal{C}$ conserve particle
number, momentum, and energy. Since the collision operator includes
velocity-space derivatives and integrals, but no coupling in $Z$, it
consists of a dense $(N_{\lambda}N_{E}) \times (N_{\lambda}N_{E})$ matrix. 

Discretization and inversion of the operator $1-\Delta t\mathcal{C}$
are done carefully to minimize computational expense and to preserve
numerically the analytic conservation properties. First, we discretize
the differential operators ${\mathcal C}_{\mathrm{L}}$ and ${\mathcal
C}_{\mathrm{D}}$ 
on a three-point stencil using a novel discretization scheme that
guarantees exact conservation properties on \T{AstroGK}'s nonuniform
velocity-space grids. This is accomplished by incorporating the
Gaussian integration weights in the discretization, leading to a
first-order accurate scheme across most of the velocity-space domain.

Next, we apply another Godunov splitting so that we can invert two
reduced matrices: $1-\Delta t \left({\mathcal C}_{\mathrm{L}}+{\mathcal
U}_{\mathrm{L}}\right)$ and $1-\Delta t\left({\mathcal
C}_{\mathrm{D}}+{\mathcal U}_{\mathrm{D}}\right)$. Both ${\mathcal
C}_{\mathrm{L}}$ and ${\mathcal C}_{\mathrm{D}}$ can be 
written as tridiagonal matrices, but ${\mathcal U}_{\mathrm{L}}$ and
${\mathcal U}_{\mathrm{D}}$ are in general dense. However, both
${\mathcal U}_{\mathrm{L}}$ and ${\mathcal U}_{\mathrm{D}}$ can be
expressed as tensor products, so that we can use the Sherman-Morrison
formula~\cite{ShermanMorrison_49,ShermanMorrison_50} to reduce greatly
the 
numerical expense of the matrix inversions. Therefore, the inversion 
 $\left(1-\Delta t\mathcal{C}\right)^{-1}$ is ultimately computed by
inverting a small number of tridiagonal matrices.

\subsection{Parallelization scheme}
\label{sec:alg_parallel}

Decomposition of the gyrokinetic distribution function is accomplished
using a flexible parallelization scheme that allows for several
different memory layouts. In principle, the gyrokinetic distribution
function is
\begin{equation}
  g(k_x, k_y, Z, \lambda, E, \sigma, s),
\end{equation}
comprising a scalar function defined on seven-dimensional discrete
space. Since field quantities, $\phi$, $A_{\parallel}$, and $\delta
B_{\parallel}$, are three-dimensional and therefore have much smaller
memory sizes than the distribution function, a copy of the full
three-dimensional field information is held by each
processor.\footnote{
This can be a problem if we consider extremely large
simulations. We may take a large number of processors $N_{\mathrm{proc}}$,
such that $N_{k_{x}}N_{k_{y}}N_{\lambda}N_{E}N_{s}/N_{\mathrm{proc}} \ll
N_{k_{x}}N_{k_{y}}$, as long as the work load on each processor is large
enough to scale up computational efficiency. In such cases, data for
the fields rather than the distribution function dominate memory usage.
}
Thus we focus on the parallelization scheme of the distribution function
in this section.

The update of the distribution function using the gyrokinetic
equation, as described in Section~\ref{sec:gk_solver}, is accomplished in
several stages, including the main stage (Section~\ref{sec:alg_green}),
collision term stage (Section~\ref{sec:alg_coll}), and nonlinear term
stage.  Each of these stages performs operations over different
dimensions of the seven-dimensional distribution function data. For
example, the linear terms of the main stage involve compact finite
differencing in $Z$; the collision term employs velocity-space
derivatives involving $\lambda$, $\sigma$, and $E$; and the nonlinear
term employs Fourier transforms requiring information along the
components of the perpendicular wavenumber, $k_x$ and $k_y$. We refer
to the dimensions required for the current operation as the
\emph{active} dimensions, while the remaining dimensions are
\emph{inactive}.

The basic parallelization strategy is to place the seven-dimensional
distribution function data into an array in memory such that the
indices of the array associated with the active dimensions come first.
Then all of the data associated with the inactive dimensions are
combined into a single, final array index in the order specified by
the variable \T{layout}. The data array is then decomposed across
$N_{\mathrm{proc}}$ processors by splitting it up evenly over the
final index containing all of the inactive dimensions. After one stage
of the calculation is completed, the distribution function data is
redistributed in preparation for the next stage of the calculation.
This redistribution is accomplished by a series of one-to-one
communications, carefully designed to perform only necessary
communications and to avoid communication deadlocks. During the
initialization for each run, the code determines this redistribution
scheme.  

For the gyrokinetic solver (Section~\ref{sec:gk_solver}), the
redistribution of the seven-dimensional distribution function data
into a three-dimensional data array is performed as follows. Since
compact finite differences are used along the mean magnetic field
($Z$), the first index of the data array corresponds to the $Z$
dimension. The second index is associated with $\sigma$ since most
operations are common to the same $|v_{\parallel}|$. The remaining
five dimensions of the distribution function data are combined into
the third index of the data array. The order of this combination may
be chosen by the user to yield the best parallel performance by
setting the character input variable \T{layout}, which consists of
five alphabetical characters indicating the order of the dimensions.
At present, the following options are available for \T{layout}:
`\T{lxyes}', `\T{lyxes}', `\T{lexys}', `\T{yxels}', and
`\T{yxles}'.  For example, if \T{layout} = `\T{yxles}', the data
are combined into the final index such that $k_y$ changes first, then
$k_x$, then $\lambda$, then $E$, and finally $s$. Such a layout is
advantageous, especially for collisionless simulations, when
redistribution to calculate the collision term is unnecessary. If the
number of processors $N_{\mathrm{proc}}$ is chosen such that all $k_x$
and $k_y$ indices are on the local processor [requiring
$\mod(N_{\lambda}N_{E}N_{s},N_{\mathrm{proc}})=0$], then
redistribution for the nonlinear term is unnecessary.

For the nonlinear term, if the number of processors
$N_{\mathrm{proc}}$ or the \T{layout} is chosen so that all $k_x$ and
$k_y$ indices are not on the local processor, then the code
redistributes the distribution function data over the processors into
a two-dimensional data array with $k_x$ (or $k_y$) as the first index
and all remaining inactive dimensions into a second index to be split
over processors. In this case, $Z$ and $\sigma$ are the first
dimensions to be combined, and then remaining dimensions follow the
layout specified in \T{layout}, except for $k_x$ (or $k_y$).  It then
performs the one-dimensional fast Fourier transform in $k_x$ (or
$k_y$) to compute the 
nonlinear term.  For the collision term, the code redistributes the
data with $\lambda$ and $\sigma$ first to obtain the pitch angle $\xi$
for the pitch-angle scattering term, and with $E$ first for the energy
diffusion term.  Therefore, the more physics that is included in the
calculation, the more redistribution that occurs, and the performance
of the code diminishes as the complexity of the problem is
increased. For large simulations employing $N_{\mathrm{proc}}>1000$,
one must choose the number of processors and \T{layout} carefully to
prevent the computational effort required for data redistribution from
becoming a bottleneck.

\subsection{Additional features}
\label{sec:alg_others}

We list here some additional features implemented in \T{AstroGK} which
have not yet been described.

\subsubsection{Driven simulation}
\label{sec:alg_drive}

One might want to drive the system
externally. For instance, turbulence simulation is quite often driven
externally at large scale to achieve a driven-dissipative system
where stationary turbulence inertial range spectrum will be observed.

In \T{AstroGK}, we can add $A_{\parallel}^{\mathrm{antenna}}$ in
\eqref{eq:field_matrix_definition_q2} to drive an {\it antenna}
current in the direction parallel to the mean magnetic field.
In this case, the normalized parallel Amp\`ere's law 
\eqref{eq:ampparallel_normalized} is modified as
\begin{equation}
 \frac{k_{\perp}^{2}}{2\beta_{0}} 
  \left(A_{\parallel,\bm{k}_{\perp}}
   + A_{\parallel,\bm{k}_{\perp}}^{\mathrm{antenna}}\right)
  = \sum_{s} q_{s} n_{0s} {\mathcal
  M}^{(1)}(g_{\bm{k}_{\perp},s}).
  \label{eq:ampparallel_normalized_drive}
\end{equation}
Property of the antenna is specified by its amplitude $A_{\parallel0}$,
	    frequency $\omega_{0}$, and wavenumber $\bm{k}_{0}$:
\begin{equation}
 A_{\parallel,\bm{k}_{\perp0}}^{\mathrm{antenna}}
  = A_{\parallel0} e^{-\imag (\omega_{0}t - k_{\parallel0}z)}.
\end{equation}

\subsubsection{Timestep}
\label{sec:alg_timestep}

The timestep $\Delta t$ is variable in
\T{AstroGK}. For linear runs, the numerical scheme is unconditionally
stable as long as $r_{t}\leq1/2$, so adjustment of the timestep is
necessary only to achieve an acceptable accuracy and does not affect
the stability. For nonlinear runs, the nonlinear term is handled
explicitly and therefore the code automatically adjusts the timestep
to meet the CFL condition according to the nonlinear drift velocity in
the plane perpendicular to the mean magnetic field.

We estimate an acceptable timestep based on the CFL condition as
\begin{equation}
 \Delta t_{\mathrm{CFL}} = C_{\mathrm{CFL}}
  \min\left(
       \frac{\Delta x}{\max\left(v_{\mathrm{D},x}^{\mathrm{NL}}\right)},
       \frac{\Delta y}{\max\left(v_{\mathrm{D},y}^{\mathrm{NL}}\right)}
      \right)
\end{equation}
where the nonlinear drift velocity $\bm{v}_{\mathrm{D}}^{\mathrm{NL}}$
	   is given by the third term in 
	   \eqref{eq:drift_velocity}, $\Delta x=2\pi/k_{x,\mathrm{max}}$,
	   $\Delta y=2\pi/k_{y,\mathrm{max}}$, and $C_{\mathrm{CFL}}$
	   is a user input constant. We check if the current time step
	   is greater than $\Delta t_{\mathrm{CFL}}$ at every timestep,
	   and divide $\Delta t$ by a constant factor (2 by default) if
	   $\Delta t> \Delta t_{\mathrm{CFL}}$. We also increase $\Delta
	   t$ by multiplying the same factor if $\Delta t$ is
	   substantially smaller than $\Delta t_{\mathrm{CFL}}$.

If $\Delta t$ changes, some matrices must be updated accordingly, so
the initialization routine for the matrices is called at this
timing. The first two steps for the nonlinear term after the timestep
change are again the Euler method and AB2.

\subsubsection{Diagnostics}
\label{sec:alg_diag}

\T{AstroGK} can write out the full data of the fields and the
distribution functions. Frequent output of the full data is undesired
due to 
the consumption of
large amounts of disk space.  Instead, the code computes
reduced diagnostic data on the fly, including the following:
\begin{itemize}
    \item Macroscopic quantities, such as the entropy,
		   energies, heating rates, and certain averages of the
		   fields and the
	  moments~\cite{HowesCowleyDorland_06,TatsunoDorlandSchekochihin_09}.
     \item Spectra in position and velocity
	   spaces~\cite{HowesDorlandCowley_08,TatsunoDorlandSchekochihin_09,TatsunoBarnesCowley_10}.
     \item Transfer functions of energies in position and velocity
       spaces~\cite{TatsunoBarnesCowley_10}.
     \item Linear frequency and growth rate~\cite{HowesCowleyDorland_06,NielsonHowesTatsuno_10}.
     \item Error estimate of velocity-space
	   resolution~\cite{BarnesAbelDorland_09,BarnesDorlandTatsuno_10}.
\end{itemize}
   Some of the output utilize \T{NetCDF} or \T{HDF5} library to make
	    structured binary data for portability.


\section{Code verification}
\label{sec:examples}

In this section, we provide some examples of \T{AstroGK} test runs,
ranging from the simple linear, electrostatic ($A_{\parallel}=\delta
B_{\parallel}=0$) problems to fully nonlinear, electromagnetic
problems.  Validity of the linear physics within the code is
demonstrated by the comparison of linear \T{AstroGK} results with
analytic solutions. For nonlinear problems, comparison to an analytic
solution is generally not simple, so here we compare with the
nonlinear results of an independent reduced magnetohydrodynamics (MHD)
code. In the large-scale limit $k_{\perp}\rho_{\mathrm{i}}\ll 1$, the
gyrokinetic-Maxwell equations simplify to the equations of reduced MHD~\cite{SchekochihinCowleyDorland_09}, so such a  comparison is useful to
validate the nonlinear behavior of \T{AstroGK}. It is worth noting that the
gyrokinetic simulations contain much richer physics and show
interesting phenomena, but a detailed analysis of such behavior is
beyond the scope of this paper and is left for future studies.

Unless otherwise stated, the plasma is assumed to be quasi-neutral
with
$n_{0\mathrm{i}}/n_{0\mathrm{e}}=-q_{\mathrm{i}}/q_{\mathrm{e}}=1$. Free
parameters are the ion-to-electron mass ratio
$m_{\mathrm{i}}/m_{\mathrm{e}}$, the equilibrium ion-to-electron
temperature ratio $T_{0\mathrm{i}}/T_{0\mathrm{e}}$, and the plasma
beta $\beta_{s}$ of the reference species $s$.

We comment on the number of grid points in the perpendicular plane of
position space compared to the corresponding number of Fourier modes
kept in the pseudo-spectral method.  The number of grid points in the
perpendicular plane $N_{x}$ and $N_y$ are specified by
user. Dealiasing requires that Fourier modes must be discarded
according to the $2/3$ rule~\cite{Orszag_71}, giving the number of
perpendicular wave 
modes in \T{AstroGK} $N_{k_{x}}\approx2/3N_{x}$ and
$N_{{k}_{y}}\approx1/3N_{y}$. Note that the reality constraint on the
complex Fourier coefficients means that the modes in the lower half
($k_y <0$) of the $(k_{x}, k_{y})$ plane are not independent and need
not be kept. In addition to the standard operating mode that
nonlinearly evolves many Fourier modes, \T{AstroGK}
also allows a very useful linear mode where $N_{x}=N_{y}=1$ and only
$|k_{\perp}|$ needs to be specified.
  
\subsection{Linear physics of \Alfven waves}
\label{sec:linear_response}

Testing the ability of \T{AstroGK} to model accurately the linear
physics of \Alfven waves, including their dispersion and collisionless
damping at sub-Larmor radius scales, is critical to demonstrating the
validity of the code. The linear collisionless gyrokinetic dispersion
relation~\cite{HowesCowleyDorland_06} describes the physical behavior
and provides analytical solutions for comparison to numerical results.
It is worth noting here that the results of the linear collisionless
gyrokinetic dispersion relation agree with the full Vlasov--Maxwell hot
plasma dispersion relation~\cite{Stix_92} in the gyrokinetic limit, as
has been demonstrated by Howes et al.~\cite{HowesCowleyDorland_06}.

The tests in this subsection explore the linear plasma response in a
uniform plasma with a straight equilibrium magnetic field, so that
$L_{\Bg}^{-1}=L_{n_{0s}}^{-1}=L_{T_{0s}}^{-1}=\kappa= 0$.  All tests use a
realistic ion-to-electron mass ratio for protons,
$m_{\mathrm{i}}/m_{\mathrm{e}}=1836$. The linear frequencies and
damping rates presented here are normalized by the MHD \Alfven frequency
$\omega_{\mathrm{A}}$ (defined later on), denoted by the overbar,
$\overline{\omega} \equiv \omega/\omega_{\mathrm{A}}$.
 With these simplifications, the normalized complex
eigenfrequency of the linear collisionless gyrokinetic dispersion
relation is a function of only three parameters, $\overline{\omega} =
\overline{\omega}(k_{\perp}
\rho_{\mathrm{i}}, \beta_{\mathrm{i}}, T_{0\mathrm{i}}/T_{0\mathrm{e}})$. 

\subsubsection{Linear Laplace--Fourier transform solution}
\label{sec:ex_lft}

\Alfven waves can be driven in \T{AstroGK} by applying a parallel antenna 
current throughout the simulation domain (see Section~\ref{sec:alg_others}).
The parallel vector potential due to the antenna drives a single
wavevector $\V{k}_0= k_{\perp} \xhat + k_{\parallel} \zhat$ at a
frequency $\omega_{0}$ with amplitude $A_{\parallel 0}$.
Using the parallel wave number of the
drive, we can define the Alfv\'en frequency for the normalization as
$\omega_{\mathrm{A}}=k_{\parallel} v_{\mathrm{A}}$, where the Alfv\'en
velocity is defined by
$v_{\mathrm{A}}=B_{0}/\sqrt{\mu_{0}n_{0\mathrm{i}}m_{\mathrm{i}}}$.
 
\begin{figure}[htbp]
 \begin{center}
  \includegraphics[scale=0.6]{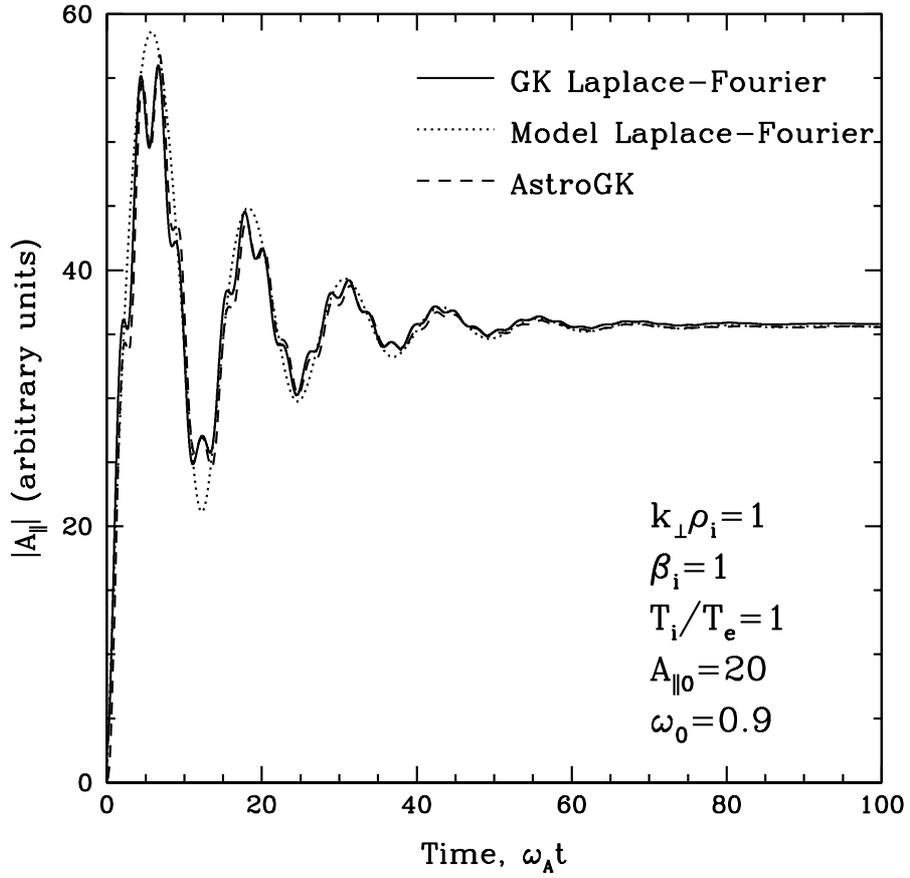}
  \caption{\label{fig:linear_laplace}
  Evolution of the amplitude $|A_\parallel|$ vs.~time from a linear
  \T{AstroGK} run (dashed line) with $\delta B_{\parallel}=0$ for
  parameters $k_\perp \rho_{\mathrm{i}}=1$, $\beta_{\mathrm{i}}=1$, and
  $T_{0\mathrm{i}}/T_{0\mathrm{e}}=1$. The Laplace--Fourier solutions of
  the gyrokinetic system (solid line) and a simple model system (dotted
  line) are given for comparison.
  }
 \end{center}
\end{figure}

The amplitude response of the driven linear gyrokinetic system can be
compared with an analytical Laplace--Fourier transform solution, given by
\eqref{eq:lfgk} in \ref{sec:laplace_fourier_solution}. For
the example presented here, the parallel magnetic field perturbation
was forced to be zero, $\delta B_\parallel=0$. In
Fig.~\ref{fig:linear_laplace}, we plot the analytical Laplace--Fourier
solution of the gyrokinetic system (solid line) and compare it to the
output of \T{AstroGK} (dashed line).  Parameters for the run are
$k_\perp \rho_{\mathrm{i}}=1$, $\beta_{\mathrm{i}}=1$, and
$T_{0\mathrm{i}}/T_{0\mathrm{e}}=1$; the system is driven with
amplitude $A_{\parallel 0}=20$ (in arbitrary unit) at frequency
$\overline{\omega}_{0}=0.9$ from zero initial conditions for the
fields and perturbed distribution functions $g_s$. From the linear
gyrokinetic dispersion relation, the linear eigenvalue for these
parameters is $\overline{\omega}=\overline{\omega}_{\mathrm{R}}+\imag
\overline{\omega}_{\mathrm{I}}=1.4057-0.073004\imag$.
The numbers of grids are $(N_{Z}, N_{\lambda}, N_{E})=(32,8,32)$.
These choices achieve a converged result with minimal computational
effort. Low collision frequencies for the same-species collisions are
chosen to yield weakly collisional behavior, with
$\nu_{\mathrm{i}}/\omega_{\mathrm{A}}=\nu_{\mathrm{e}}/\omega_{\mathrm{A}}=2 \times 10^{-2}$.
The close agreement of the solid and dashed lines in
Fig.~\ref{fig:linear_laplace}---faithfully reproducing even the small
amplitude, high-frequency oscillation---demonstrates the ability of
\T{AstroGK} to model accurately a driven linear gyrokinetic system.

To determine the effective frequency $\overline{\omega}_{\mathrm{R}}$
and damping rate $-\overline{\omega}_{\mathrm{I}}$ of the linear
response to the driving in an \T{AstroGK} simulation, the
Laplace--Fourier transform solution can be found for a model linear
system. The model system treats the time evolution of $ A_\parallel$
as a linear operator with a complex eigenvalue given by $- \imag
(\omega_{\mathrm{R}}+ \imag \omega_{\mathrm{I}})$ and includes the
driving term,
\begin{equation}
 \frac{\partial A_\parallel}{\partial t} = - \imag \omega  A_\parallel + 
  A_{\parallel 0}e^{-\imag \omega_{0}t}.
  \label{eq:linear_response_model_eq}
\end{equation}
The Laplace--Fourier transform solution yields the time evolution of the
amplitude:
\begin{equation}
|A_\parallel(t)| = A_{\parallel 0} \left\{\frac{  1+ 
e^{2 \omega_{\mathrm{I}} t} - 2e^{ \omega_{\mathrm{I}} t} \cos [(\omega_{\mathrm{R}}- \omega_0)t]}
{(\omega_{\mathrm{R}}- \omega_0)^2 + \omega_{\mathrm{I}}^2} \right\}^{1/2}.
\label{eq:lfmod}
\end{equation}
After normalizing frequencies appropriately, when
$\overline{\omega}_{0}=0.9$, this function yields the fit (dotted
line) in Fig.~\ref{fig:linear_laplace} with $ A_{\parallel 0}=18.2$
and $\overline{\omega}=1.406-0.073\imag$. Note that fitting the
oscillation of the solution at the beat frequency
$(\omega_{\mathrm{R}}-\omega_0)$ allows a precise
determination of the resonant frequency. The fractional error in the
damping rate is larger, in particular due to the difficulty of fitting
the exponential decay in the presence of the higher frequency
oscillations arising in the gyrokinetic system---compare the
difference between the simple model (dotted line) and the gyrokinetic
solution (solid line). Estimating the error in the frequency and
damping rate based on these fits, we determine values of
$\overline{\omega}_{\mathrm{R}}=1.406 \pm 0.004 $ and
$-\overline{\omega}_{\mathrm{I}}=0.073 \pm 0.003$.

\subsubsection{Linear frequencies and damping rates}
\label{sec:ex_lft2}

\begin{figure}[htbp]
 \begin{center}
  \includegraphics[scale=0.33]{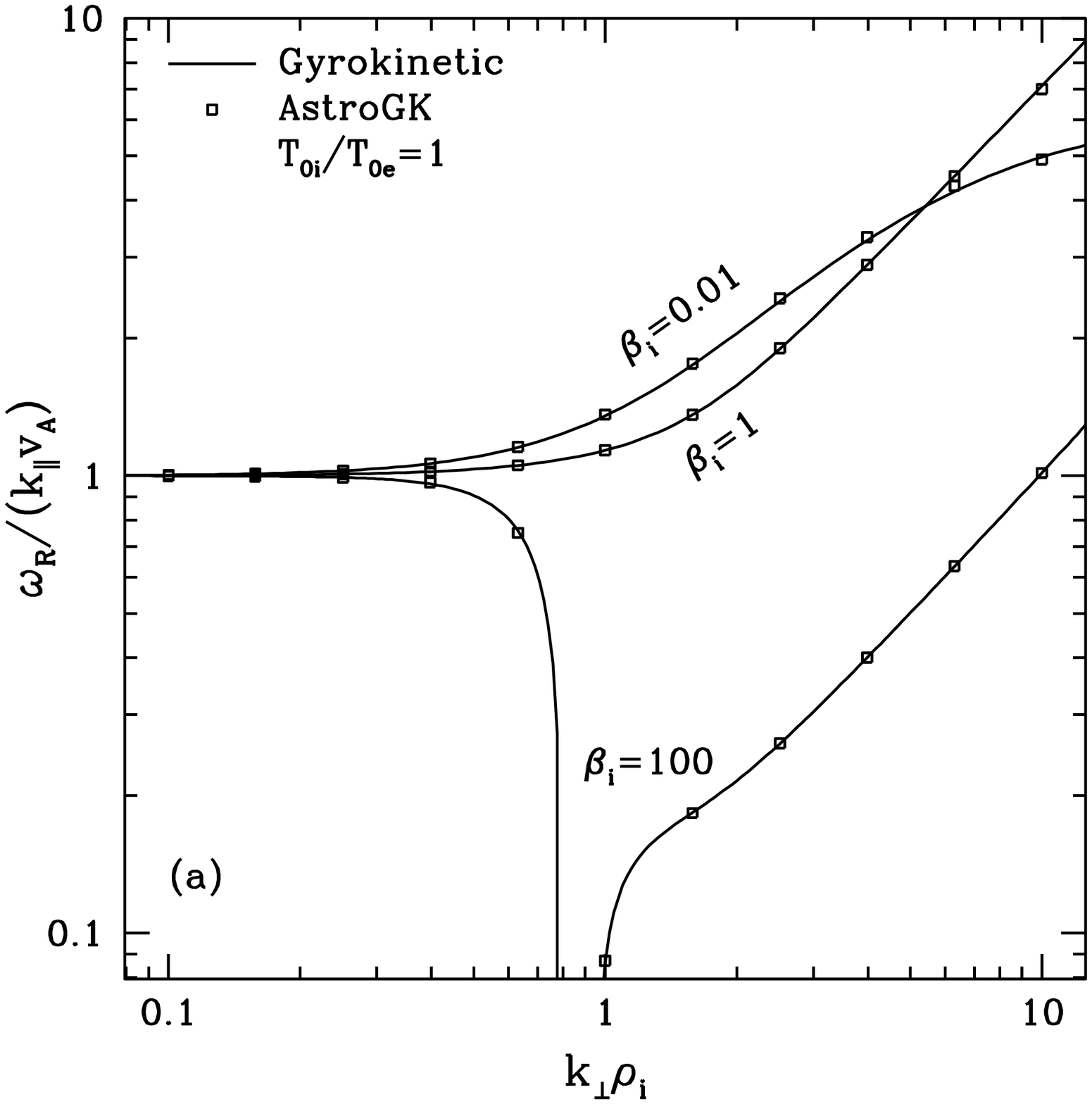}
  \includegraphics[scale=0.33]{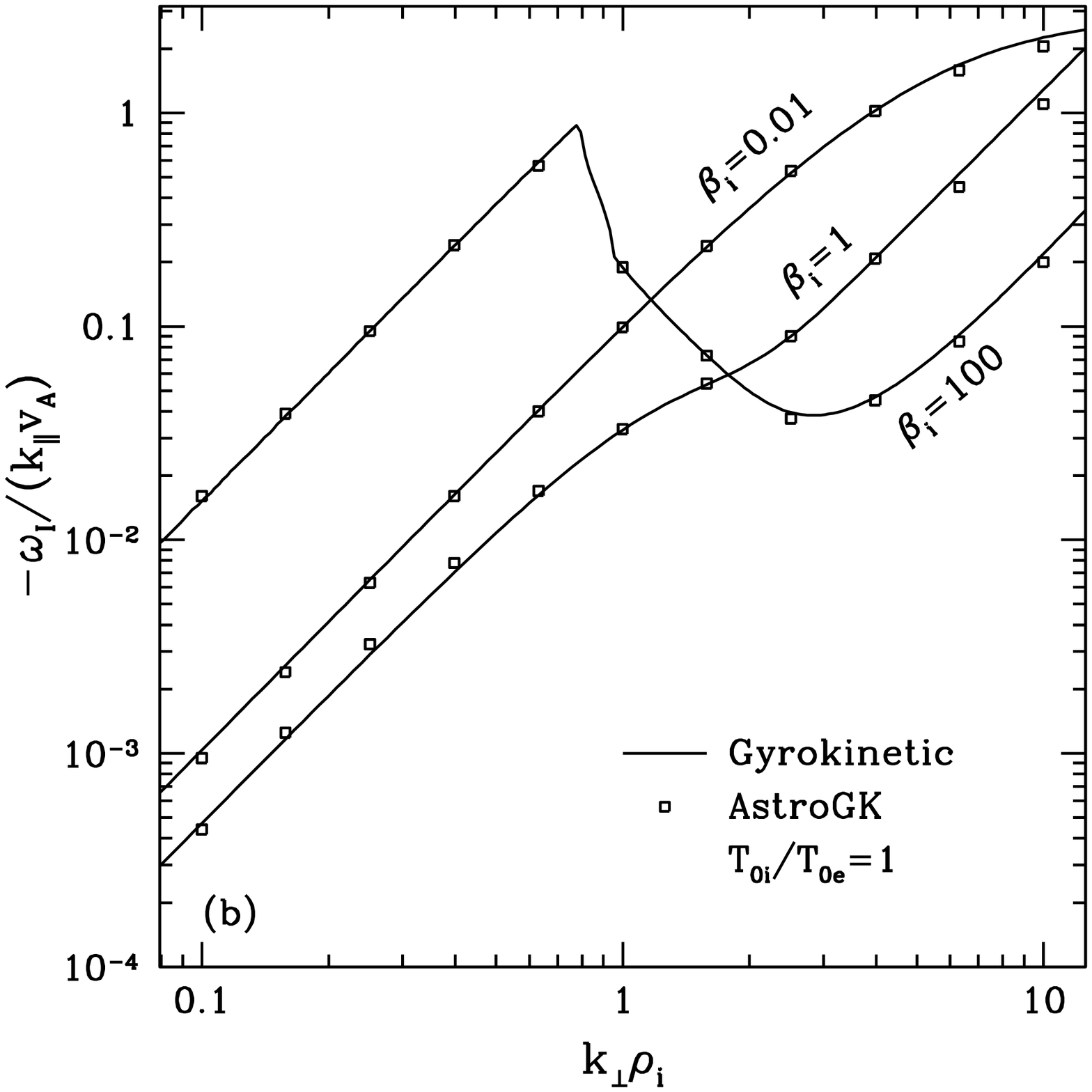}
  \includegraphics[scale=0.33]{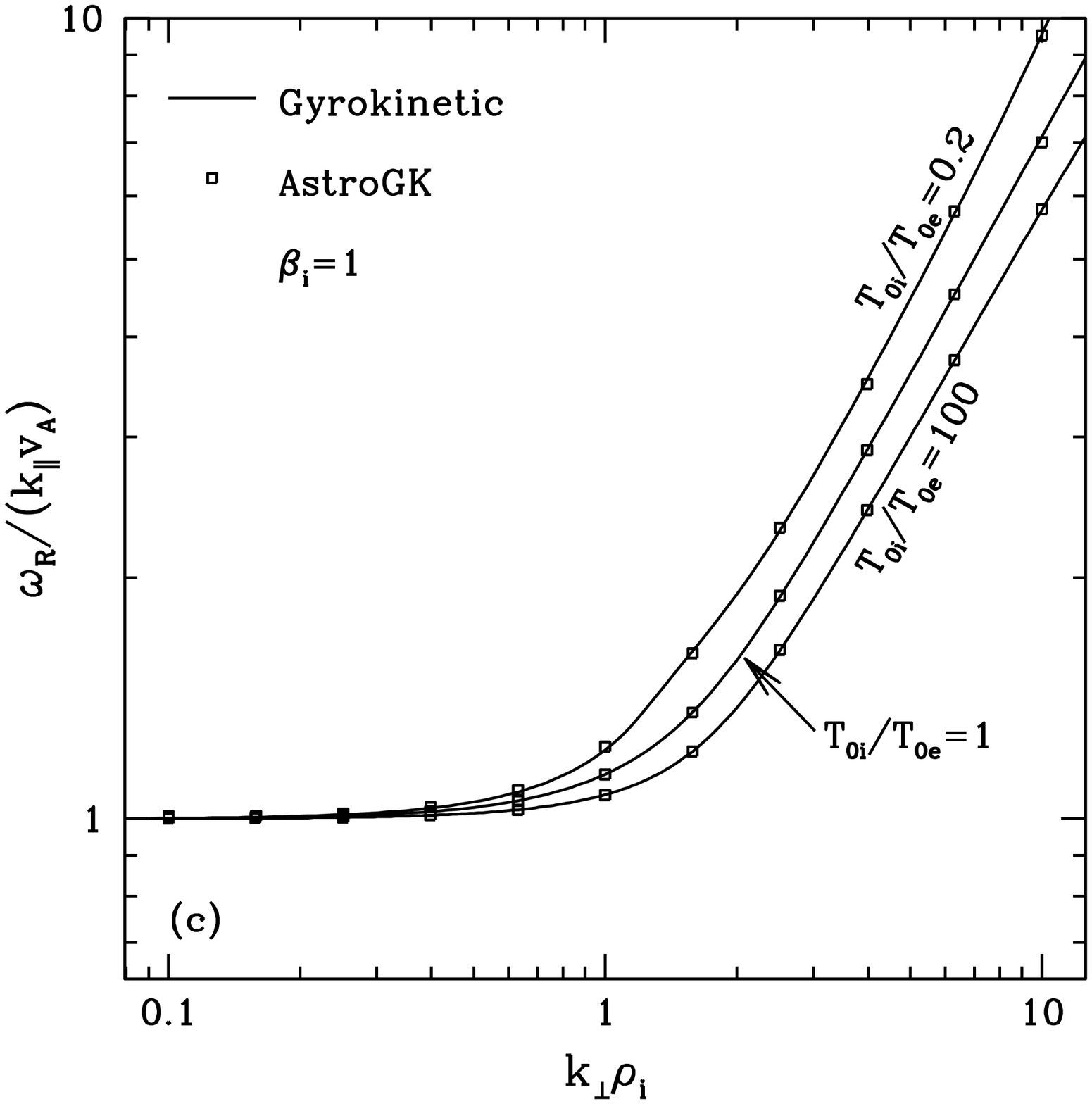}
  \includegraphics[scale=0.33]{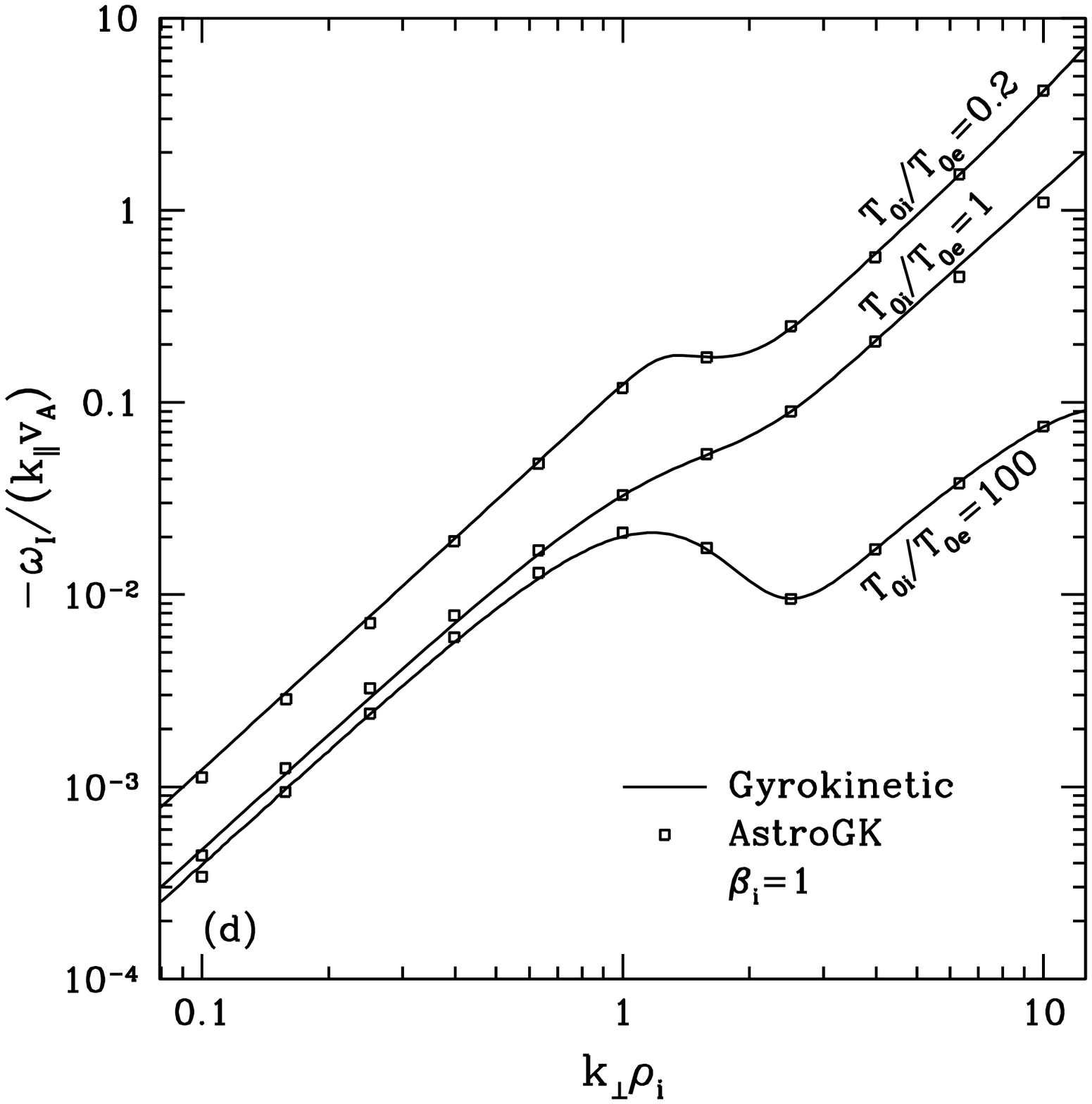}
  \caption{\label{fig:linear_wg}
  The normalized (a and c) real frequency
  $\overline{\omega}_{\mathrm{R}}=\omega_{\mathrm{R}}/\omega_{\mathrm{A}}$
  and (b and d) damping rate
  $-\overline{\omega}_{\mathrm{I}}=-\omega_{\mathrm{I}}/\omega_{\mathrm{A}}$
  vs.~$k_{\perp} \rho_{\mathrm{i}}$
  for varied ion plasma beta $\beta_{\mathrm{i}}=0.01,1,100$ with
  fixed $T_{0\mathrm{i}}/T_{0\mathrm{e}}=1$ (a and b) and for varied
  ion-to-electron temperature ratio
  $T_{0\mathrm{i}}/T_{0\mathrm{e}}=0.2,1,100$ with fixed
  $\beta_{\mathrm{i}}=1$ (c and d) from the gyrokinetic dispersion
  relation (solid line) and linear \T{AstroGK} simulations (open
  squares).  }  \end{center}
\end{figure}

The model Laplace--Fourier transform solution given by \eqref{eq:lfmod}
is used to determine the normalized frequencies ($\overline{\omega}_{\mathrm{R}}$) and
damping rates ($-\overline{\omega}_{\mathrm{I}}$) of linear modes on the \Alfven
branch, including the kinetic \Alfven wave at $k_\perp
\rho_{\mathrm{i}} \gg 1$, over a wide range of plasma parameters.  In 
the upper panels of Fig.~\ref{fig:linear_wg}, we present the
normalized (a) frequencies and (b) damping rates of the \Alfven
solution vs. perpendicular wavenumber $k_{\perp} \rho_{\mathrm{i}}$ for varied
values of ion plasma beta $\beta_{\mathrm{i}} = 0.01, 1, 100$ at
$T_{0\mathrm{i}}/T_{0\mathrm{e}}=1$; in the lower panels are the normalized (c)
frequencies and (d) damping rates of the \Alfven solution for varied
values of ion-to-electron temperature ratio
$T_{0\mathrm{i}}/T_{0\mathrm{e}}=0.2,1,100$ with $\beta_{\mathrm{i}}=1$.
The numbers of grids are $(N_{Z}, N_{\lambda}, N_{E})=(32,8,32)$.
To ensure that
structure in velocity space does not reach the velocity grid Nyquist
frequency, the same species collisionalities are set in the range to
$0.1 |\omega_{\mathrm{I},s}| \lesssim \nu_{s} \lesssim
|\omega_{\mathrm{I},s}|$ for each run.

\subsubsection{Linear ion-to-electron heating ratios}
\label{sec:ex_heating}

\begin{figure}[htbp]
 \begin{center}
  \includegraphics[scale=0.5]{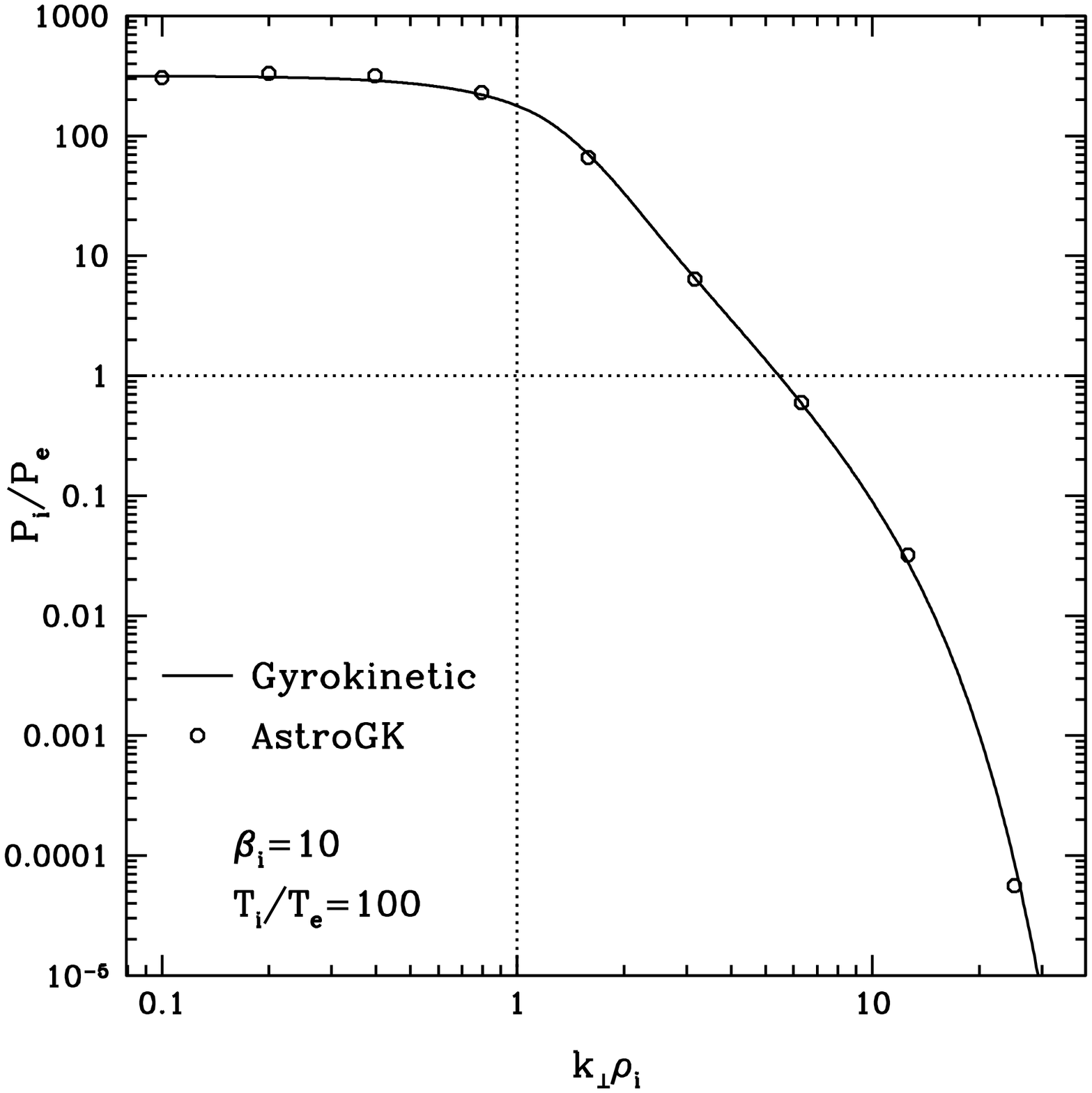}
  \caption{\label{fig:linear_pipe}
  Ratio of the ion-to-electron heating
  $P_{\mathrm{i}}/P_{\mathrm{e}}$. Results are shown for
  $\beta_{\mathrm{i}}=10$ and $T_{0\mathrm{i}}/T_{0\mathrm{e}}=100$. The
  analytical results from the linear collisionless gyrokinetic
  dispersion relation (solid line) are compared to \T{AstroGK} results
  (circles), showing excellent agreement over seven orders of magnitude
  in $P_{\mathrm{i}}/P_{\mathrm{e}}$.  }
 \end{center}
\end{figure}

Using the heating equations for
gyrokinetics~\cite{HowesCowleyDorland_06}, we calculate the
ion-to-electron heating
ratio $P_{\mathrm{i}}/P_{\mathrm{e}}$ from the linear collisionless
gyrokinetic dispersion relation and compare it to the results of
\T{AstroGK}. The results are presented in Fig.~\ref{fig:linear_pipe} for 
parameters $\beta_{\mathrm{i}}=10$ and
$T_{0\mathrm{i}}/T_{0\mathrm{e}}=100$, chosen to give the heating ratio
that varies by many orders of magnitude around $k_\perp
\rho_{\mathrm{i}} \sim 1$. The resolution for these runs is
$(N_{Z},N_{\lambda},N_{E})=(64,64,32)$. The results of \T{AstroGK} 
show excellent agreement with theory over seven orders of magnitude in 
$P_{\mathrm{i}}/P_{\mathrm{e}}$.




\subsection{Linear Ion-Temperature-Gradient (ITG) instability}
\label{sec:ex_itg}

Here we describe linear drift-wave dynamics in the collisionless limit~\cite{RudakovSagdeev_61,AntonsenCoppiEnglade_79}
to validate the electrostatic calculation of \T{AstroGK} with a
Boltzmann electron response. We assume
$L_{B_{0}}^{-1}=\kappa=L_{n_{0s}}^{-1}=0$. Considering an ion distribution
function with $\exp\left[\imag (\bm{k}\cdot\bm{r}-\omega t)\right]$ dependence,
from \eqref{eq:gkeq} we find:
\begin{equation}
  h_{\bm{k}_{\perp},\mathrm{i}} = 
   \frac{q_{\mathrm{i}} \phi_{\bm{k}_{\perp}}}{T_{0\mathrm{i}}}
   \frac{\omega}{\omega - \kpar V_{\parallel}}
   \left[ 1 +
    \frac{\omega_{\ast T}}{\omega}
    \left(
     \frac{V^2}{v_{\mathrm{th},\mathrm{i}}^2} - \frac{3}{2}
    \right)
   \right]
   J_0(\alpha_{i}) f_{0i},
   \label{eq:itg_ion_linear_reponse}
\end{equation} 
where we have introduced the drift frequency:
\begin{equation}
 \omega_{*T}
  = \frac{1}{2} \frac{k_{y} \rho_{\mathrm{i}}
  v_{\mathrm{th},\mathrm{i}}}{L_{T_{0\mathrm{i}}}} 
  = \frac{k_{y} T_{0\mathrm{i}}}
  {L_{T_{0\mathrm{i}}} q_{\mathrm{i}} \Bg}.
  \label{eq:itg_itg_drift_frequency}
\end{equation}
Plugging this into the quasi-neutrality condition with the Boltzmann
electron response $h_{\mathrm{e}}=0$, we obtain the following dispersion
relation: 
\begin{equation}
 \frac{q_{\mathrm{e}}^{2} n_{0\mathrm{e}}}{T_{0\mathrm{e}}}
 + \frac{q_{\mathrm{i}}^{2} n_{0\mathrm{i}}}{T_{0\mathrm{i}}}
 = - \frac{q_{\mathrm{i}}^{2} n_{0\mathrm{i}}}{T_{0\mathrm{i}}}
 \Gamma_{0\mathrm{i}} \zeta_{\mathrm{i}} \Xi(\zeta_{\mathrm{i}})
 + \frac{q_{\mathrm{i}}^{2} n_{0\mathrm{i}}}{T_{0\mathrm{i}}}
 \frac{\omega_{\ast T}}{\omega}
 \left[
 \left(
 \frac{1}{2} \Gamma_{0\mathrm{i}}
 + b_{\mathrm{i}} \Gamma_{1\mathrm{i}}
 \right)
 \zeta_{\mathrm{i}} \Xi(\zeta_{\mathrm{i}})
 - \zeta_{\mathrm{i}}^{2} \Gamma_{0\mathrm{i}}
 \left( 1 + \zeta_{\mathrm{i}} \Xi(\zeta_{\mathrm{i}}) \right)
 \right],
 \label{eq:itg_dispersion_relation}
\end{equation}
where $\zeta_{\mathrm{i}} = \omega /(\kpar v_{\mathrm{th},\mathrm{i}})$,
and $\Xi$ is the plasma dispersion function~\cite{FriedConte_61}.

\begin{figure}
 \begin{center}
  \includegraphics[scale=0.25,angle=270]{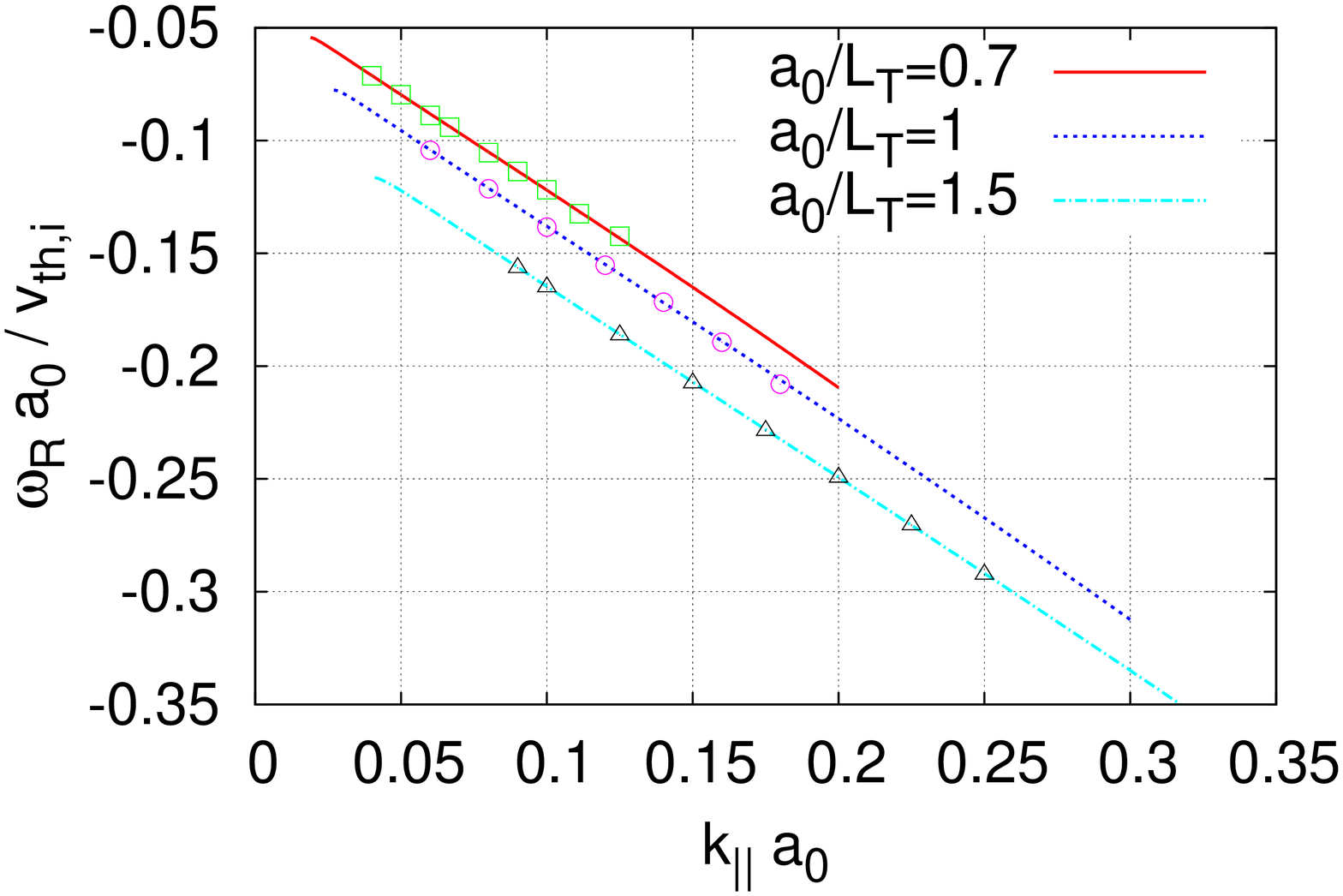}
  \includegraphics[scale=0.25,angle=270]{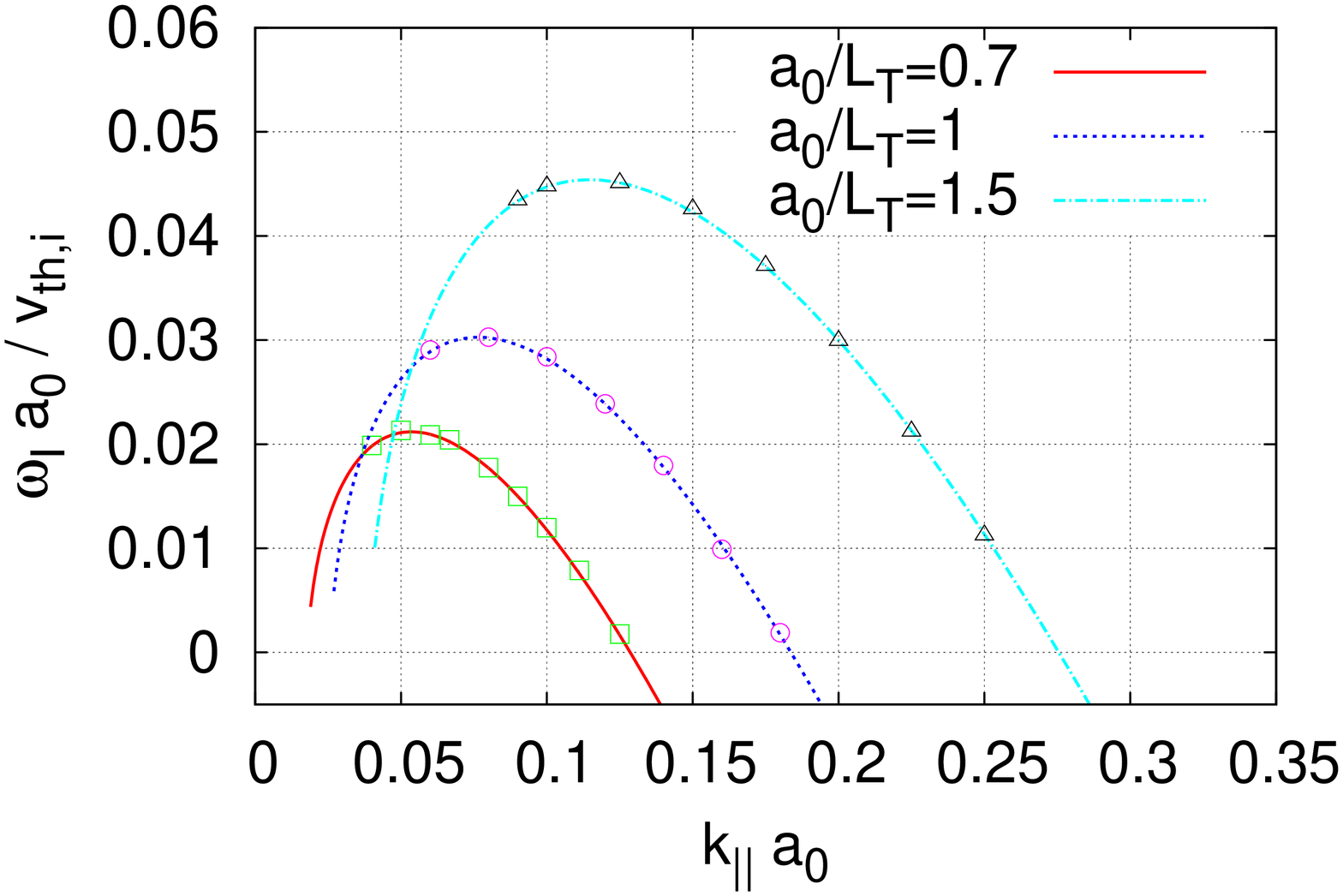}
  \caption{\label{fig:ITG_dispersion}
  Dispersion relation of the slab ITG instability under the code normalization
  (see \ref{sec:normalization}).
  Left: frequency
  ($\omega_{\mathrm{R}}$), Right: growth rate ($\omega_{\mathrm{I}}$).
  Lines are from dispersion relation \eqref{eq:itg_dispersion_relation} and
  symbols are from \T{AstroGK}.}
 \end{center}
\end{figure}

We excite a perturbation of $h_{\mathrm{i}}\propto f_{0\mathrm{i}}$
which generates an electrostatic potential perturbation. $A_{\parallel}$
and $\delta B_{\parallel}$ are forced to be zero throughout the
simulation.
For the following parameters, $T_{0\mathrm{i}}/T_{0\mathrm{e}}=1$,
$(k_{x}\rho_{\mathrm{i}},k_{y}\rho_{\mathrm{i}})=(0,1)$,
the numerical solution to the dispersion relation
\eqref{eq:itg_dispersion_relation} and the eigenvalues $\omega$ obtained
from \T{AstroGK} are shown in Fig.~\ref{fig:ITG_dispersion}.
We take $(N_{Z}, N_{\lambda}, N_{E})=(32, 16, 16)$ for all runs.
The figure shows perfect agreement between the numerical solution and
the theory, even though reproducing a slowly growing mode with small
$\kpar$ is particularly challenging since \T{AstroGK} uses a
finite-difference scheme in the $Z$ direction.

\subsection{Collisions and velocity-space resolution}
\label{sec:ex_collision}

Here we demonstrate the accuracy of the collision model employed in
\T{AstroGK} and the spectral convergence of the velocity-space integration.
For the convergence studies, we again consider the electrostatic plasma slab 
with a background ion temperature gradient and a Boltzmann response for
the electrons. We focus on the case with 
$T_{0\mathrm{i}}/T_{0\mathrm{e}}=k_{\parallel}
L_{T_{0\mathrm{i}}}=k_{\perp}\rho_{\mathrm{i}}=1$, and we use 
$(N_{Z}, N_{\lambda}, N_E)=(256,128,256)$ 
as our base case resolution. We then independently vary $N_E$ and
$N_{\lambda}$ and calculate the relative error in the ITG mode frequency
and growth rate.

\begin{figure}[htbp]
 \begin{center}
  \includegraphics[scale=0.3]{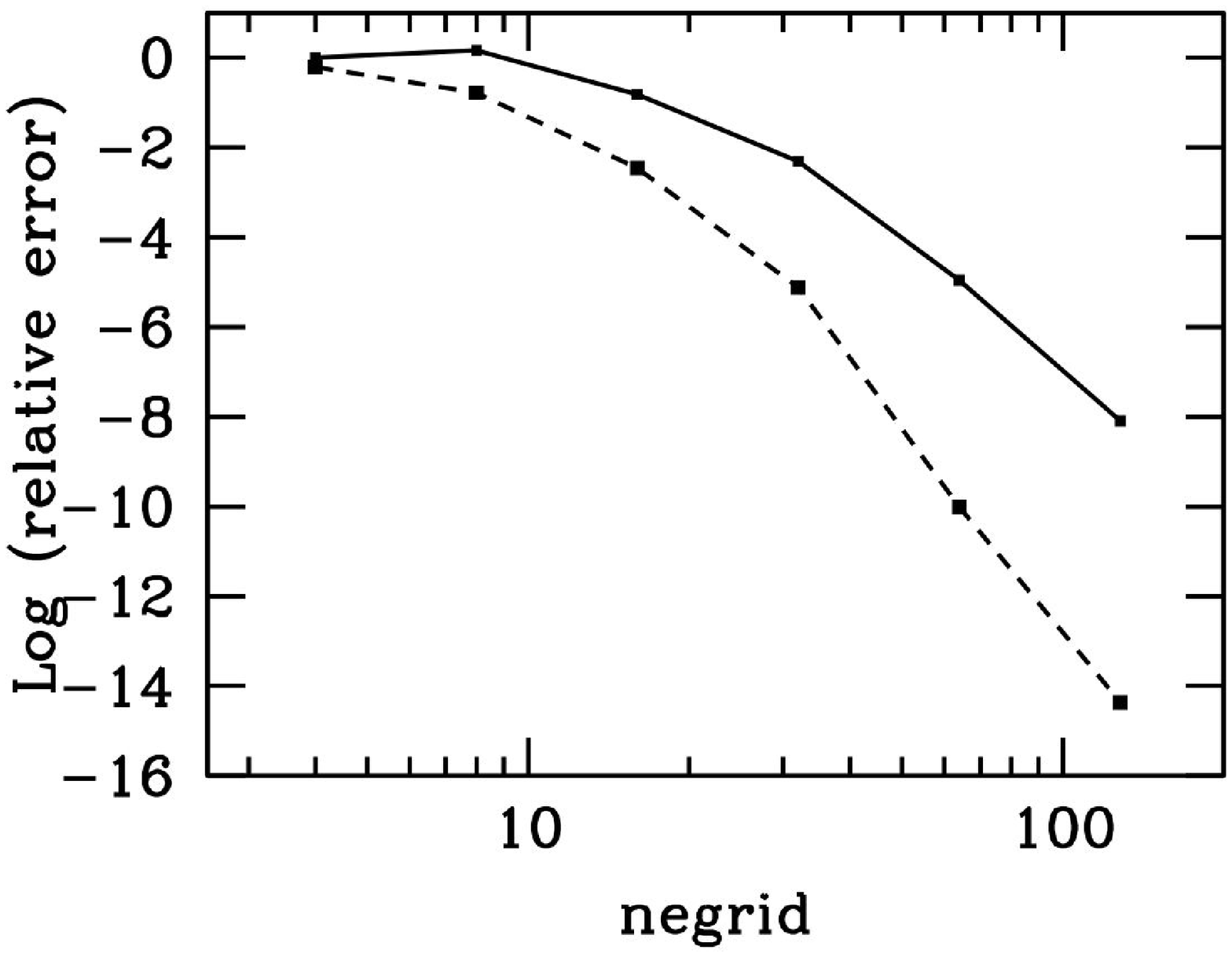}
  \includegraphics[scale=0.3]{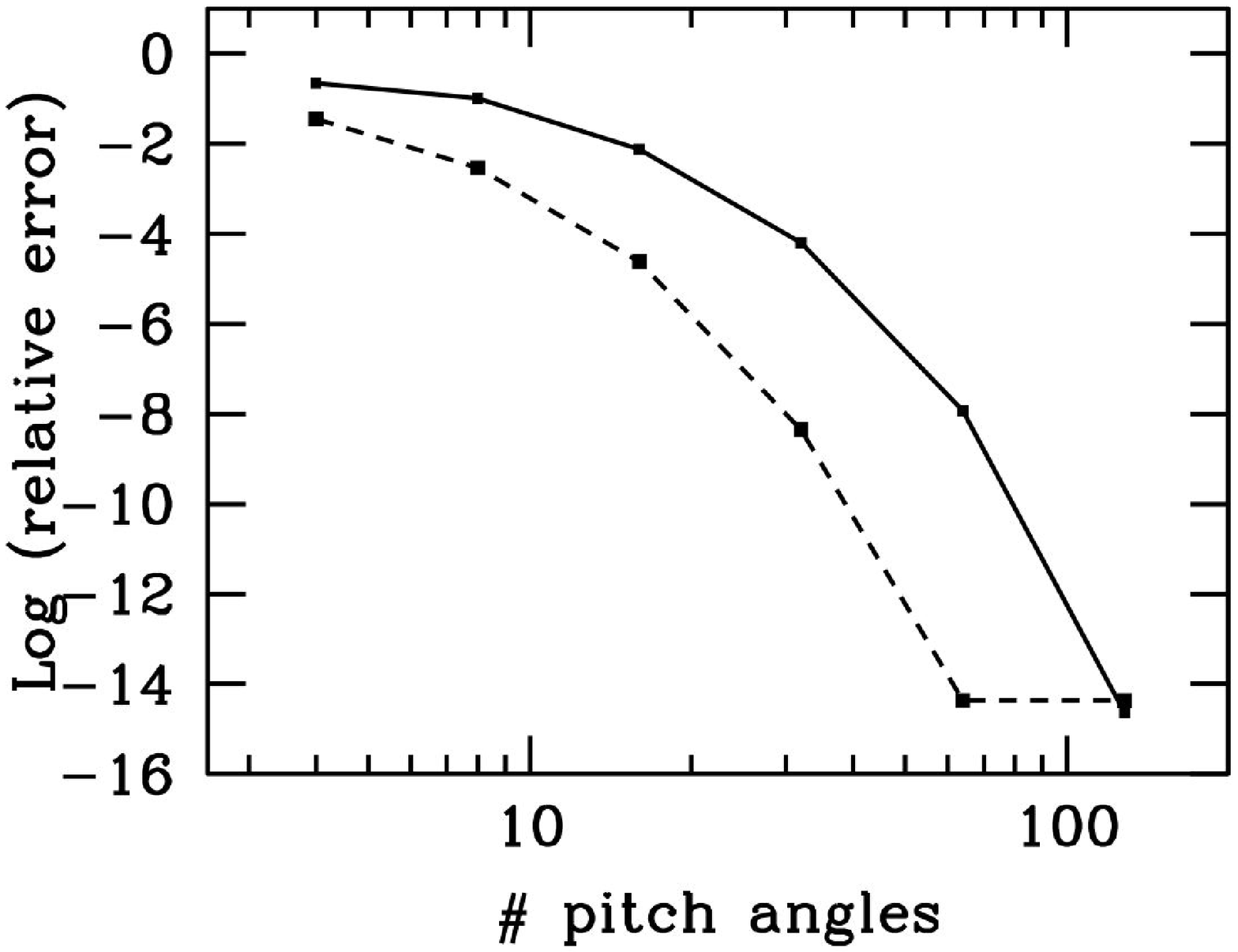}
  \caption{\label{fig:itgconv} 
  Error estimates in ITG frequency and growth rate as the number of
  velocity-space grid points is varied. Left: $N_{E}$ dependence, Right:
  the number of pitch angles (=$2N_{\lambda}$) dependence. The solid line
  is obtained from the 
  \T{AstroGK} velocity-space error diagnostics described in the text,
  and the dashed line is the relative error given by
  \eqref{eq:relative_vel_error}. The error decreases faster than a power
  law, a result of the spectral accuracy of the velocity-space integration. 
  }
 \end{center}
\end{figure}

Results from the convergence study are given in Fig.~\ref{fig:itgconv}.
The relative error, $\epsilon$, is defined as
\begin{equation}
\epsilon =
 \frac{\left|\omega_{\mathrm{agk}}-\omega\right|}{\left|\omega\right|}, 
 \label{eq:relative_vel_error}
\end{equation}
where $\omega_{\mathrm{agk}}$ is the complex frequency computed in
\T{AstroGK} and $\omega$ is the analytic frequency. We see that
$\epsilon$ (dashed line) is less than $1\%$ for $N_E\gtrsim10$ and
$N_{\lambda}\gtrsim3$. Also, the relative error exhibits the
exponential convergence indicative of a spectrally accurate
discretization scheme. Accuracy at large number of velocity-space
grid points is limited only by computational precision, as we see for 
$N_{\lambda}=64$ (double precision is used here).

Also shown in Fig.~\ref{fig:itgconv} is an error estimate calculated
on the fly as a velocity-space resolution diagnostic in \T{AstroGK}
(solid line).  A detailed description of the diagnostic is given
in~\cite{BarnesDorlandTatsuno_10}. The basic idea is to 
calculate the fields using both the standard velocity-space integration
scheme and a less accurate integration scheme that involves dropping a
velocity-space grid point and recalculating integration weights.  By
using the reduced grid and corresponding weights, the spectral accuracy 
is lost. Consequently, the error estimate is quite conservative: it is
essentially computing the error for a simulation with half the order of
accuracy (similar to using half as many grid points).  In fact, we see
in Fig.~\ref{fig:itgconv} that the error estimate agrees rather 
well with the actual error at half the number of grid points. The
qualitative trend is correct, but one must realize that the quantitative
error estimate is conservative.

\subsubsection{Slow mode damping}
\label{sec:ex_slowmode}

For the collision operator verification, we consider slow mode damping in
the low $k_{\perp}\rho_{\mathrm{i}}$, high $\beta_{\mathrm{i}}$ limit.
Here, analytic expressions can be obtained in both the strongly
collisional ($k_{\parallel}\lambda_{\mathrm{mfp}} \ll1$) and collisionless 
($k_{\parallel}\lambda_{\mathrm{mfp}}\gg1$) regimes, where
$\lambda_{\mathrm{mfp}}=v_{\mathrm{th,i}}/\nu_{\mathrm{i}}$ is the ion
mean free path. The complex frequencies are given by
\begin{equation}
\omega =
 -\imag\frac{\left|k_{\parallel}\right|v_{\mathrm{A}}}{\sqrt{\pi\beta_{\mathrm{i}}}} 
\label{eq:slow_mode_collisionless}
\end{equation}
for $k_{\parallel}\lambda_{\mathrm{mfp}}\gg 1$, and
\begin{equation}
\omega = \pm k_{\parallel}v_{\mathrm{A}}
 \sqrt{1-\left(\frac{\mu_{\parallel,\mathrm{i}}
	  k_{\parallel}}{2v_{\mathrm{A}}}\right)^{2}}
 -\imag\frac{\mu_{\parallel,\mathrm{i}}k_{\parallel}^{2}}{2}
\label{eq:slow_mode_collisional}
\end{equation}
for $k_{\parallel}\lambda_{\mathrm{mfp}}\ll 1$, with
$\mu_{\parallel,\mathrm{i}}\propto v_{\mathrm{th,i}}\lambda_{\mathrm{mfp}}$
being the parallel ion viscosity. From these expressions, we
see that the damping in the strongly collisional regime
\eqref{eq:slow_mode_collisional} is due primarily to viscosity, while
the collisionless regime \eqref{eq:slow_mode_collisionless}
is dominated by Barnes damping. 

In order to isolate the slow mode from the Alfv\'en wave, we take
$\phi=A_{\parallel}=0$ throughout the simulation. The electron dynamics
can be neglected because of the high $\beta_{\mathrm{i}}$. Consequently,
the system is described by the ion dynamics coupled with the $\delta
B_{\parallel}$ fluctuation (see Section~6 in~\cite{SchekochihinCowleyDorland_09}).
We initially launch a perturbation of the form,
$\delta f_{\mathrm{i}} \propto (V_{\perp}^{2}/v_{\mathrm{th,i}}^{2}-1)
f_{0\mathrm{i}}$, which generates $\delta B_{\parallel}$ perturbation,
and measure the damping rate of $\delta B_{\parallel}$.

\begin{figure}[htbp]
\begin{center}
\includegraphics[height=3.3in]{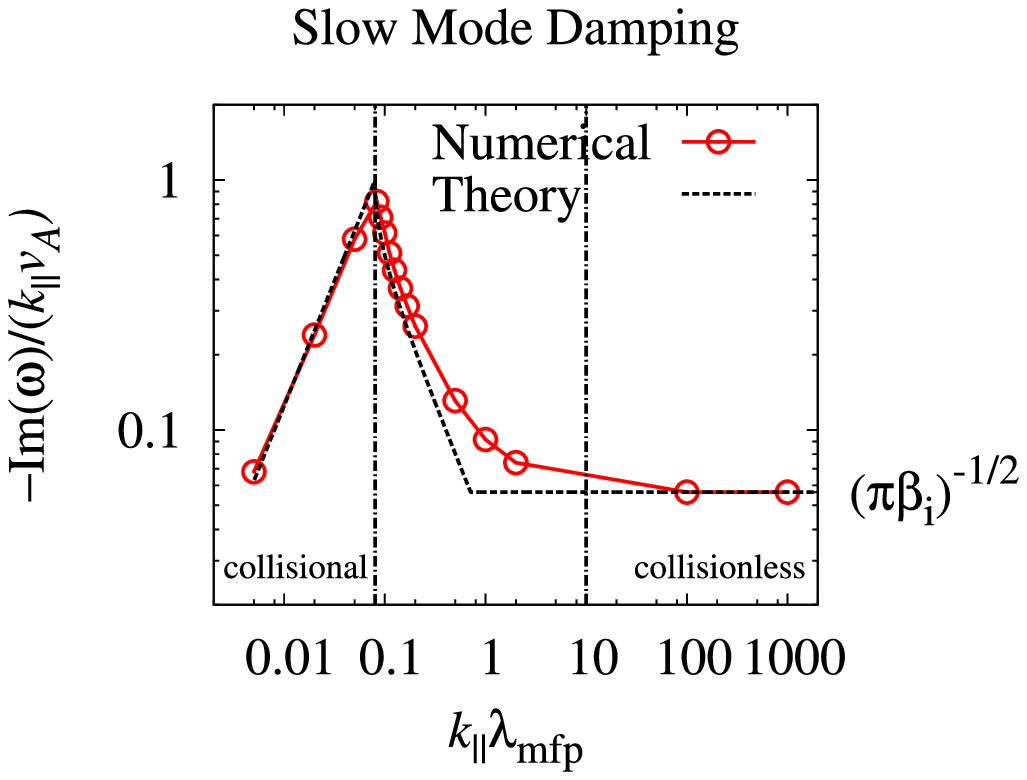}
\end{center}
\caption{\label{fig:nuscan}
 Damping rate of the slow mode for a range of collisionalities
 spanning the collisionless to strongly collisional regimes. Dashed
 lines correspond to the theoretical prediction for the damping rate in
 the collisional ($k_{\parallel}\lambda_{\mathrm{mfp}}\ll 1$) and
 collisionless ($k_{\parallel}\lambda_{\mathrm{mfp}}\gg 1$) limits.
 The ion parallel viscosity is estimated by fitting numerical data with
 the theory as $\mu_{\parallel,\mathrm{i}}\approx 2.5
 v_{\mathrm{th,i}}\lambda_{\mathrm{mfp}}$.
 The solid line is the result obtained numerically with \T{AstroGK}.
 Vertical dot-dashed lines denote approximate regions (collisional and
 collisionless) for which the analytic theory is valid.}
\end{figure}

In Fig.~\ref{fig:nuscan}, we plot the collisional dependence of the
numerically obtained slow mode damping rate for
$k_{\perp}\rho_{\mathrm{i}}=10^{-5}$, $\beta_{\mathrm{i}}=100$. For the
most collisionless case which requires the highest resolution in
velocity space, we take $(N_{Z}, N_{\lambda}, N_{E})=(32, 32, 64)$.
We find quantitative agreement with the analytic
expressions (\ref{eq:slow_mode_collisionless}) and
(\ref{eq:slow_mode_collisional}) in the appropriate regimes. In
particular, we recover the correct 
viscous behavior in the $k_{\parallel}\lambda_{\mathrm{mfp}}\ll 1$ limit
(damping rate proportional to $\mu_{\parallel,\mathrm{i}}$), the correct
collisional damping in the $k_{\parallel}\lambda_{\mathrm{mfp}}\sim 1$
limit (damping rate inversely proportional to
$\mu_{\parallel,{\mathrm{i}}}$), and the correct collisionless
(i.e. Barnes) damping in the $k_{\parallel}\lambda_{\mathrm{mfp}}\gg 1$
limit.
The proporionality constant $c\equiv\mu_{\parallel,\mathrm{i}}/
(v_{\mathrm{th,i}}\lambda_{\mathrm{mfp}})$ estimated from the
simulation is $c\approx2.5$, while $c\approx0.9$ from the Braginskii's
analysis~\cite{Braginskii_65}. 

\subsection{Linear tearing instability}
\label{sec:ex_tearing}

A tearing instability~\cite{FurthKilleenRosenbluth_63} is one kind of
magnetic reconnection process driven by the free magnetic energy
stored in the current sheet configuration, and is a fascinating example to
study in the gyrokinetic framework.

In collisional plasmas, inter-species collisions producing the
resistivity allow topological changes of the magnetic field lines. The
singularity occurs around the magnetic neutral line in the ideal limit,
which is regularized by the finite resistivity. This is a standard
boundary layer problem. The growth rate and boundary layer width
scalings with respect to the Lundquist number $S$ are obtained in~\cite{FurthKilleenRosenbluth_63}.

As the collisionality is decreased, the boundary layer becomes
narrower until kinetic effects come into play, such as the Hall effect
(ion inertia), FLR effects, and electron inertial effects. Therefore,
the scaling law should be altered in such weakly collisional
plasmas. In most situations of interest in fusion and astrophysical
plasmas, these effects can play a crucial role in the problem.

\T{AstroGK} includes full collision physics, and therefore correctly
captures the macroscopic resistivity. (The resistivity is given by
$\eta\approx0.38\mu_{0}\nu_{\mathrm{e}}d_{\mathrm{e}}^{2}$, where
$\nu_{\mathrm{e}}$ is the electron collision frequency and $d_\mathrm{e}$ is
the electron skin depth~\cite{Spitzer_56}.) We provide here a scaling
study of the problem 
in purely two-dimensional setting ($\p/\p Z=0$)
as the collisionality is varied, effectively
varying the 
Lundquist number $S$. We consider an equilibrium distribution function of
electrons as a shifted Maxwellian $\delta f_{\mathrm{e}} \propto
V_{\parallel} f_{0\mathrm{e}}$ to give the following current sheet
configuration:
\begin{equation}
 A_{\parallel}^{\mathrm{eq}} = A_{\parallel0}^{\mathrm{eq}} \cosh^{-2}
  \left(\frac{x-L_{x}/2}{a}\right) S_{\mathrm{h}}(x)
  \label{eq:tearing_background}
\end{equation}
where
\begin{equation}
   S_{\mathrm{h}}(x) =
   \frac{
  \tanh^{2}\left(\frac{2\pi}{L_{x}} x\right)
  + \tanh^{2}\left(\frac{2\pi}{L_{x}} x-2\pi\right)
  - \tanh^{2}(2\pi)
  }{
  2 \tanh^{2}\pi - \tanh^{2}(2\pi)
  }
  \nonumber
\end{equation}
is a shape function to ensure periodicity in the box sized $L_{x}=3.2\pi
a$, $A_{\parallel0}^{\mathrm{eq}}$ is a constant. We take
$a=50\rho_{\mathrm{i}}$ so that the kinetic effect is relatively
weak. We perturb the system with $k_{y}a=0.8$
\footnote{
The initial distribution function of electrons in total becomes:
\begin{equation}
 \delta f_{\mathrm{e}} = 
  \frac{q_{\mathrm{e}}}{T_{0\mathrm{e}}} V_{\parallel}
  f_{0\mathrm{e}}
  \sum_{\bm{k}} (k_{\perp}d_{\mathrm{e}})^{2}
  (A_{\parallel,\bm{k}}^{\mathrm{eq}}+\epsilon A_{\parallel0}^{\mathrm{eq}}
  \delta(k_{x}a=0,k_{y}a=0.8)) e^{k_{\perp}^{2}\rho_{\mathrm{e}}^{2}/4} 
  e^{\imag \bm{k}_{\perp}\cdot\bm{R}_{e}},
\end{equation}
where $A_{\parallel,\bm{k}}^{\mathrm{eq}}$ is a Fourier representation of
\eqref{eq:tearing_background}, the second term gives a small
sinusoidal perturbation with $\epsilon$ being a small constant, and
$\delta$ is the Dirac's delta function.
}
(giving a standard stability index~\cite{FurthKilleenRosenbluth_63}
$\Delta'a\approx23.2$), and observe the
linear growth rate. We fix $(N_{\lambda},N_{E})=(20,16)$, and $N_{x}$
varies from 512 to 2048 to resolve the current layer.

\begin{figure}[htbp]
 \begin{center}
  \includegraphics[scale=0.5]{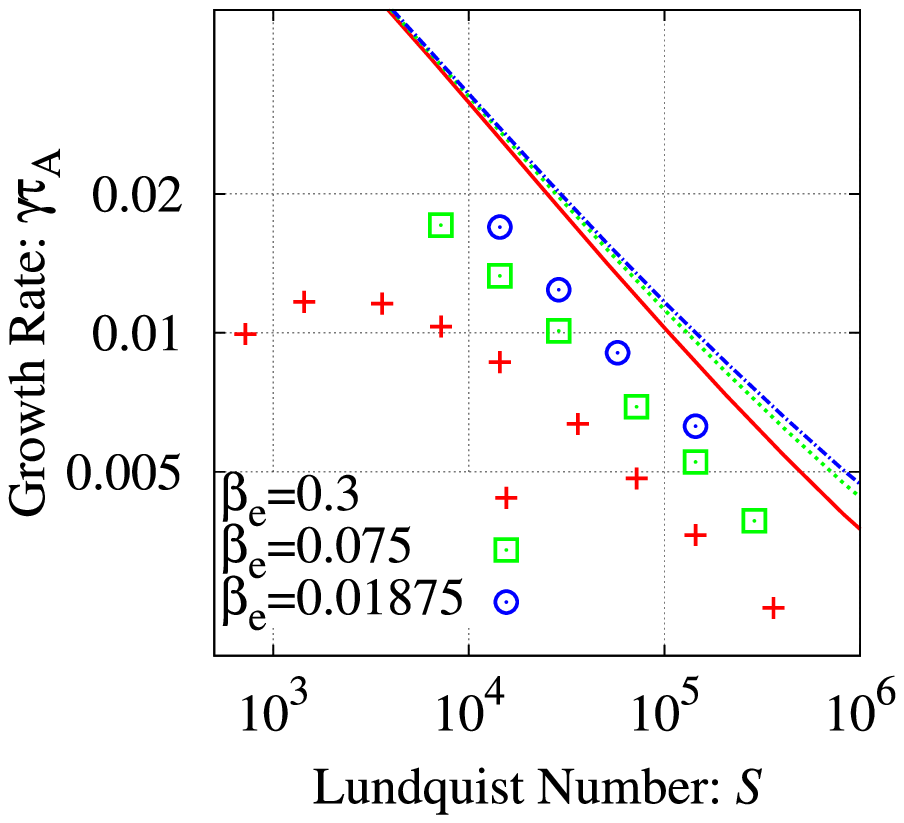}
  \includegraphics[scale=0.5]{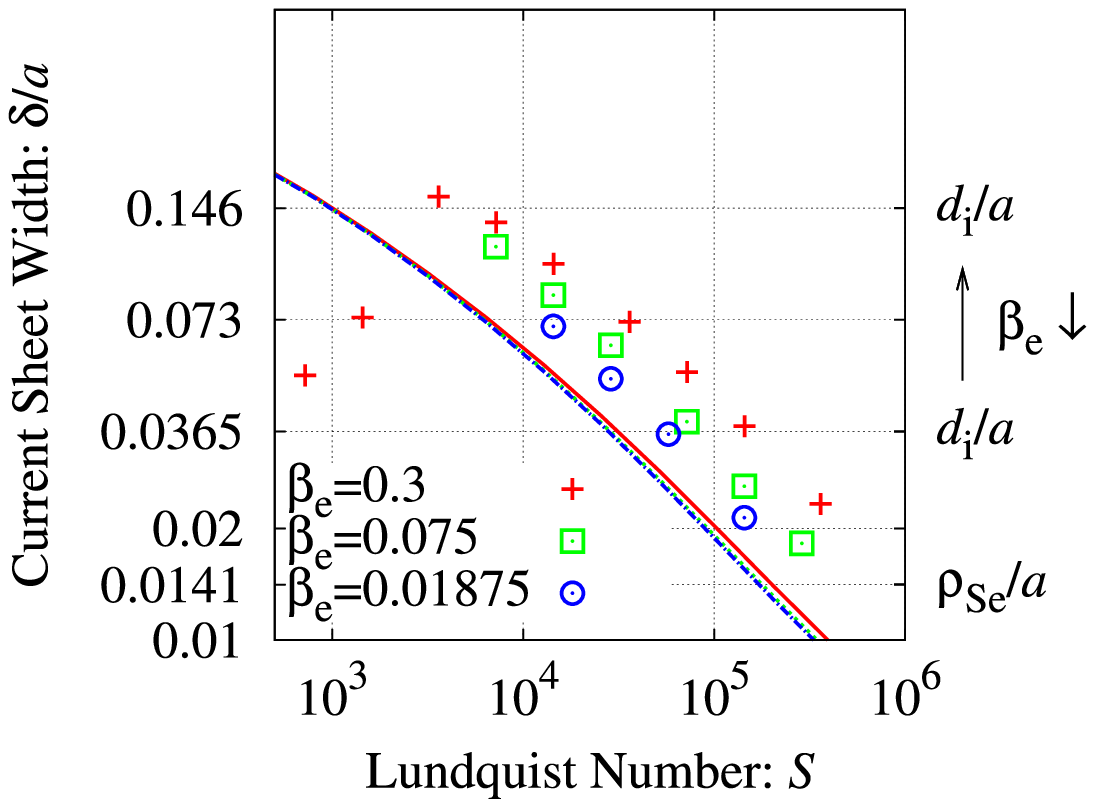}
  \caption{\label{fig:tearing}(Color online)
  Normalized growth rate (left) and current layer width (right) scaling against the
  Lundquist number $S$. Those obtained from \T{AstroGK} (symbols) and
  from the two-fluid model (lines) are shown for three sets of
  $(\beta_{\mathrm{e}},m_{\mathrm{i}}/m_{\mathrm{e}})$.
  }
 \end{center}
\end{figure}

Figure~\ref{fig:tearing} shows the scaling of the growth rate and
current layer width as the Lundquist number $S$ is increased. The
Lundquist number is defined by
$S=\mu_{0}av_{\mathrm{A}\perp}/\eta=2.63(\nu_{\mathrm{e}}\tau_{\mathrm{A}})^{-1}(d_{\mathrm{e}}/a)^{-2}$
with $\tau_{\mathrm{A}}=a/v_{\mathrm{A}\perp}$, so
increased Lundquist number corresponds to decreased electron
collisionality $\nu_\mathrm{e}$.  The growth rate is normalized by the standard
MHD time scale $\tau_{\mathrm{A}}$. The \Alfven
velocity $v_{\mathrm{A}\perp}$ is measured by the peak value of the
background magnetic field given by \eqref{eq:tearing_background}.

For a fixed value of $T_{0\mathrm{i}}/T_{0\mathrm{e}}=1$, we evaluate
the Lundquist number dependence for three sets of parameters:
$(\beta_{\mathrm{e}}, m_{\mathrm{i}}/m_{\mathrm{e}})=$ $(0.3,100)$,
$(0.075,400)$, and $(0.01875,1600)$. For the given parameters, the
characteristic scale lengths of the kinetic effects are the ion sound
Larmor radius
$\rho_{\mathrm{S}e}\equiv\sqrt{T_{0\mathrm{e}}/m_{\mathrm{i}}}/\Omega_{\mathrm{i}}=0.0141a$,
and the ion inertial skin depth $d_{\mathrm{i}}/a=0.365, 0.73,
0.146$. We also show the scaling obtained from the two-fluid MHD model
by Fitzpatrick and Porcelli~\cite{FitzpatrickPorcelli_04} for
reference. For the given parameters, the two-fluid scaling is almost
equal to that from the single-fluid model.

For the small $\beta_{\mathrm{e}}$ case, we observe that the gyrokinetic
scaling is close to that from the fluid model as expected. However, as
$\beta_{\mathrm{e}}$ is increased, the scaling deviates and the growth rate
decreases as kinetic effects become non-negligible. The major
difference between the two-fluid model and the kinetic simulation
using \T{AstroGK} originates from the treatment of the second-order
velocity moment (temperature or pressure). The fluid model assumes
adiabatic ions and isothermal electrons, whereas the gyrokinetic
pressure is generally tensorial and  thus contains far richer
physics~\cite{NumataDorlandHowes_10}.


\subsection{Orszag--Tang vortex problem} 
\label{sec:ex_ot}

To validate that results of \T{AstroGK} for a nonlinear
electromagnetic problem, we present here the well-known MHD vortex problem of
Orszag and Tang~\cite{OrszagTang_79}. Their original simulation solves incompressible
reduced MHD equations for the stream function $\varphi$ ($= -\phi /
B_{0}$) and flux function $\psi$ ($= A_{\parallel}$) defined in the
plane perpendicular to the mean magnetic field:
\begin{align}
 \pdf{}{t} \nabla_{\perp}^{2} \varphi
 + \left\{\varphi, \nabla_{\perp}^{2} \varphi \right\}
 = &
 \frac{1}{\mu_{0}n_{0\mathrm{i}}m_{\mathrm{i}}} 
 \left\{\psi, \nabla_{\perp}^{2} \psi\right\}
 + \frac{\mu}{n_{0\mathrm{i}}m_{\mathrm{i}}} \nabla_{\perp}^{4} \varphi,
 \label{eq:ot_reducemhd_vor}\\
 \pdf{}{t} \psi + \left\{\varphi, \psi\right\}
 = & \frac{\eta}{\mu_{0}} \nabla_{\perp}^{2} \psi,
 \label{eq:ot_reducemhd_psi}
\end{align}
where $\mu$ and $\eta$ are the viscosity and resistivity.
The nonlinear evolution of a system governed by the reduced MHD
equations provides a useful comparison for the nonlinear evolution of
the GK-M equations using \T{AstroGK}. Since GK-M equations in the
large-scale limit, $k_{\perp} \rho_{\mathrm{i}} \ll 1$, simplify to the
equations of reduced MHD~\cite{SchekochihinCowleyDorland_09}, the
results of \T{AstroGK} in 
this limit should be similar to the results of a reduced MHD code,
which does not contain the small scale physics that occurs when
$k_{\perp} \rho_{\mathrm{i}} \gtrsim 1$.

Given an initial condition:
\begin{align}
 \varphi = & -2 a v_{\mathrm{A}\perp} \left(\cos 2\pi \frac{x}{L_{x}} + \cos 2\pi
 \frac{y}{L_{y}}\right), 
 \label{eq:ot_initial_condition_MHD_phi} \\
 \psi = & a B_{\perp 0} \left(\cos 4\pi \frac{x}{L_{x}} + 2 \cos 2\pi
 \frac{y}{L_{y}} \right), 
 \label{eq:ot_initial_condition_MHD_psi}
\end{align}
where $v_{\mathrm{A}\perp}=B_{\perp
0}/\sqrt{\mu_{0}n_{0\mathrm{i}}m_{\mathrm{i}}}$ and $L_{x}=L_{y}=2\pi
a$,
we have made a simulation of the initial value problem using an
independent reduced MHD code. The initial conditions clearly define the
standard MHD units for normalization: namely, system size $a$ and
Alfv\'en transit time $\tau_{\mathrm{A}} = a/v_{\mathrm{A}\perp}$. The
dimensionless dissipation parameters are the Reynolds number $R_{\mathrm{e}}\equiv a
v_{\mathrm{A}\perp} n_{0}m_{\mathrm{i}}/\mu$ and the Lundquist number 
$S\equiv\mu_{0}av_{\mathrm{A}\perp}/\eta$. 
We use $(N_{x}, N_{y}) = (256, 256)$ grid points with the dissipation
coefficients $1/R_{\mathrm{e}} = 1/S = 1.5 \times 10^{-3}$.

For the \T{AstroGK} simulation, we set initial conditions on the
distribution function for each species  $h_{s}$ to give the
corresponding fields. Therefore, we define the distribution functions by
\begin{align}
 h_{\mathrm{i}} = & C_{\phi} \left(\cos 2\pi \frac{x}{L_{x}} + \cos
 2\pi \frac{y}{L_{y}} \right) f_{0\mathrm{i}},
 \label{eq:ot_init_ion_dist_func} \\
 h_{\mathrm{e}} = & C_{A_{\parallel}} \left(2 \cos 4\pi \frac{x}{L_{x}}
 + \cos 2\pi \frac{y}{L_{y}}\right) V_{\parallel} f_{0\mathrm{e}},
 \label{eq:ot_init_electron_dist_func}
\end{align}
where $C_{\phi}$ and $C_{A_{\parallel}}$ are coefficients chosen such
that the resulting $\phi$ and $A_{\parallel}$ are equivalent to
\eqref{eq:ot_initial_condition_MHD_phi} and
\eqref{eq:ot_initial_condition_MHD_psi}. 
We use $(N_{x}, N_{y}, N_{\lambda}, N_E) = (256, 256, 32, 16)$ points,
$T_{0\mathrm{i}}/T_{0\mathrm{e}}=1$,
$m_{\mathrm{i}}/m_{\mathrm{e}}=1836$, $\beta_{\mathrm{i}} = 10^{-3}$,
$\rho_{\mathrm{i}}/a = 0.01$, and ignore collisions. We assume
$\delta B_{\parallel}=0$.

\begin{figure}
 \begin{center}
 \includegraphics[height=8cm,angle=270]{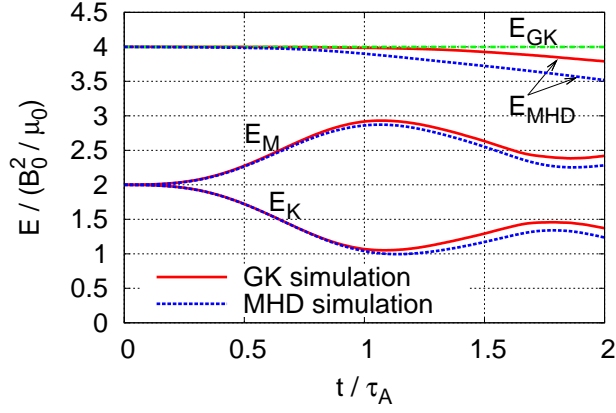}
 \caption{\label{fig: kOT energy}Time evolution of
 various energies. 
 Red solid lines represent values from \T{AstroGK} and blue dashed
 from reduced MHD simulations.
 Note that $E_{\rm GK}$ is only defined for \T{AstroGK} simulation.
}
\end{center}
\end{figure}

Time evolutions of various energies are shown in Fig.~\ref{fig: kOT energy}
for both codes.
Time and the energy are normalized to the MHD units,
i.e. $\tau_{\mathrm{A}}$ and $B_{\perp 0}^{2}/\mu_{0}$. The result from 
\T{AstroGK} simulation also shows the gyrokinetic energy:
\begin{align}
 E_{\mathrm{GK}} = &
 L_{x}L_{y}\sum_{\bm{k}_{\perp}}
 \left[
 \sum_{s} \int 
 \frac{T_{0s}|h_{s,\bm{k}_{\perp}}|^{2}}{2f_{0s}}
 \diff \bm{v}
 - \left(\sum_{s}\frac{q_{s}^{2}n_{0s}}{2T_{0s}}\right)
 |\phi_{\bm{k}_{\perp}}|^{2}
 + \frac{k_{\perp}^{2} |A_{\parallel,\bm{k}_{\perp}}|^{2}}{2\mu_{0}}
 \right]
 \nonumber \\
 = &
 L_{x}L_{y}\sum_{\bm{k}_{\perp}}
 \left[
 \sum_{s} \int
 \frac{T_{0s}|g_{s,\bm{k}_{\perp}}|^{2}}{2f_{0s}}
 \diff \bm{v}
 + \left(\sum_{s} \frac{q_{s}^{2} n_{0s}}{2T_{0s}}\right)
 (1 - \Gamma_{0s}) |\phi_{\bm{k}_{\perp}}|^{2}
 + \frac{k_{\perp}^{2} |A_{\parallel,\bm{k}_{\perp}}|^{2}}{2\mu_{0}}
 \right],
 \label{eq:ot_energy_gk_phys_unit}
\end{align}
  normalized
into the MHD units.
The kinetic energy ($E_{\mathrm{K}}=L_{x}L_{y}
\sum_{\bm{k}_{\perp}}\left(n_{0\mathrm{i}}m_{\mathrm{i}}/2\right)|k_{\perp}\varphi_{\bm{k}_{\perp}}|^{2}$)
and magnetic energy ($E_{\mathrm{M}}=L_{x}L_{y}\sum_{\bm{k}_{\perp}}
|k_{\perp}\psi_{\bm{k}_{\perp}}|^{2}/(2\mu_{0})$) evolve 
similarly in the two models. It is noted that 
the leading contribution of the ion part of the second 
term in \eqref{eq:ot_energy_gk_phys_unit} yields the kinetic
energy. Thus the difference between the gyrokinetic energy
$E_{\mathrm{GK}}$ and the MHD energy
$E_{\mathrm{MHD}}=E_{\mathrm{K}}+E_{\mathrm{M}}$ comes from $g_{s}^{2}$
and the electron part of the second term in
\eqref{eq:ot_energy_gk_phys_unit}. 
The apparent agreement of $E_{\mathrm{GK}}$ and $E_{\mathrm{MHD}}$ in
the initial phase is not strange since the energy contained in the fields
is much bigger than that in $g$ by a factor of $\beta_{\mathrm{i}}^{-1}$. 
This is true for the Alfv\'enic dynamics---the pure linear Alfv\'en
wave propagates with approximately zero $g$~\cite{SchekochihinCowleyDorland_09}. 

\begin{figure}
 \begin{center}
  \includegraphics[height=6.5cm,angle=270]{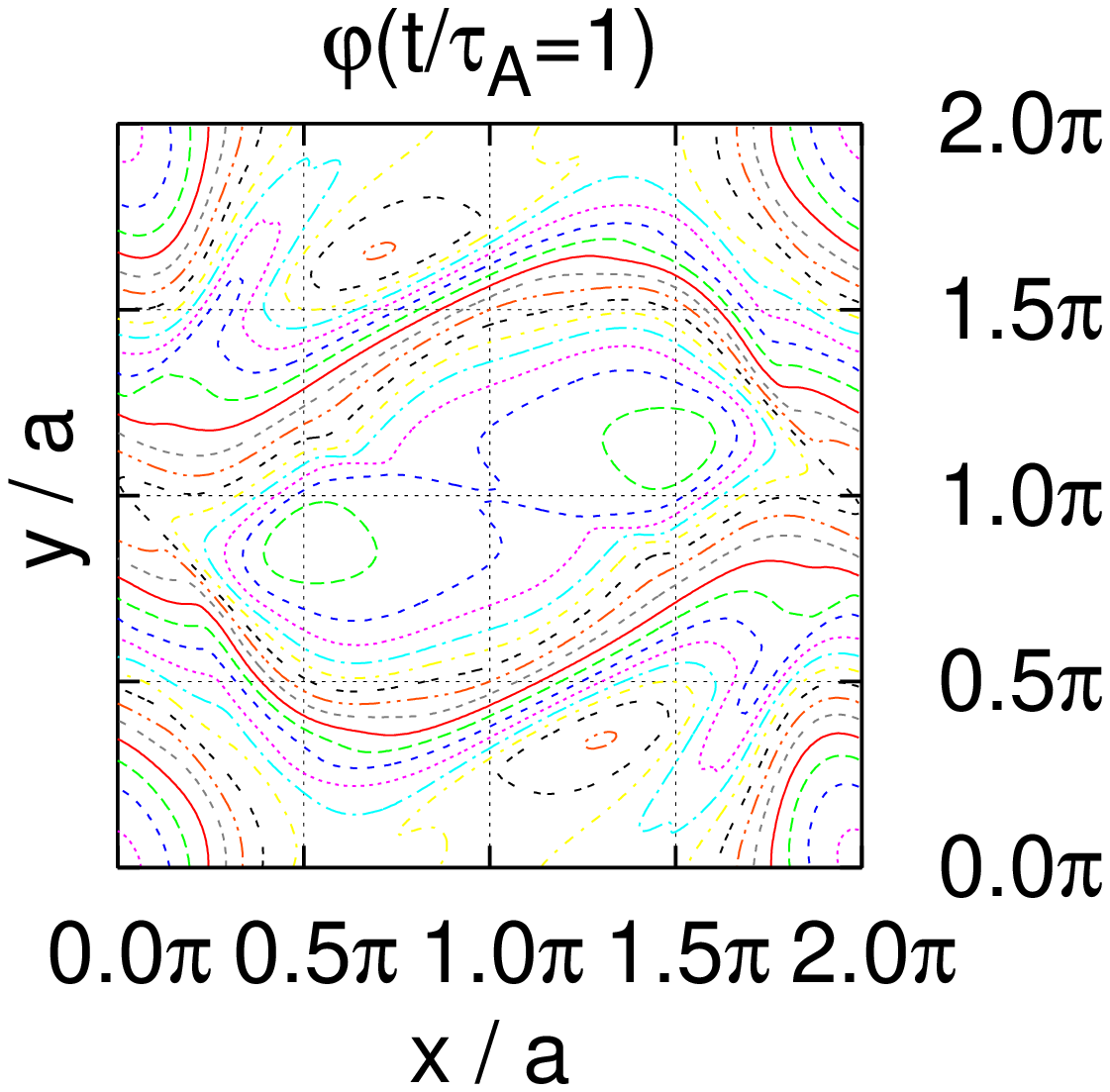}
  \includegraphics[height=6.5cm,angle=270]{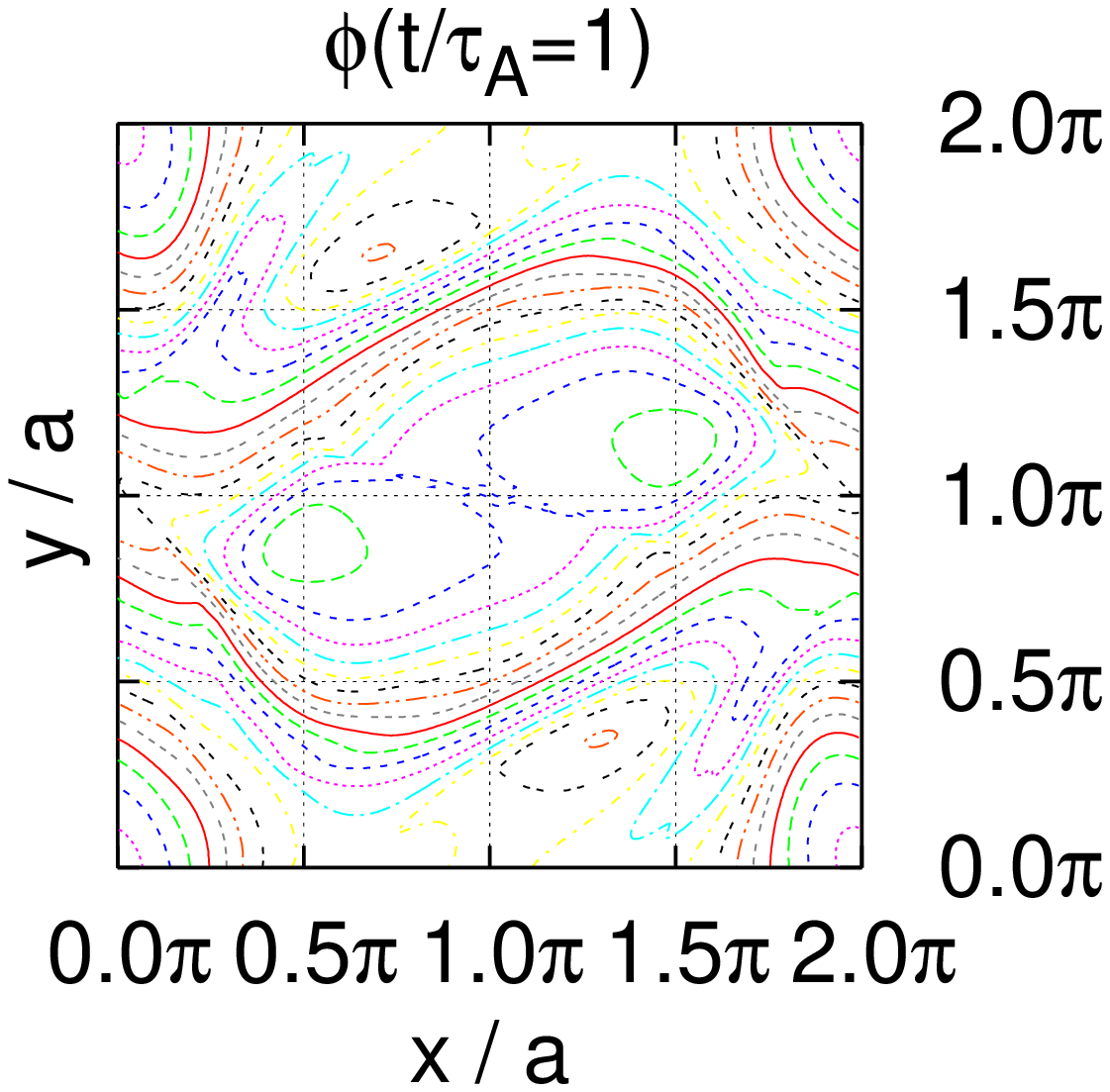}
  \includegraphics[height=6.5cm,angle=270]{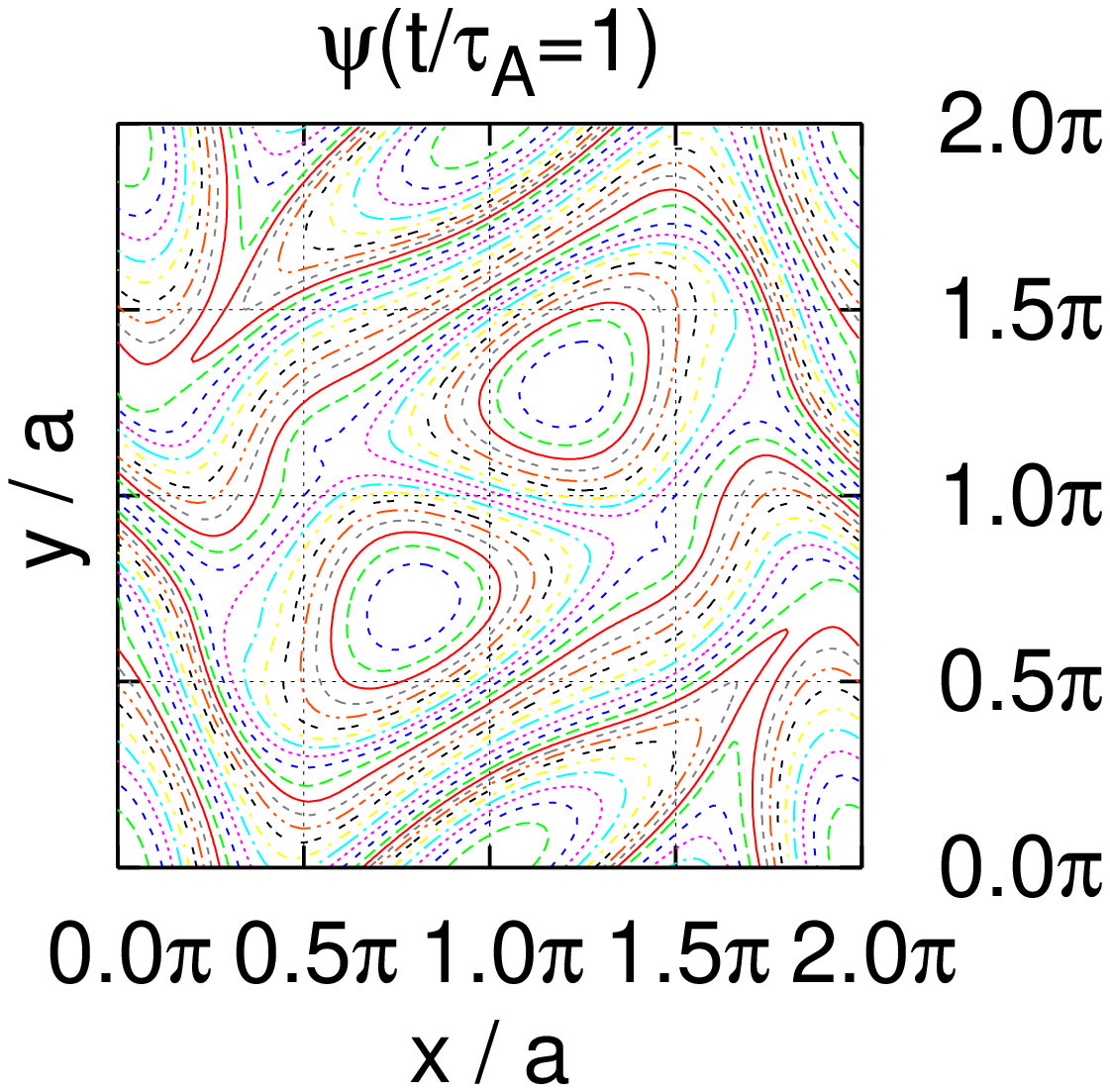}
  \includegraphics[height=6.5cm,angle=270]{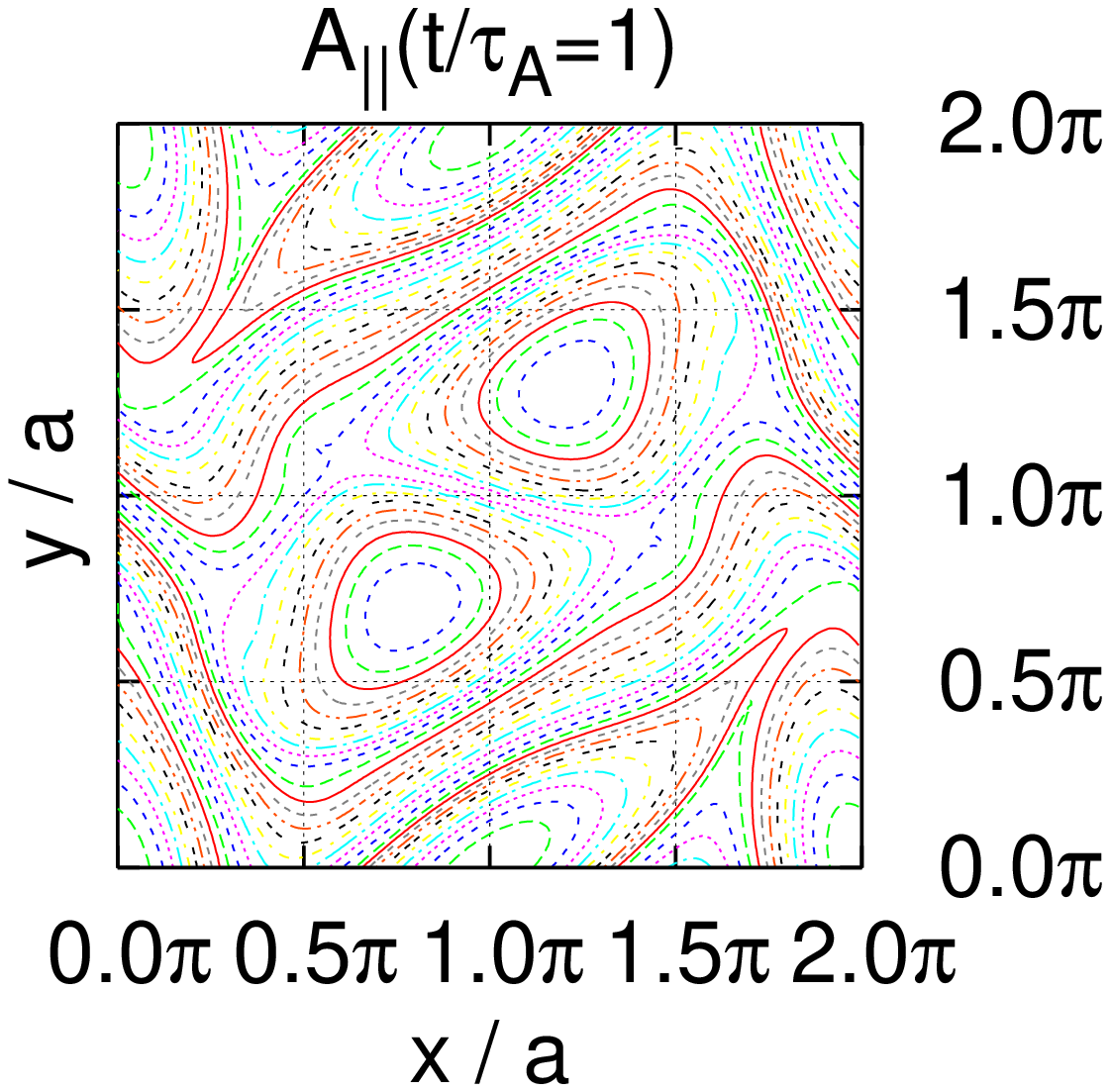}
  \caption{\label{fig: kOT contours}
  Stream and flux functions at $t/\tau_{\mathrm{A}}=1$ taken from reduced MHD (left) and
  \T{AstroGK} (right) simulations.
  }
 \end{center}
\end{figure}

Contour plots of the stream and flux functions are also shown in
Fig.~\ref{fig: kOT contours}.
The overall agreement is almost perfect, which implies that \T{AstroGK}
correctly reproduces MHD results. The small differences in these plots,
especially at the small scales, are due to the kinetic effects resolved by 
\T{AstroGK} and are therefore physically meaningful. A more thorough discussion
of these differences will be discussed elsewhere.


\section{Performance}
\label{sec:performance}

In this section, we determine the scaling of \T{AstroGK} with each of
the problem dimensions and evaluate the strong and weak scalings of
the parallel performance.  The scaling performance tests in this
section employ a nonlinear simulation of driven turbulence with plasma
parameters
$\beta_{\mathrm{i}}=T_{0\mathrm{i}}/T_{0\mathrm{e}}=n_{0\mathrm{i}}/n_{0\mathrm{e}}=-q_{\mathrm{i}}/q_{\mathrm{e}}=1$,  
$m_{\mathrm{i}}/m_{\mathrm{e}}=1836$. The simulation is stirred by an
antenna at the smallest wavenumber in the box corresponding to
$k_{\perp} \rho_{\mathrm{i}}=1$ and collisions are turned off.

\subsection{Single processor scaling with problem dimensions}
\label{sec:perf_single}

\begin{figure}[htbp]
 \begin{center}
  \includegraphics[scale=0.7]{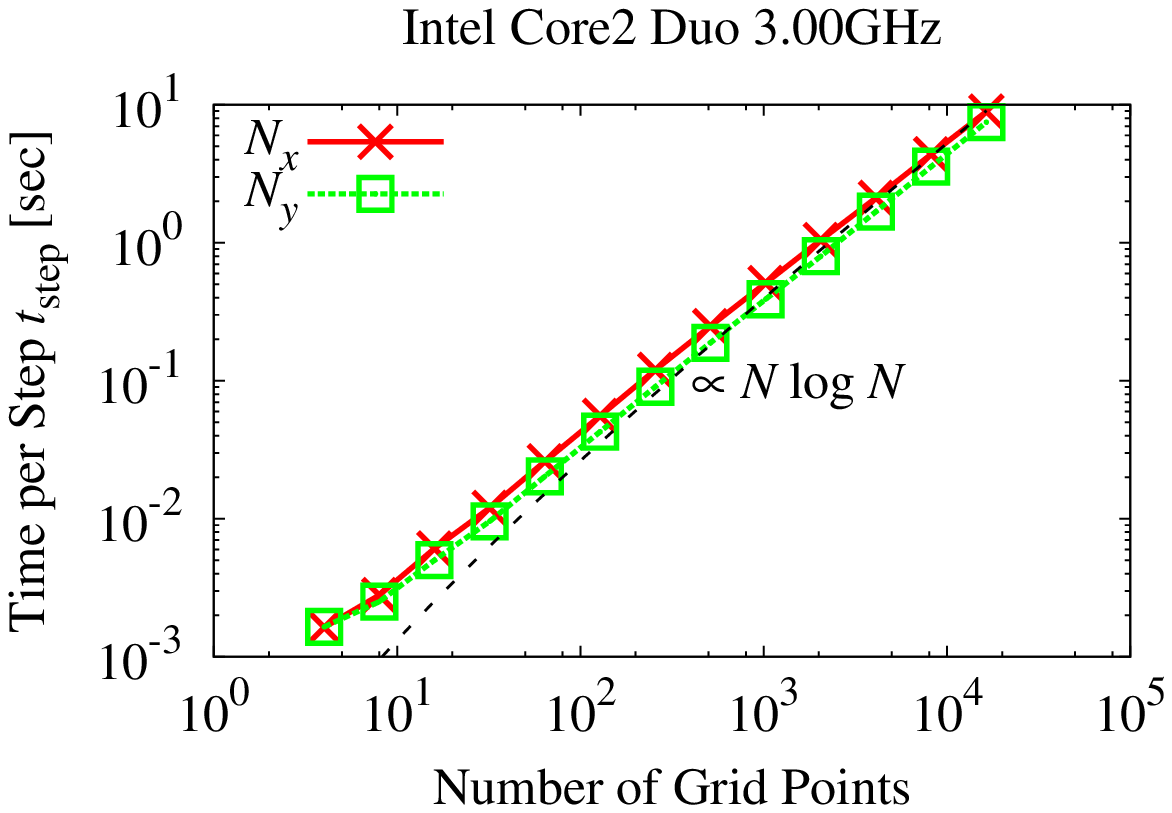}
  \includegraphics[scale=0.7]{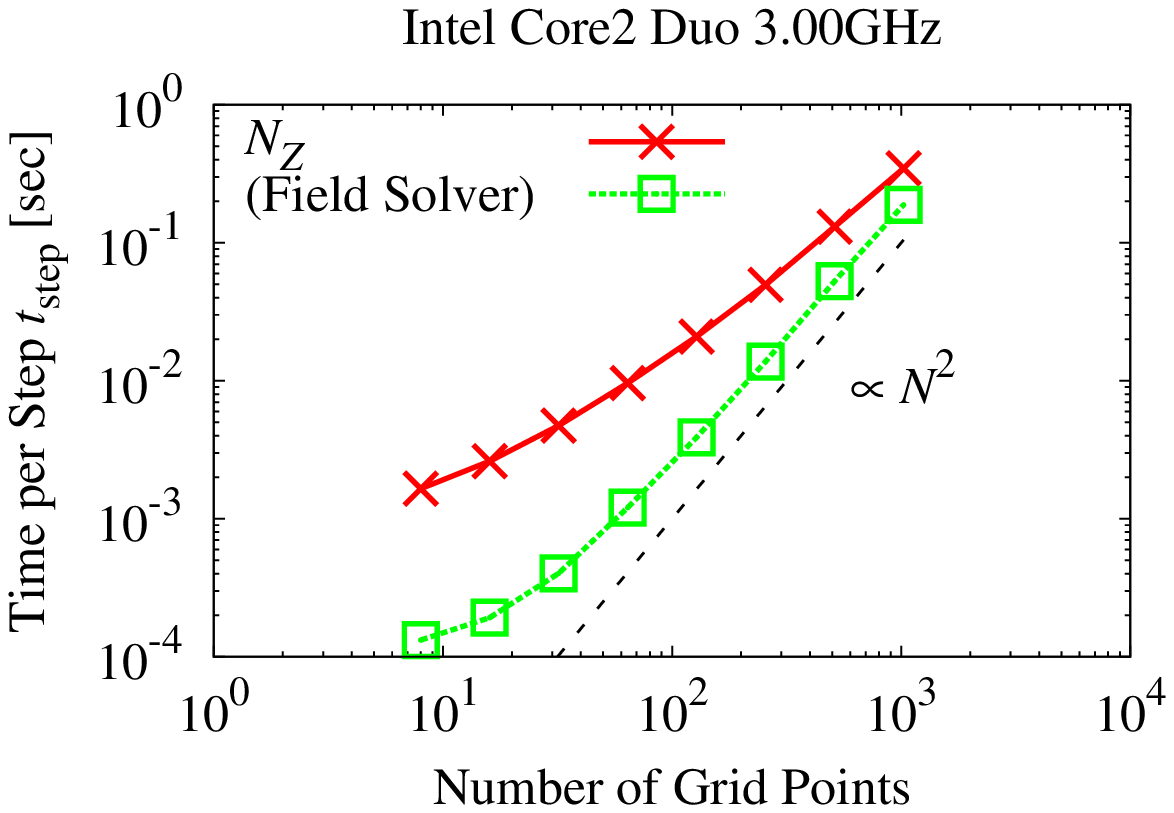}
  \includegraphics[scale=0.7]{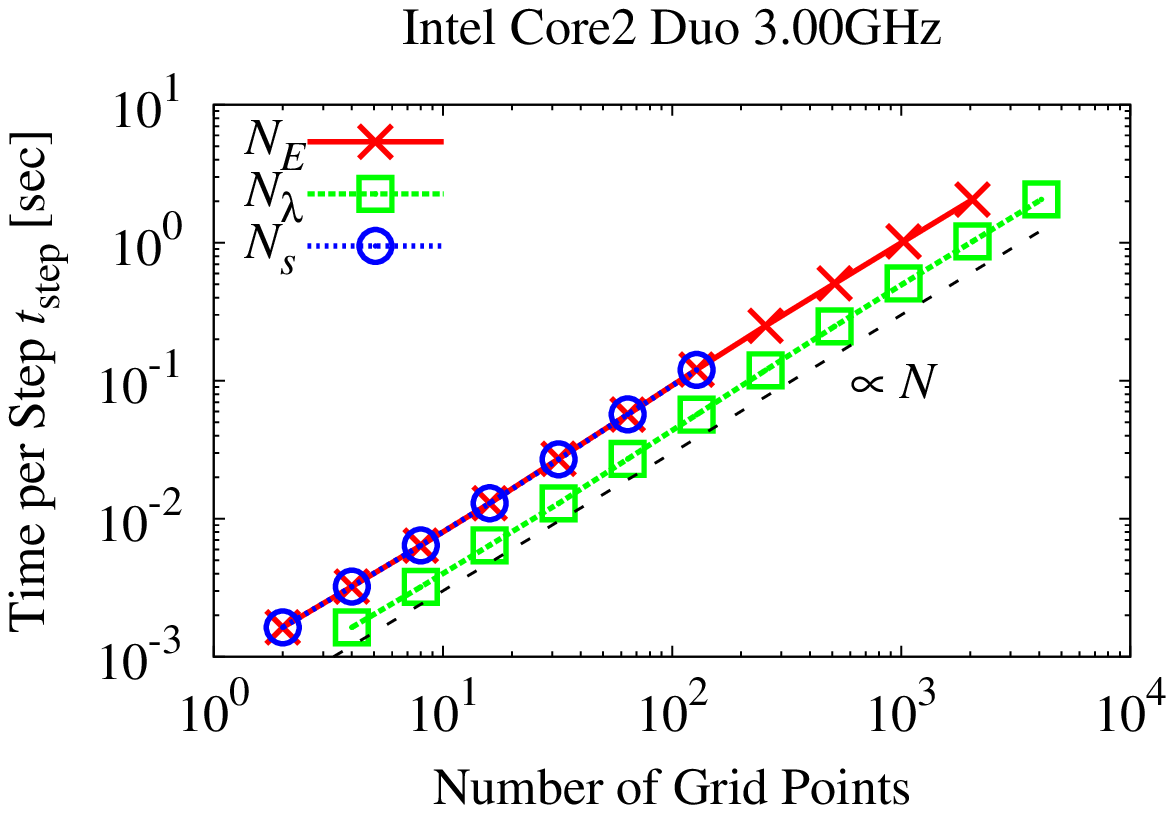}
  \caption{\label{fig:spnxy}Single processor scaling of the time per step
  $t_{\mathrm{step}}$ vs.~$N$, where $N$ corresponds to; either the
  number of grid points $N_{x}$ (crosses) or $N_{y}$ (squares) in the
  top panel, $N_{Z}$ in the middle panel, and either $N_{\lambda}$
  (crosses) or $N_{E}$ (squares) or $N_{s}$ (circles) in the bottom
  panel. The asymptotic scalings are achieved for $N_{x,y}$ as $N \log
  N$ (top), and for $N_{E,\lambda,s}$ as $N$ (bottom). For $N_{Z}$
  scaling (middle), the asymptotic scaling is not achieved. The field
  solver scales as $N_{Z}^{2}$, but is still subdominant for the
  practical problem size.
  }
 \end{center}
\end{figure}

We can determine the scaling of the time per step in \T{AstroGK} as
each of the dimensions of the problem is increased. These tests are
performed on a single processor to eliminate the requirement for
communications between processors. We begin with a small nonlinear run
with the following problem dimensions:
$(N_{x},N_{y},N_{Z},N_{\lambda},N_{E},N_{s})=(4,4,8,4,2,2)$. To
test the scaling of the time per computational step for a given
dimension, we increase only that dimension successively by a factor of
two until the problem is too large for available memory; the time per
step is measured for each of these runs. The results are presented in
Fig.~\ref{fig:spnxy}. The $N_{x}$ and $N_{y}$ dimensions scale
asymptotically as $N \log N$, as expected for fast Fourier
transforms. The timestep is expected to scale with the number of grid
points along the mean magnetic field as $N_{Z}^{2}$ because of the field 
solver. However, for practical problem sizes ($N_{Z}\lesssim1000$), the
field solver is still subdominant compared with the gyrokinetic
solver. Therefore, we observe $N_{Z}^{c}$ dependence with $c<2$. In each
of the dimensions $N_{\lambda}$, $N_{E}$, and $N_{s}$, the scaling is 
linear as anticipated.  Thus, the wallclock time per step on a single
processor scales as
\begin{equation}
 t_{\mathrm{step}} \propto (N_{x} \log N_{x})(N_{y} \log
  N_{y})(N_{Z}^{c})(N_{\lambda})(N_{E})(N_{s}). 
\end{equation}

\subsection{Parallel performance scaling}
\label{sec:parallel_performace}

Parallel performance of \T{AstroGK} is measured by taking the weak and
strong scalings: The weak scaling is probed by holding the computational
work per processing core constant while the number of cores, thus the
total problem size, is increased. On the other hand, the strong scaling is
probed by holding the problem size constant while the number of
processing cores is increased. Both tests are performed on \T{Kraken}
Cray XT5 system at the National Institute for Computational Sciences at
the University of Tennessee. \T{Kraken} consists of 8256 compute nodes
each having 12 processing cores, resulting in 99,072 compute cores in
total.

The number of grid points is chosen such that parallelization is
achieved efficiently in the layout `{\tt yxles}'.  In fully developed
kinetic turbulence, fine structure develops in velocity space as well as
in position space, thus, it is required to take the same order of
grid points in both spaces. The referenced maximum number of
total grids is $256^{2}\times128^{2}$ for the current highest
resolution runs~\cite{TatsunoDorlandSchekochihin_09}. Here, however,
position-space resolution is taken relatively small in the strong
scaling because of the memory requirement for small number of
processors.

\subsubsection{Weak scaling}
\label{sec:perf_weak}

\begin{figure}[htbp]
 \begin{center}
 \includegraphics[scale=1.]{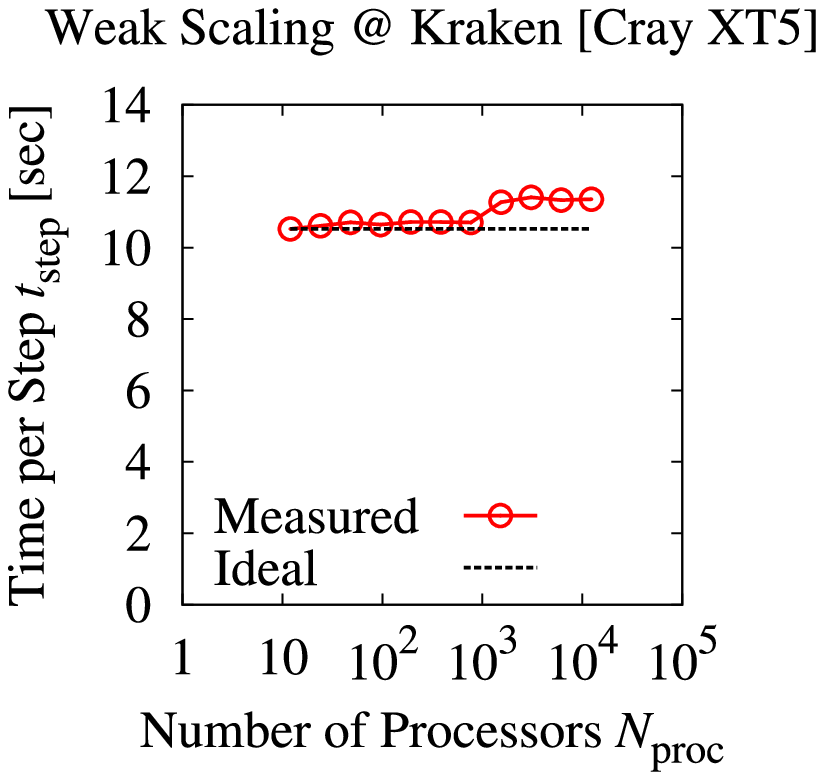}
  \caption{\label{fig:ws} 
  Weak scaling of \T{AstroGK} determined by holding the
  computational work per processor constant while the number of
  processors is increased. The time per step $t_{\mathrm{step}}$
  vs. the number of processing cores $N_{\mathrm{proc}}$
  is plotted.
  }
 \end{center}
\end{figure}

The initial problem uses $(N_{x},N_{y},N_{Z},N_{\lambda},N_{E}, N_{s})=
(160,160,36,6,4,2)$ on 12 processing cores. Each time the processing
core count is doubled, the problem size is doubled by alternately
doubling first $N_{E}$ and then $N_{\lambda}$---since both of these
dimensions scale linearly with the problem size, doubling one of these
dimensions effectively doubles the computational work, leading to fair
assessment of the weak scaling.
The weak scaling behavior of the wallclock time per step
$t_{\mathrm{step}}$ vs. the number of processing cores up to
$N_{\mathrm{proc}}=$ 12,288 is plotted in Fig.~\ref{fig:ws}. 
\T{AstroGK} follows the ideal scaling until
$N_{\mathrm{proc}}=$ 12,288 with slight degradation of performance ($\sim
5\%$) due to the increase of communication for $N_{\mathrm{proc}}>1000$.
The layout specified for the parallel communication for this test is 
`\T{yxles}'.


\subsubsection{Strong scaling}
\label{sec:perf_strong}

\begin{figure}
 \begin{center}
  \includegraphics[scale=1.]{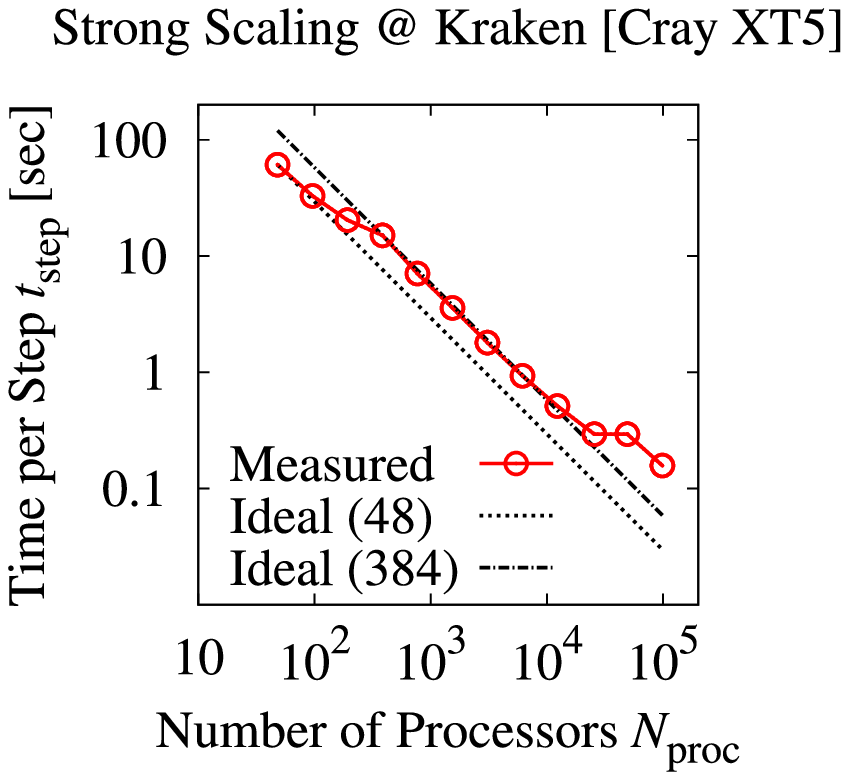}
  \caption{\label{fig:ss}
  Plot of strong scaling taken by increasing the number of
  processors for the fixed problem size. The time per step
  $t_{\mathrm{step}}$ vs. the number of processors $N_{\mathrm{proc}}$
  is shown. Ideal linear scaling lines compared with
  $N_{\mathrm{proc}}=48$ and $N_{\mathrm{proc}}=384$ are depicted.
  }
 \end{center}
\end{figure}

The dimensions of the nonlinear turbulence problem employed for this
scaling are
$(N_{x},N_{y},N_{Z},N_{\lambda},N_{E},N_{s})=(32,32,24,192,256,2)$.
The strong scaling behavior of the wallclock time per step
$t_{\mathrm{step}}$ vs. the number of processors $N_{\mathrm{proc}}$
is plotted from $N_{\mathrm{proc}}=48$ to $N_{\mathrm{proc}}=$ 98,304 in
Fig.~\ref{fig:ss}. Again, the layout specified for the parallel
communication is `\T{yxles}'.

To accommodate this large computational problem on a small number of
processors requires more memory per core than is available when all 12
cores on a compute node are used.  Therefore, for the lowest four data
points on the scaling curve (up to 384 processors), only 1, 2, 4, and
8 core(s) are utilized per node. The rest of the runs all utilize 12
cores per node. As the number of cores per node increases, the
computation time deviates from the ideal linear scaling (``Ideal
(48)'' line in the figure). The sharing of communication and memory
bandwidth between multiple cores lead to a factor of two degradation
of performance. If the number of cores per node is fixed at 12, we
observe a nearly ideal strong scaling from $N_{\mathrm{proc}}=384$ up
to $N_{\mathrm{proc}}=$ 24,576, as indicated by ``Ideal (384)'' line in
the figure. Significant performance loss occurs only at
$N_{\mathrm{proc}}=$ 49,152.





\section{Summary}
\label{sec:summary}

We have presented detailed descriptions of the gyrokinetic-Maxwell
equations solved in \T{AstroGK} and the algorithms adopted in the
code. It employs an unconditionally stable implicit method for the
linear terms and a third-order explicit  multistep method for the
nonlinear term. To reduce the computational cost for the implicit solve of
the linear term, it utilizes the Godunov splitting method. Together
with the Green's function approach developed by Kotschenreuther et
al., \T{AstroGK} solves an $N_{Z} \times N_Z$ size linear system at
each velocity grid point 
for each species. For collisionless runs, the overall accuracy of the
algorithm is second order in $\Delta t$ and second order in $\Delta
Z$, with spectral convergence for the velocity-space integration; for 
collisional runs, the velocity-space derivatives lead to a drop
in the overall accuracy to first order in $\Delta t$ and first order
in the velocity-space integration.

The computational cost on a single processor follows the theoretical
scaling except for the $N_{Z}$ dependence. The time per computational
step is expected to follow $N_{Z}^{2}$ because of the field solver for
large $N_{Z}$. However, for practical problem sizes $N_{Z}\lesssim
1000$, the cost of the gyrokinetic solver is still larger or
comparable with that of the field solver, and we do not observe the
asymptotic scaling.  Excellent parallel performance has also been
demonstrated for both weak and strong scaling tests, showing the ideal
scaling up to about 10,000 processors.

The algorithms at the heart of \T{AstroGK} are the same as those in
\T{GS2}, but the \T{AstroGK} code has been streamlined by the removal 
of the magnetic geometry and trapped particle effects. Therefore, it
is optimized for the study of fundamental, low-frequency kinetic
effects in simple plasma geometries and for the exploration of the
dynamics in many astrophysical plasmas of interest, in which  the mean
magnetic field at Larmor radius scales can often be well approximated
as straight and uniform.

\T{AstroGK} is also an ideal developmental testbed, both for novel
computational approaches and application to new physical plasma
systems. One idea currently under development is the 
treatment of electrons as a fluid, since certain situations are
indifferent to the kinetic behavior of the electrons.  For instance,
if one focuses on ion kinetic effects at scales
$k_{\perp}\rho_{\mathrm{i}}\sim1$, the large ion-to-electron mass ratio
$m_{\mathrm{i}}/m_{\mathrm{e}} \gg 1$ \emph{may} lead 
to negligible electron kinetic effects in some cases.  In~\cite{SchekochihinCowleyDorland_09}, the isothermal electron fluid
equations are given. Implementation of this alternative treatment of
the electron dynamics in \T{AstroGK} will lead to dramatic
improvements in computational speed for the study of ion kinetic effects.

One important aspect of \T{AstroGK} is its portability. It is designed
to work on a wide range of computing environments, from individual
desktop PCs to petascale supercomputers, and new users can easily port
it to their particular computing environment. It also supports major
FORTRAN 95 compilers. The potential drawback of this emphasis on
portability is that it is not necessarily optimized to any particular
architectures. Further efforts to optimize the code will be made to
enhance both serial and parallel efficiency of the code.


\section*{Acknowledgments}
This work is supported in part by the DOE Center for Multiscale Plasma
Dynamics (Fusion Science Center Cooperative Agreement ER54785), the DOE
Maryland Fusion Theory Research Program (DOE Grant No. DEFG0293-ER54197),
the Leverhulme Trust Network for Magnetised Turbulence in Astrophysical
and Fusion Plasmas, the Aspen Center for Physics, the Wolfgang Pauli Institute
in Vienna, the National Science Foundation through TeraGrid resources
provided by National Institute for Computational Sciences under Grant
No. TG-PHY090084 and by the Texas Advanced Computing Center under
Grant Nos. TG-AST030031N and TG-PHY090080.

\appendix


\section{Symbols, definitions, coordinate, and normalization}
\label{sec:symbols}

In this Appendix, we present complementary explanations of the symbols,
coordinate systems, and normalizations used in
Section~\ref{sec:gkeqns}. Table~\ref{tbl:symbols} lists the symbols and
their explanations, definitions, and normalizations.
\begin{table}
 \begin{center}
 \caption{
  \label{tbl:symbols}
  Symbols in the gyrokinetic-Maxwell equations and their explanations,
  definitions and normalizations. The subscript $s$ signifies the symbol
  is dependent on species, and Boltzmann's constant is absorbed  to give 
  temperature in units of energy.}
 \begin{tabular}{cll} \hline\hline
  Symbol & Explanation & Normalization \\ \hline
  $\bm{B}_0$ & Mean magnetic field & \\
  $\hat{\bm{b}}_0 = \bm{B}_0/B_0$ & Unit vector in mean field direction& \\
  $f_{0s}$ & Background Maxwellian dist. func. & \\
  $h_{s}$ & Non-Maxwellian part of dist. func. & 
	  $\varepsilon \hat{h}_{s} f_{0s}$\\
  $\phi$ & Electrostatic potential
      & $\varepsilon\hat{\phi} (T_{00}/q_{0})$ \\ 
  $A_{\parallel}$ & Parallel component of vector potential
      & $\varepsilon \hat{A}_{\parallel}
	  \left(v_{\mathrm{th}0}T_{00}/q_{0}\right)$ \\
  $\delta B_{\parallel}$ & Parallel component of magnetic field
      & $\varepsilon \delta \hat{B}_{\parallel} \Bg$ \\
  $m_{s}$ & Mass & $\hat{m}_{s} m_{0}$ \\
  $q_{s}$ & Electric charge & $\hat{q}_{s} q_{0}$ \\
  $n_{0s}$ & Density of the background & $\hat{n}_{0s} n_{00}$\\
  $T_{0s}$ & Temperature of the background & $\hat{T}_{0s} T_{00}$ \\
  $\nu_{s}$ & Collision Frequency & $\hat{\nu}_{s}
	  (v_{\mathrm{th},0}/a_{0})$ \\
  $L_{\Bg}$, $L_{n_{0s}}$, $L_{T_{0s}}$& Scales of the background
      & $\hat{L}_{\Bg,n_{0s},T_{0s}} a_{0}$ \\
  $\kappa$ & Curvature of the mean field & $\hat{\kappa}/a_{0}$\\
  $v_{\mathrm{th},s}\equiv\sqrt{2T_{0s}/m_{s}}$
  & Thermal velocity 
      & $\sqrt{\hat{T}_{0s}/\hat{m}_{s}} v_{\mathrm{th}0}$ \\
  $v_{\mathrm{th}0}\equiv\sqrt{2T_{00}/m_{0}}$ &
      Reference thermal velocity & \\
  $\Omega_{s}\equiv |q_{s}|\Bg/m_{s}$
  & Cyclotron frequency 
      & $\left(|\hat{q}_{s}|/\hat{m}_{s}\right) \Omega_{0}$ \\
  $\Omega_{0}\equiv q_{0}\Bg/m_{0}$
  & Reference cyclotron frequency  & \\
  $\rho_{s}\equiv v_{\mathrm{th},s}/\Omega_{s}$
  & Thermal Larmor radius 
      &
	  $\left(\sqrt{\hat{m}_{s}\hat{T}_{0s}}/|\hat{q}_{s}|\right)\rho_{0}$ \\
  $\rho_{0}\equiv\sqrt{2m_{0}T_{00}}/(q_{0}\Bg)$ 
  & Reference thermal Larmor radius & \\
  $d_{s}\equiv\sqrt{m_{s}/(\mu_{0}n_{0s}q_{s}^{2})}$ 
  & Inertial skin depth
      &
	  $\sqrt{\hat{m}_{s}/(\beta_{0}\hat{n}_{0s}\hat{q}_{s}^{2})}\rho_{0}$ \\
  $\beta_{s}\equiv2\mu_{0}n_{0s}T_{0s}/\Bg^{2}$
  & Plasma beta & $\hat{n}_{0s}\hat{T}_{0s}\beta_{0}$ \\
  $\beta_{0}\equiv 2\mu_{0}n_{00}T_{00}/\Bg^{2}$
  & Reference plasma beta & \\
  $\alpha_{s}=k_{\perp}V_{\perp}/\Omega_{s}$ & Argument of
      the Bessel functions \\
  $b_{s}=(k_{\perp}\rho_{s})^{2}/2$ & Argument of
      $\Gamma_{n}$ \\
  $\mu_{0}$ & Vacuum permeability &
	  \\ \hline
 \end{tabular}
 \end{center}
\end{table}

\subsection{Coordinate}
\label{sec:coordinate}

In gyrokinetics, it is convenient to describe dynamics of the
distribution function in the gyro-center coordinate
$(\bm{R}_{s},\bm{V}_{s})$ rather than the particle coordinate
$(\bm{r},\bm{v})$. The following equations:
\begin{equation}
 \bm{R}_{s} = \bm{r} + \frac{\bm{v} \times
  \zhat}{\Omega_{s}}, ~~~
  \bm{V}_{s} = \bm{v},
\end{equation}
define linear transformation of the particle coordinate to the
gyro-center coordinate, called the {\it Catto transform}~\cite{Catto_78}. We consider the Cartesian coordinate in $\bm{r}$ and
$\bm{R}_{s}$:
\begin{align}
 \bm{r} = & x\xhat+y\yhat+z\zhat, &
 \bm{R}_{s} = & X_{s}\Xhat+Y_{s}\Yhat+Z_{s}\Zhat,
\end{align}
where $\xhat,\yhat$ and $\Xhat,\Yhat$ are unit vectors that span the
plane perpendicular to the mean field in $\zhat=\Zhat$.

Derivatives with respect to the position coordinate are equivalent
($\p/\p \bm{r}=\p/\p \bm{R}_{s}$), while those with respect to
the velocity coordinate are different, which appears in the collision
operators (See \ref{sec:collisions} and
Refs.~\cite{AbelBarnesCowley_08,BarnesAbelDorland_09}).

The velocity coordinate is written in two different ways in the text. In
the gyrokinetic equation, we mainly use the polar coordinates $(V_\parallel, V_\perp, \Theta)$: the
relations to the Cartesian components are given by
\begin{align}
 V_{\perp} = & \sqrt{V_{x}^{2}+V_{y}^{2}}, &
 V_{\parallel} = & V_{z}, &
 \tan \Theta = & \frac{V_{y}}{V_{x}},
\end{align}
and $|\bm{V}|=V=\sqrt{V_{\perp}^{2}+V_{\parallel}^{2}}$. It is
convenient to write in energy and pitch-angle coordinates when the
collision operator and velocity space integrals are considered:
\begin{align}
 E= &V_{\perp}^{2}+V_{\parallel}^{2}, &
 \lambda = & \frac{V_{\perp}^{2}}{V^{2}B_0}.
\end{align}

\subsection{Normalization}
\label{sec:normalization}

The GK-M equations solved by \T{AstroGK} are cast into dimensionless
form through the normalization of all quantities with respect to the
parameters of a reference species
(denoted by the subscript $0$), the mean
magnetic field strength $\Bg$, and the parallel length scale $a_0$.

The presence of a mean magnetic field leads to different
characteristic temporal and spatial scales in the parallel and
perpendicular directions. The length scale in the perpendicular plane
is characterized by the thermal Larmor radius of the reference
species, $\rho_{0}$. The ratio of the perpendicular and parallel
scales defines the small expansion parameter
$\varepsilon\equiv\rho_{0}/a_{0}\ll1$ in gyrokinetic theory~\cite{HowesCowleyDorland_06,SchekochihinCowleyDorland_09}.
Denoting  normalized quantities with the
``hat'' symbol, we define the normalized  length scales in perpendicular and
parallel directions by $\hat{k}_{\perp} =  k_{\perp}\rho_{0}$ and
$\hat{k}_{\parallel} =  k_{\parallel} a_{0}$.  The time scale is
normalized by the thermal crossing time of the reference species in
the parallel direction, $\hat{t} = t
/(a_0/v_{\mathrm{th}0})$.

Species dependent quantities retain their species subscript $s$ after
normalization to the reference species, for example mass $\hat{m}_{s}=
m_{s}/m_{0}$. The two-dimensional velocity space of distribution
function $g_s$ employs a species dependent normalization so that
integrations over velocity space remain efficient even when
characteristic thermal velocities of the plasma species differ by a
large factor. The coordinates used in velocity space by \T{AstroGK}
are the energy $E_{s}=(1/2) m_{s} v_{s}^2$ and pitch angle
$\lambda_{s}= v_{\perp,s}^{2}/(v_{s}^{2} \Bg)$, which are related to
the magnetic moment by $\lambda_{s} E_{s}= m_{s}
v_{\perp,s}^{2}/(2\Bg)$. Normalizing the velocity to the species
thermal velocity $\hat{\bm{v}}_{s} = \bm{v}/ v_{\mathrm{th},s}$, the
dimensionless velocity space coordinates are given by $\hat{E}_{s}=
\hat{v}_{s}^{2}$ and
$\hat{\lambda}_{s}=\hat{v}_{\perp,s}^2/\hat{v}_{s}^2$.

The first-order fluctuating quantities in the GK-M equations are the
distribution function for each species $g_{s}$ and the electromagnetic field
variables: the scalar potential $\phi$, the parallel component of the
vector potential $A_\parallel$, and the parallel component of the
magnetic field $\delta B_\parallel$. The distribution function is
normalized by
\begin{equation}
\hat{g}_{s}  = \frac{g_{s}}{ f_{0s}} \frac{a_{0}}{\rho_{0}},
\end{equation}
where
$f_{0s}/f_{00}=\hat{n}_{0s}
\exp\left(-\hat{v}_{s}^2\right)/\left(\pi^{3/2} \hat{v}_{\mathrm{th},s}^{3} \right) $ with
$f_{00}=n_{00}/v_{\mathrm{th}0}^{3}$. The fields are normalized by
\begin{align}
\hat{\phi}= & \frac{a_{0}}{\rho_{0}} \frac{q_{0}\phi}{T_{00}},& 
  \hat{A}_{\parallel} = & \frac{a_{0}}{\rho_{0}} v_{\mathrm{th}0}
 \frac{q_{0}A_{\parallel}}{T_{00}}, &
\delta \hat{B}_{\parallel}=& \frac{a_{0}}{\rho_{0}} \frac{\delta B_{\parallel}}{\Bg} .
\end{align}
Notice that the dimensionless normalizations for all fluctuating,
first-order quantities are also multiplied by a factor $a_0/\rho_0$ so
that all normalized terms have unity order of magnitude.

An illustration of the normalization of velocity space integrals in
Maxwell's equations follows for the integral $\int g_{s} \diff
\bm{v}$. To normalize the integral, we multiply by
$1/(v_{\mathrm{th}0}^3 f_{00})(a_{0}/\rho_{0})$ to obtain $\hat{n}_{0s}
\int \left(e^{-\hat{v}_{s}^{2}}/\pi^{3/2}\right) \hat{g}_{s} \diff
\hat{\bm{v}}_{s}$.


\section{Model collision operator}
\label{sec:collisions}

In this Appendix, we present the model collision operator employed
in \texttt{AstroGK}.  This complements the overview of the numerical
implementation given in Section~\ref{sec:alg_coll}.

\texttt{AstroGK} uses the model Fokker--Planck collision operator given in~\cite{AbelBarnesCowley_08,BarnesAbelDorland_09},
which includes the effects of pitch-angle scattering and energy diffusion while satisfying Boltzmann's $H$-Theorem
and conserving particle number, momentum, and energy.  Upon gyro-averaging, the same-species collision operator
is written in the spectral representation as
\begin{equation}
 {\mathcal C}_{k_{\perp}}(h_{\bm{k}_{\perp}}) =
  {\mathcal C}_{\mathrm{L}}(h_{\bm{k}_{\perp}}) +
  {\mathcal C}_{\mathrm{D}}(h_{\bm{k}_{\perp}}) +
  {\mathcal U}_{\mathrm{L}}(h_{\bm{k}_{\perp}}) +
  {\mathcal U}_{\mathrm{D}}(h_{\bm{k}_{\perp}}),
\end{equation}
where
\begin{equation}
 {\mathcal C}_{\mathrm{L}}(h_{\bm{k}_{\perp}}) =
  \frac{\nu_{\mathrm{D}}(v/v_{\mathrm{th}})}{2}
  \left(
   \pdf{}{\xi} \left(1-\xi^{2}\right)\pdf{h_{\bm{k}_{\perp}}}{\xi}
   - \frac{k_{\perp}^{2}v^{2}}{2\Omega^{2}}
   \left(1+\xi^{2}\right)h_{\bm{k}_{\perp}}\right),
\end{equation}
and
\begin{equation}
 {\mathcal C}_{\mathrm{D}}(h_{\bm{k}_{\perp}}) =
  \frac{1}{2v^{2}}\pdf{}{v}
  \left(\nu_{\parallel}(v/v_{\mathrm{th}})v^{4} f_0
   \pdf{}{v}\frac{h_{\bm{k}_{\perp}}}{f_0}\right) - 
  \nu_{\parallel}(v/v_{\mathrm{th}})\frac{k_{\perp}^{2} v^{2}}{4\Omega^{2}}
  \left(1-\xi^{2}\right)h_{\bm{k}_{\perp}},
\end{equation}
are the gyro-averaged Lorentz and energy diffusion operators,
respectively. Together, these form the exact test-particle piece of the
linearized Landau operator. The velocity-dependent collision
frequencies $\nu_{\mathrm{D}}$ and $\nu_{\parallel}$ are given by
\begin{align}
\nu_{\mathrm{D}}(x) &= \nu \frac{\Phi(x)-G(x)}{x^3}, &
\nu_{\parallel}(x) &= \frac{2\nu G(x)}{x^3},
\end{align}
with $\Phi(x)=(2/\sqrt{\pi})\int_{0}^{x} \exp(-y^{2})\diff y$ the error
function, $G(x)=\left(\Phi(x)-x \diff\Phi/\diff x\right)/(2x^{2})$ the
Chandrasekhar function, and $\nu=\sqrt{2}\pi n_0 q^4 \ln
\Lambda/\left(m^{1/2}T_0^{3/2}\right)$ the same-species collision
frequency, which is an input parameter.

The test-particle operator given above does not conserve particle
momentum and energy, so the additional terms ${\mathcal U}_{\mathrm{L}}$
and ${\mathcal U}_{\mathrm{D}}$ are added to recover conservation
properties.  Care is taken in choosing the form of these conservation
terms so that Boltzmann's $H$-Theorem is respected.  With these
constraints, one obtains:
\begin{equation}
 {\mathcal U}_{\mathrm{L}}(h_{\bm{k}_{\perp}}) =
  \nu_{\mathrm{D}} f_{0}
  \left(
   J_{0}(\alpha) v_{\parallel}
   \frac{\int \nu_{\mathrm{D}} v_{\parallel} J_0(\alpha)
   h_{\bm{k}_{\perp}} \diff \bm{v}}
   {\int \nu_{\mathrm{D}} v_{\parallel}^{2} f_{0} \diff \bm{v}}
   + J_{1}(\alpha) v_{\perp}
   \frac{\int \nu_{\mathrm{D}} v_{\perp} J_{1}(\alpha)
   h_{\bm{k}_{\perp}} \diff \bm{v}}
   {\int \nu_{\mathrm{D}} v_{\parallel}^{2} f_{0} \diff \bm{v}}\right),
\end{equation}
and
\begin{align}
 {\mathcal U}_{\mathrm{D}}(h_{\bm{k}_{\perp}}) = &
 - \Delta \nu f_{0} 
 \left(
 J_{0}(\alpha) v_{\parallel}
 \frac{\int \Delta\nu v_{\parallel} J_{0}(\alpha)
 h_{\bm{k}_{\perp}} \diff \bm{v}}
 {\int \Delta\nu v_{\parallel}^{2} f_{0} \diff \bm{v}}
 + J_{1}(\alpha) v_{\perp}
 \frac{\int \Delta\nu v_{\perp} J_{1}(\alpha)
 h_{\bm{k}_{\perp}} \diff \bm{v}}
 {\int \Delta\nu v_{\parallel}^{2} f_{0} \diff \bm{v}}
 \right)
 \nonumber \\
 & + \nu_{\mathrm{E}} v^{2} J_{0}(\alpha) f_{0}
 \frac{\int\nu_{\mathrm{E}} v^{2} 
 J_{0}(\alpha) h_{\bm{k}_{\perp}} \diff \bm{v}}
 {\int \nu_{\mathrm{E}} v^{4} f_{0} \diff \bm{v}},
\end{align}
where the additional collision frequencies $\Delta\nu$ and
$\nu_{\mathrm{E}}$ are defined as
\begin{align}
\Delta\nu &= \nu_{\mathrm{D}} - 2 (v/v_{\mathrm{th}})^{2} \nu_{\parallel},\\
\nu_{\mathrm{E}} &= - \left(\nu_{\parallel}+2\Delta\nu\right).
\end{align}
The terms ${\mathcal U}_{\mathrm{L}}$ and ${\mathcal U}_{\mathrm{D}}$
are treated separately in \texttt{AstroGK} so that the Lorentz and
energy diffusion operators can be split with the conservation properties
and $H$-Theorem maintained within each splitting. When combined, these
conserving terms constitute an approximation to the field-particle piece
of the linearized Landau operator.

The effect of ion--electron collisions are neglected in \texttt{AstroGK} because they are small in the electron--ion
mass ratio.  However, electron--ion collisions are comparable in size to same-species collisions, so they are retained.
Consequently, the electron collision operator has the following additional term:
\begin{equation}
 {\mathcal C}^{\mathrm{ei}}(h_{\bm{k}_{\perp},\mathrm{e}}) =
  {\mathcal C}_{\mathrm{L}}^{\mathrm{ei}}(h_{\bm{k}_{\perp},\mathrm{e}})
  +
  \nu_{\mathrm{D}}^{\mathrm{ei}}
  \frac{2 v_{\parallel} u_{\parallel,\mathrm{i}}}
  {v_{\mathrm{th},\mathrm{e}}^{2}} J_{0}(\alpha_{\mathrm{e}}) f_{0\mathrm{e}},
\end{equation}
where $u_{\parallel,{\mathrm{i}}}$ is the perturbed ion parallel flow
velocity, and ${\mathcal C}_{\mathrm{L}}^{\mathrm{ei}}$ and
$\nu_{\mathrm{D}}^{\mathrm{ei}}$ are obtained from their same-species
counterparts by replacing the same-species collision frequency $\nu$ 
with the inter-species collision frequency.




\section{Laplace--Fourier solution for driven gyrokinetics}
\label{sec:laplace_fourier_solution}



We derive a Laplace--Fourier solution for time dependence of the
 $A_{\parallel}$ amplitude for a driven gyrokinetic system in this 
 Appendix. 
 The solution presented here is for linear,
 collisionless gyrokinetics without $\delta B_\parallel$, the nonlinear 
 term, and the linear driving terms,
and with an external driving force given by
\begin{equation}
 A_{\parallel}^{\mathrm{antenna}} =
  \left\{
   \begin{matrix}
    A_{\parallel0} e^{-\imag (\omega_{0} t - \bm{k}_{0}\cdot\bm{x})}
    & t \ge 0 \\
    0 & t < 0
   \end{matrix}
  \right.
  .
\end{equation}
We also set
$n_{0\mathrm{i}}/n_{0\mathrm{e}}=-q_{\mathrm{i}}/q_{\mathrm{e}}=1$ for
simplicity.
Given the driving parameters used in the code, the solution should agree
with a result obtained from the code without free parameters.

Performing a Laplace transform in time and a Fourier transform in
space on this system of equations, we can solve for the
Laplace--Fourier transformed distribution function for the driven
Fourier mode (without the species index $s$):
\begin{equation}
 \tilde{g}_{\bm{k}_{\perp0}} =
  \frac{g_{\bm{k}_{\perp0}}(0)}{p+\imag k_{\parallel0} V_{\parallel} }
  - \frac{q f_{0}}{T_{0}} J_{0}(\alpha_{0})
  \left[
   \frac{\imag k_{\parallel0} V_{\parallel}
   \tilde{\phi}_{\bm{k}_{\perp0}}}
   {p+\imag k_{\parallel0}V_{\parallel}}
   + \frac{V_{\parallel}
   \left(
    p \tilde{A}_{\parallel,\bm{k}_{\perp0}}
    -A_{\parallel,\bm{k}_{\perp0}}(0)\right)
  }
  {p+\imag k_{\parallel0}V_{\parallel}}
 \right]
\end{equation}
where $\alpha_{0}=k_{\perp0}V_{\perp}/\Omega$.
Setting the initial conditions to zero, $g_{\bm{k}_{\perp0}}(0) =
A_{\parallel,\bm{k}_{\perp0}}(0)=0$, and
substituting into Maxwell's equations, we obtain:
\begin{equation}
 \begin{pmatrix}
  H & K \\ H-K & b_{\mathrm{i0}}/\overline{\omega}^{2}
 \end{pmatrix}
 \begin{pmatrix}
  \overline{E}_{\parallel} \\ \overline{A}_{\parallel}
 \end{pmatrix}
 = 
 \begin{pmatrix}
  0 \\ - \left(b_{\mathrm{i0}}/\overline{\omega}^{2}\right)
  \overline{S}
 \end{pmatrix}
\end{equation}
where we have used the following definitions to simplify the notation:
\begin{align}
 \overline{E}_{\parallel} = & \tilde{\phi}_{\bm{k}_{\perp0}}
 - \frac{\imag p \tilde{A}_{\parallel,\bm{k}_{\perp0}}}{k_{\parallel0}},
 & 
 \overline{A}_{\parallel} = & 
 \frac{\imag p \tilde{A}_{\parallel,\bm{k}_{\perp0}}}{k_{\parallel0}},
 \\
 H = & \sum_{s} \frac{T_{0\mathrm{i}}}{T_{0s}}
 \left(1+\Gamma_{0s} \zeta_{s} \Xi(\zeta_{s})\right), 
 &
 K = & \sum_{s} \frac{T_{0\mathrm{i}}}{T_{0s}}
 \left(1-\Gamma_{0s}\right),
 \\
 \overline{S} = & \frac{\imag p A_{\parallel0}/k_{\parallel0}}{p+\imag\omega_{0}},
 & 
 \overline{\omega} = & \frac{\imag p}{k_{\parallel0}v_{\mathrm{A}}},
\end{align}
$b_{\mathrm{i}0}=(k_{\perp0}\rho_{\mathrm{i}})^{2}/2$,
$v_{\mathrm{A}}=B_{0}/\sqrt{\mu_{0}n_{0\mathrm{i}}m_{\mathrm{i}}}$,
$\zeta_{s}=\imag p/(k_{\parallel0}v_{\mathrm{th},s})$, and $\Xi$ is the
plasma dispersion function~\cite{FriedConte_61}.
The dispersion relation:
\begin{equation}
 p^{2}+\left[Q(p)\right]^{2} =
  p^{2} + \frac{b_{\mathrm{i}0} H k_{\parallel0}^{2}
  v_{\mathrm{A}}^{2} }{HK-K^2}=0
  \label{eq:drbpar0}
\end{equation}
leads to the Alfv\'en wave solutions.

Now we will focus on the solution for
 $A_{\parallel,\bm{k}_{\perp0}}(t)$.
The Laplace--Fourier solution is
\begin{equation}
\tilde{A}_{\parallel,\bm{k}_{\perp0}}(p)=
 -\frac{\left[Q(p)\right]^{2}
 A_{\parallel0}}{(p^{2}+\left[Q(p)\right]^{2})
 (p + \imag\omega_{0})}.
\end{equation}
To proceed further, we make the approximation:
\begin{equation}
 p^{2} + Q^{2} \simeq (p+\imag\omega_{1})(p+\imag\omega_{2}),
\end{equation}
where $\omega_{1,2}$ are the complex eigenfrequencies independent of
$p$; we know for this system these solutions typically have the form
$\omega_{1}=\omega_{\mathrm{r}}+\imag\omega_{\mathrm{i}}$,
$\omega_{2}=-\omega_{\mathrm{r}}+\imag\omega_{\mathrm{i}}$ with a
negative growth rate $\omega_{\mathrm{i}}<0$.
With this simplification, the inverse Laplace transform is easily found
by application of the residue theorem to find:
\begin{equation}
 \frac{A_{\parallel,\bm{k}_{\perp0}}(t)}{A_{\parallel0}}
  = 
  \frac{\left[Q(-\imag\omega_{0})\right]^{2} e^{-\imag\omega_{0}t}} 
  {(\omega_{1}-\omega_{0})(\omega_{2}-\omega_{0})}
  +\frac{\left[Q(-\imag\omega_{1})\right]^{2} e^{-\imag\omega_{1}t}} 
  {(\omega_{0}-\omega_{1})(\omega_{2}-\omega_{1})}
  +\frac{\left[Q(-\imag\omega_{2})\right]^{2} e^{-\imag\omega_{2}t}} 
  {(\omega_{0}-\omega_{2})(\omega_{1}-\omega_{2})}.
  \label{eq:lfgk}
\end{equation}
Note that the second and third term will decay with time. This solution
is computed numerically for comparison to code results, as presented in
Section~\ref{sec:ex_lft}.

\bibliography{abbrev,astrogk}
\providecommand{\noopsort}[1]{}

\end{document}